\documentclass[
 amssymb, amsmath,
 reprint,%
aps, prl,
superscriptaddress,%
floatfix, showkeys
]{revtex4-2}

\pdfoutput=1

\usepackage{graphicx}
\graphicspath{{images/}}

\usepackage{amssymb}
\usepackage{amsthm}
\usepackage{physics}
\usepackage{bm}

\usepackage[T1]{fontenc} 
\usepackage[utf8]{inputenc}
\usepackage{lmodern}

\usepackage{subcaption}
\usepackage{multirow}
\usepackage{etoolbox}
\usepackage{booktabs}
\usepackage{multirow}
\usepackage[colorlinks=true,linkcolor=blue]{hyperref}

\usepackage{siunitx}
\usepackage{chemgreek}
\usepackage{chemmacros}

\DeclareSIUnit{\cal}{cal}
\DeclareSIUnit{\kcal}{\kilo\cal\per\mol}
\DeclareSIUnit{\atm}{atm}
\DeclareSIUnit{\debye}{debye}
\DeclareSIUnit{\angstrom}{\text{\AA}}

\chemsetup{
 formula = mhchem,
 greek   = upgreek,
 modules = thermodynamics,
 modules = redox,
 modules = reactions
}

\newcommand{\etal}{\textit{et al.}}
\newcommand{\expe}{Expt.}

\begin{document}

\title{Structural and thermodynamics properties of pure phase alkanes, monoamides and alkane/monoamide mixtures with an \textit{ab initio} based force-field model}

\author{Abdelmounaim Failali}
\affiliation{CEA, DES, ISEC, DMRC, Univ Montpellier, Marcoule, F-30207, Bagnols-sur-Ceze Cedex, France}
\affiliation{Univ. Lille CNRS, UMR 8523 -- PhLAM -- Physique des Lasers Atomes et Molecules, F-59000, Lille, France}

\author{Eléonor Acher}
\affiliation{CEA, DES, ISEC, DMRC, Univ Montpellier, Marcoule, F-30207, Bagnols-sur-Ceze Cedex, France}

\author{Dominique Guillaumont}
\affiliation{CEA, DES, ISEC, DMRC, Univ Montpellier, Marcoule, F-30207, Bagnols-sur-Ceze Cedex, France}

\author{Valérie Vallet}
\affiliation{Univ. Lille CNRS, UMR 8523 -- PhLAM -- Physique des Lasers Atomes et Molecules, F-59000, Lille, France}

\author{Florent Réal}
\email{florent.real@univ-lille.fr}
\affiliation{Univ. Lille CNRS, UMR 8523 -- PhLAM -- Physique des Lasers Atomes et Molecules, F-59000, Lille, France}
\date{Received: \today}

\begin{abstract}
A polarizable force-field (FF) model for short- and long-alkane chains and amide derivatives was constructed based solely on accurate quantum chemical (QC) calculations. First, the FF model accuracy was accessed by performing molecular dynamics (MD) simulations to calculate liquid-phase thermodynamic and structural properties for alkanes, for which  experimental data are available. Second, The FF was then used to perform molecular dynamics simulations to calculate thermodynamic, structural and excess properties of monoamide/dodecane mixtures, namely DEHiBA/dodecane and DEHBA/dodecane. Aggregation phenomena appear for both types of mixtures and monoamide pure phases. A detailed structural analysis revealed, at small monoamide mole fraction the formation of dimers, while trimerization at larger monoamide concentrations and in their pure phases. Analysis of the relative orientation of the dimers have also been performed and showed a small difference for both phases.
\end{abstract}

\maketitle







\section{Introduction}
Alkane and amide derivatives molecules are used and quite important in many research areas, such as biology, medicine, nuclear field and petrology. In these latter, the structural, thermodynamics and transport properties are of interest for the design and study of artificial or biological membranes and also play a major role in the recovery and refining of crude oil~\cite{papavasileiou2019molecular,goodsaid1981,yeagle2004membrenes,murgich2003molecular,chunming2004applications,ungerer2006applications,chen2020transport,petrov2012petroleum}.

In the nuclear field, amide derivatives can be used as extracting molecules for actinide ions for nuclear fuel solutions as alternatives to the TBP molecules, with advantages highlighted in literature~\cite{prabhu1997,recycling}. Alkanes such as dodecane, TPH (Tetra Propylene Hydrogenated), kerosene, and so forth~\cite{schulz1987science,U-Pu-extraction-Amides} are hydrocarbons suitable as solvents to dilute TBP and/or amide derivatives and to achieve  the extraction of U and Pu from the irradiated nuclear fuel. 

One way to obtain insights into the properties of such solutions (alkanes + amides derivatives) at the molecular level,  involves the use of atomistic molecular dynamics simulations. In this regard, several force fields have been developed over the years, such as CHARMM, GROMOS, MM4, OPLS and AMBER~\cite{gromos,MM4,charmm,amber,opls}. Generally, the hydrocarbons (alkanes) parameters are used to describe alkyl chains regardless of whether they are amide, acid, amine chains or lipid, peptide or some protein tails in biology applications. Most of these force-fields (FFs) have been derived focusing on individual aspects and based on the reproduction of experimental data, such as enthalpies of vaporization, vapor pressures and densities. However, the FF parameters are not always transferable to all the molecular group series, and they are reoptimized before use in most cases. As a result, different versions of AMBER, CHARMM, OPLS FFs for hydrocarbons have been developed over the years for a better description of the macroscopic properties~\cite{Charmm2005polarizable,OPLS2012,amber2014new}. For example, the first version of OPLS FF for hydrocarbons (labeled OPLS-AA) is quite successful for short hydrocarbons but not for long~\cite{opls}, some studies in literature have reported the gel-phase formation at room temperature for long-chain hydrocarbons of more than eight carbons (including dodecane) using the OPLS-AA force field~\cite{vo2015computational,opls2013effect}. This comes from the fact that they used experimental data measured for small molecules to develop such FF, and hence inducing a lack of physical meaning of the interactions for longer chains. Therefore, Siu~{\etal} have reoptimized the OPLS-AA parameter set for long hydrocarbons, termed L-OPLS~\cite{OPLS2012}. Generally, FF parameters obtained for small molecules do not always represent the real interactions between atoms, since the goal was to reproduce the reference experimental data in an average way (optimized empirically to match liquid properties). In summary, in the literature exists a zoo of FFs, the latter are mostly derived for individual molecules and specific cases, which makes the prediction of new features of newly designed molecules challenging and questionable.

\begin{figure}
    \centering
    \includegraphics[width=\linewidth]{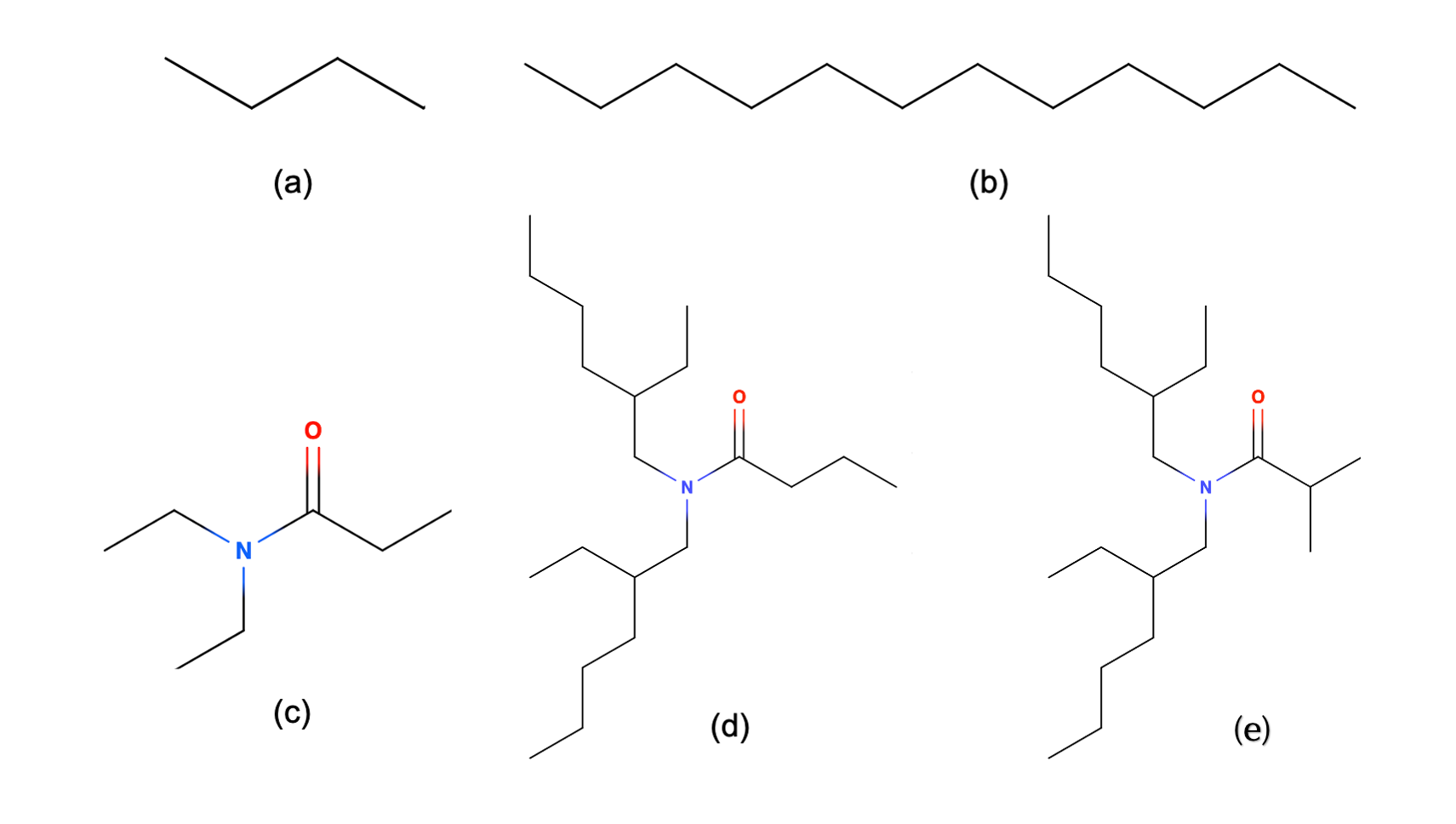}
    \caption{Schematic representations of (b) butane, (b) dodecane, (c) N,N-Diethylpropionamide (DEPA), (d) N,N-di-2-ethylhexyl-butyramide (DEHBA) and (e) di-2-ethylhexylisobutyramide (DEHiBA) molecules.}
    \label{fig:molecules}
\end{figure}

In the context of nuclear fuel reprocessing, the most studied extractant are phosphorus molecules given their huge application in the field; more precisely, the TBP molecule. Over the years, many FFs have been developed (polarizable and non polarizable) to model  the behavior of such molecule whether in binary mixtures with alkanes or in full phases with uranyl and nitrates ions~\cite{servis2018role,cui2014molecular,mu2016comparativeTBP}. As for amide derivatives (shown in Figure~\ref{fig:molecules}), most available FFs have been derived for small molecules and then combined with the hydrocarbons parameters to describe the alkyl chains of the long amides~\cite{paquet2019aggregation,qiao2015molecular}. However, as in the case of hydrocarbons, there are several FFs available in the literature but they are not always transferable and accurate enough for direct use. Hence the re-optimization of the FF parameters is often required~\cite{amides1,amides2,amides3}. For amide derivatives, it was shown that the transferability problems may arise \textit{i)} from the fact that amides are polar molecules and thus FFs describing this kind of molecules should include these effects explicitly~\cite{amides3} \textit{ii)} and, of course, from the parameters used to describe the alkyl chains. 

Nowadays, the use of \textit{ab initio} data to parameterize FFs has become increasingly common, since a solid physical/mathematical foundation provides a better understanding of the physics and chemistry of the systems to be investigated. Moreover, most transferability problems encountered for force fields with empirical parameters are related to the reference data used. If the systems used in the parameterization process significantly differ from the ones being investigated and/or if the amount of data set used for parameterization is small, limited to certain kinds of data at restricted temperature and pressure conditions, the parameters may not be as trustworthy. The available experimental data are often limited to certain molecules at specific experimental conditions, while, quantum chemistry methods in combination with the availability of significant computational resources offer the possibility to generate a large number of quantum chemical reference data essential to the development of force fields (atomic/molecular data, dissociation curves, interaction energies, etc.). 

In  order  to  be  confident that the behavior observed in molecular dynamics (MD) simulations is representative of real dynamical systems, the selection of an accurate force field is essential. Hence, our group is working on the development of a new class of polarizable ab initio-based force fields with the right balance between accuracy and efficiency~\cite{tcpe,real,md-Houriez-JCP2019-151-174504}. Herein, we have developed a new set of FF for alkanes and amide derivatives based solely on quantum chemistry calculations. Since alkanes are known to be a non-polar molecules, polarization effects were neglected and only Coulomb, repulsion and dispersion interactions were considered. However, it should be mentioned that the alkanes do not induce polarization but that they are polarizable in the model. As for the amides, polar molecules for which  polarization forces play an important role, polarization was incorporated alongside Coulomb, repulsion and dispersion interactions to ensure the transferability of the parameters for longer amides, as well as the correct description of the intermolecular interactions.

First, parameters for alkane and amide model molecules have been derived using \textit{ab initio} quantum chemical data. Then, we have combined the two sets of parameters to describe the large amides, for which the alkyl chains have been described with the derived alkane parameters. The newly proposed parameter sets were validated on physical properties of interest, namely density, heat of vaporization as well as on the distributions of trans and gauche conformation for alkanes.The simulations on the alkane/monoamide mixtures allow us to predict the densities and excess quantities that, for the density and excess enthalpy can be compared to very recent experimental data~\cite{COQUIL2021}.

\section{Computational and simulation details}
\subsection{Force-field potential model}
The total potential energy $\Delta U$ corresponding to the force field presented in this work describes the intra and intermolecular in alkanes and amide derivatives. It is a sum of six energy components:
\begin{equation} \label{eq1}
     \Delta U = U_{bonds} +U_{angles}+U_{dihedrals} + U_{qq'}+U_{Buck}+U_{pol},
\end{equation}
corresponding to bond, angle, torsion, electrostatic, Buckingham and polarization potentials, respectively. 

The covalent interactions between atoms are modeled using harmonic bond stretching ($U_{bonds}$) and angle bending ($U_{angles}$) terms, while rotations around a bond are described by anharmonic 4-body torsional terms ($U_{dihedrals}$). These three terms are developed in the following equations:
\begin{equation}
\begin{aligned}
U_{bonds} & = \sum_{bonds}\frac{1}{2} {K_{r}}(r- {r_{0}})^2,\\
\end{aligned}
\end{equation}

\begin{equation}
\begin{aligned}
U_{angles} &= \sum_{angles} \frac{1}{2}  {K_{\theta}}(\theta- {\theta_{0}})^2, \\
\end{aligned}
\end{equation}

\begin{equation}
\begin{aligned}
 U_{dihedrals} & = \sum_{dihedrals}\frac{1}{2}\sum_{n}  {K_{n}}(1+cos(n\phi+ {\phi_{0}})).
\end{aligned}
\end{equation}

The bond length, equilibrium bond length, and the bond stretching constants are given as $r$, $r_0$, and $K_r$, respectively. The bond angle $\theta$ has its parameters defined similarly. The dihedral potential for every dihedral angle $\phi $ is a sum over a number of $n$ Fourier components. Each $n$ value is associated with an amplitude, $K_n$, and a phase shift $\phi_0$.  As for the 1-4 interactions, in our MD code, instead of using a scaling factor, they are treated with a correcting potential ($ {a_{ij}}\exp\left(- {b_{ij}}r_{ij}\right)$),  were $a_{ij}$ and  $b_{ij} $ are adjustable parameters (parameters are available in the ESI Table S1.).

The set of bonds, angles and dihedral parameters were obtained by minimizing the difference of the relative total potential energies calculated with the force field with respect to \textit{ab initio} calculations. The comparisons of energy profiles obtained from \textit{ab initio} and force-field calculations are shown in Figure~\ref{fig:dih} and in the ESI Figure S1 for the longer chain n-dodecane.

The non-bonded interactions are described by a sum of $i)$ Coulombic interactions between atom-centered point charges, $U_{qq'}$, $ii)$ a Buckingham potential accounting for repulsion and dispersion interactions, $U_{Buck}$ and lastly $iii)$ a polarization term.

For a system of N atoms, the additive terms $U_{qq'}$ and $U_{Buck}$ are defined as:
\begin{align}
U_{qq'} + & U_{Buck} =  \\
&\sum_{i=1}^{N}{\sum_{j,j>i}^{N}{\left[\frac{ {q_iq_j} }{4\pi\epsilon_0r_{ij}} + {A_{ij}}\exp\left(- {B_{ij}}r_{ij}\right)-\frac{ {C_{ij}}}{r_{ij}^6}\right]}},
\end{align}
where $r_{ij}$ is the distance between atoms $i$ and $j$, ${q_i}$ are the static point charges located on the atomic centers and obtained by QC calculations (See section~\ref{subsection:QCdata}). $A_{ij}$, $B_{ij}$ and $C_{ij}$ are adjustable parameters.

Lastly, polarization effects are incorporated with the induced dipole model, described by means of a set of induced dipole moments $\left\{\vb{\bm{\mu_i}}\right\}_{i=1,N_{\mu}}$ located on a subset of $N_{\mu}$ polarizable atomic centers. They obey 
\begin{equation}	
\vb{\bm{\mu_i}}  = {\alpha_i}\left( \vb{E_{i}^{q}}+\sum_{j= 1, j\neq i}^{N_{\mu}}\mathrm{T}_{ij}\vb{\bm{\mu_j}}\right).
\end{equation}
Here, $\alpha_i$ is the isotropic polarizability of the polarizable atom $i$, computed as described in Section~\ref{sec:FF-dev}-\ref{subsection:QCdata}. $\vb{E_{i}^{q}}$ is the electric field generated on $i$ by the surrounding static charges $q_j$, and $\mathrm{T}_{ij}$ is the dipolar tensor. They both include short-range Thole's like damping functions~\cite{md-Masella-JCP2003-119-1866,thole} with an adjustable damping parameter $a_{ij}$, as described in further details the ESI.

For most of the FFs available in the literature, the latter polarization contribution is neglected for the alkanes. However, for the amide derivatives, since they are polar molecules, it should be explicitly included to ensure the transferability of the FF parameters in the series of amide derivatives. In this work, all the $N_{\mu}$ non-hydrogen atoms are considered as polarizable centers, i.e., a single point polarizability is located on each non-hydrogen atomic center.

All the alkane/amide derivatives force-field parameters optimized with the procedures described in the coming section are listed in Table~S1 of the ESI.

\subsection{Force-field development}\label{sec:FF-dev}

The development of any new FFs consists on three main steps, besides the choice of a physically meaningful functional form for the potential energy. First, the choice and preparation of reference data, second, the adjustment of parameters to reference data and last the validation of the FF by computing macroscopic physical properties. Herein, all FF parameters have been derived based solely on QC calculations.

\subsubsection{QC calculations for reference data\label{subsection:QCdata}}

Before going any further, the choice of the QC level to calculate the reference data used for the development of any FF is fundamental; it has to ensure that the different interactions are accurately treated. For instance, for the alkanes, the accurate description of the interactions impacts the computed heats of vaporization that are expected to come out within chemical error, \SI{1}{\kcal}, of the experimental data. Moreover, the dihedral potentials are crucial for reproducing the (temperature-dependent) fractions of trans and gauche isomers, as the energy difference between the two conformations is small (about \SI{0.6}{\kcal}) and must be accurately described for alkanes (See Figure~\ref{fig:dih}).

In this work, all the structural and energetic data needed for the parametrization of our force field, were calculated using the Molpro quantum chemistry package~\cite{molpro} at the MP2 level (the M{\o}ller-Plesset Perturbation Theory)~\cite{mp2} with correlation consistent aug-cc-pVTZ basis sets by Dunning~\cite{dunning1989gaussian,avtz}. Only the 1s core orbitals of C, N and O were kept frozen. This level of theory was chosen based on two arguments (1) the geometries of simple alkanes are known to be less sensitive to the size of the basis set than the energies themselves~\cite{klauda2005ab} (2) by performing a benchmarking of MP2 interaction energies with respect to Coupled Cluster "CCSD(T)"~\cite{ccsdt} "the gold standard" (see Supporting Information). In conclusion, the MP2 approach proved to provide a good compromise between the computational cost and the accuracy of the computed interaction energies. In the present work, electrostatic parameters and atomic polarizabilities were determined in a first stage, and the Buckingham parameters were optimized in a second stage (i.e., the electrostatic parameters were kept fixed while optimizing the Buckingham parameters).

Partial charges are calculated with CM5 method (Charge Model 5, latest update of the CMx series), a method developed by Cramer, Truhlar, and co-workers~\cite{cm5}. It uses the charges obtained from a Hirshfeld population analysis (of a wave function obtained with density functional calculation) as a starting point. The charges are then varied based on some specific parameters, derived originally by fitting to gas-phase dipole moments of several molecular structures. Jorgensen~{\etal}, developers of the OPLS-AA FF series~\cite{evcm5}, stated that CM5 charges yielded the best agreement with experiments in pure liquid simulations, with the extra advantage of being essentially basis set independent~\cite{cm5}.

As for atomic polarizabilities, in literature, there are several procedures to decompose the molecular polarizability into atomic polarizabilities~\cite{atomic-polar}. They differ on whether molecular polarizability was obtained from experimental refractive indices or from QC. In this work, we have opted for the method proposed by Marenich~{\etal}~\cite{atom-polar} for partitioning the molecular polarizability into atomic contributions by the use of Hirshfeld population analysis~\cite{hirshfeld1977bonded}, involving the numerical differentiation of the dipole moments computed for different values of the applied external electric field.

Lastly, the repulsion and van der Waals dispersion parameters have been derived through a systematic potential energy surface exploration, we produce a large set of reference data, interactions energies for several different dimer relative orientations and distances, starting from ethane to $n-$dodecane for alkanes and from N,N-DiEthyl-PropanAmide (DEPA) and N,N-diethyl-2-methyl-PropAmide DEMPA, plus dimers of alkane-amide molecules (see Figure~S2-S3-S4 in ESI). We have adjusted the FF parameters around the minimum and the repulsion wall up to \SI{15}{\kcal}. A total of 57~relative orientations and conformations were considered. For the different dimer orientations, the MP2 interaction energies ($IE$) were computed using the super-molecule approach taking into account the basis set superposition error (BSSE) with the counterpoise method~\cite{basis-Boys-MP1970-19-553}: 
\begin{equation}
\label{super-molecule}
    IE = E_{AB}(AB) - E_{AB}(A) - E_{AB}(B),
\end{equation}
where, $E_{AB}(AB)$ is the energy of their interacting assembly (dimer). $E_{AB}(A)$ and $E_{AB}(B)$ denote the total energies of monomers A and B, respectively, computed with the dimer AB basis sets.

\subsubsection{FF parameterization procedure}

The parameterization strategy used to derive the FF parameters for both the bonded and non-bonded potentials for alkanes and amides, is based solely on quantum chemistry calculations, often labeled "Bottom-Up", meaning that the parameters are derived to match the computed atomic/molecular-scale data (atomic/molecular data, dissociation curves, interaction energies, etc.).  

\begin{table}[ht!]
\centering
\caption{Atom types and their definitions in the present work}
\begin{tabular}{c|c|c}
 \toprule
No.  & Atom Type &         Description          \\
\midrule
1   & CT3       &  Alkane methyl C  \ce{-C-\textbf{C}H3}  \\
2   & CT2       &  Alkane methylene C  \ce{-C-\textbf{C}H2-C}               \\
3   & CT1       &  Alkane branched C   \ce{-C-\textbf{C}H-C}                \\
4   & CT2-N     &  Methylene C  bonded to amide N      \\
5   & HN        &  H atom bonded to CT2-N                   \\
6   & HA        &  Aliphatic H atom (for alkane and amides)       \\
7   & N         &  Amide N atom             \\
8   & O         &  Amide carbonyl O atom         \\
9   & C         &  Amide carbonyl C atom        \\
 \bottomrule
 \end{tabular}
\label{tab:atom_type}
\end{table}

\begin{figure}
    \centering
\includegraphics[width=0.5\linewidth]{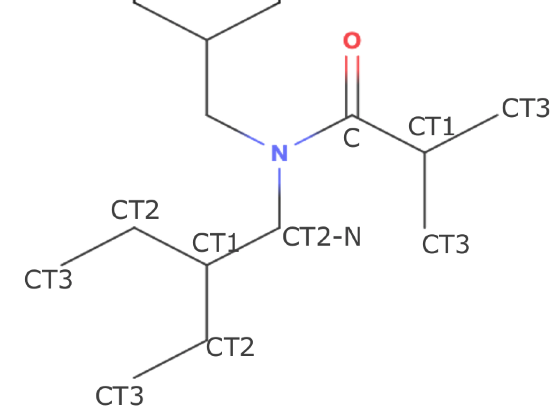}    
\caption{Illustration of atom type used in the present work}
\label{fig:atom_type}
\end{figure}

Table~\ref{tab:atom_type} summarizes the atom types used in this work (See also Figure~\ref{fig:atom_type}). We define HA as the hydrogens in the aliphatic compounds and HN the hydrogen atoms bonded to the specific carbon atom CT2-N. For carbons, we introduce four types: methyl carbon (CT3), methylene carbon (CT2), alkane branched carbon (CT1) and (CT2-N) for methylene carbon bonded to amide nitrogen atom. The oxygen of the carbonyl group is denoted \ce{O}, the carbon atom double-bonded to an oxygen atom (C), the nitrogen atom of the amide function (N).

In most available FFs in the literature, bond stretching and angle bending are described using harmonic potentials that contain the bond and angle force constants and equilibrium bond lengths and angles. These equilibria are taken from the optimized geometries of the molecules, and the motion force constants are derived by performing scans over the bond/angle of interest, considering how the energy of a bond/angle changes with its length/angle. In this work, we have compared our parameters with that proposed by the CHARMM and Amber FFs~\cite{amber2014new,charmm1994derivation}. It turned out that for alkanes, the CHARMM FF parameters matched very well the QC scans, and could be taken without further refinements (see ESI). 

As for repulsion and van der Waals parameters, our methodology consists in, first, computing interactions energies using the super molecule approach (Eq.~\ref{super-molecule}) for the different dimer relative orientations and distances (See Figures~S2-S3-S4-S5 in ESI for more details). Second, we have used the PEST optimization code~\cite{pest} to refine the parameters, by adjusting the interaction energies computed with the FF model to the MP2 QC values. 

In order to later speed up the MD simulations and ease the parametrization process, Coulomb, polarization, repulsion and dispersion contributions were only considered for interactions between carbon, nitrogen, oxygen atoms. As for all interactions that involve hydrogen atoms, only the Coulomb and repulsion terms were kept. The optimized parameters for used in this work are reported in the ESI (Table S1).

\begin{figure*}[ht!]
\centering
\includegraphics[width=\linewidth]{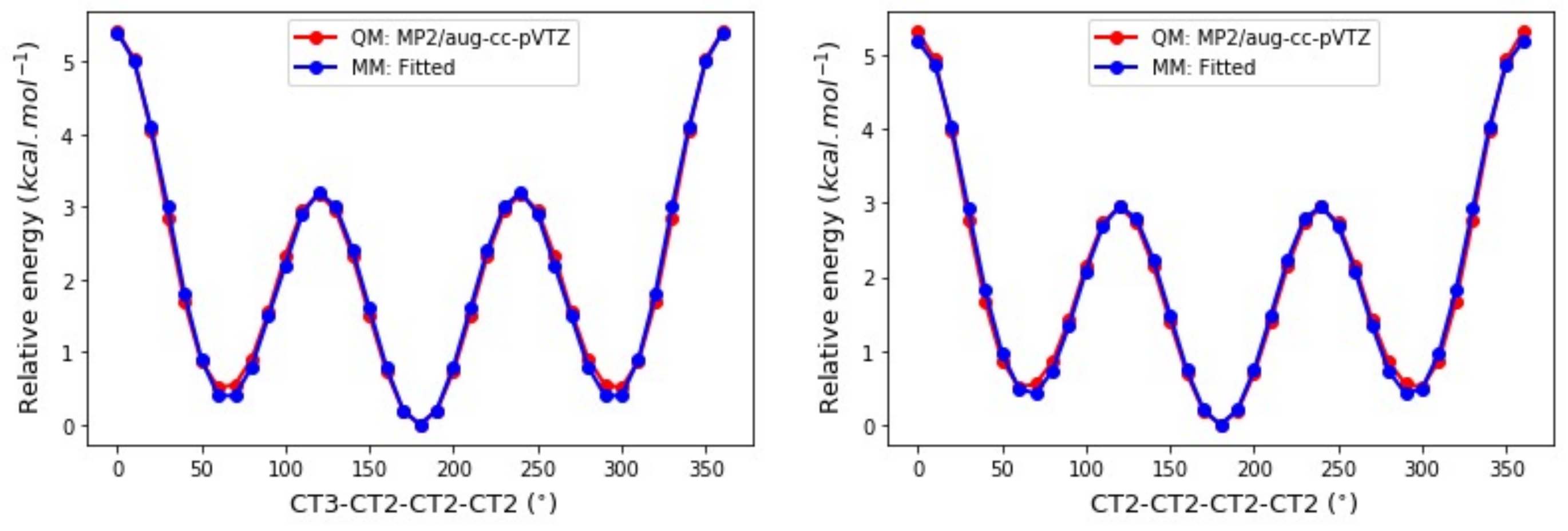}
\caption{Relative energy profiles of the dihedral angle scans in heptane (\ce{{CT2}-{CT2}-{CT2}-{CT2}} middle torsion on the left and \ce{{CT3}-{CT2}-{CT2}-{CT2}} terminal torsion on the right). The red lines represent the MP2 QC curves, while the blue ones the fitted FFs.}
\label{fig:dih}
 \end{figure*}

The dihedral parameters are chronologically the last ones to be derived, since torsion angle motions embed contributions from both the non-bonded (van der Waals, electrostatic and 1-4 interactions) terms, as well as angle bendings. The torsional parameters are therefore intimately coupled to the non-bonded and bonded parameters. In this work, complete QC scans of the dihedral torsions were performed as a sequence of constrained optimization in which the torsion angle is varied in steps of \SI{10}{\degree}. For alkanes, all possible torsions \ce{CTx-{CT2}-{CT2}-CTx} with $x={2,3}$ were scanned and resulted in quite similar torsion \textit{ab initio} energy profiles (difference between maxima $\leq$~\SI{0.2}{\kcal} as shown in Figure~\ref{fig:dih}). For most available FFs in literature, the same torsional parameters are used for both torsion involving terminal methyl and middle torsion. In this work, since we differentiate the non-bonded parameters of \ce{{CT3}} from those of \ce{{CT2}}, and since the dihedral parameters do dependent on the non-bonded ones, we have derived parameters specific to each dihedral torsion (\ce{{CT3}-{CT2}-{CT2}-{CT2}}, \ce{{CT3}-{CT2}-{CT2}-{CT3}} and \ce{{CT2}-{CT2}-{CT2}-{CT2}}). 

Figure~\ref{fig:dih} illustrates the almost prefect superposition of the classical and QC torsion curves, making us confident to explore further the different trans-gauche populations of the various species in the bulk phase. The same approach was adopted to derive torsion parameters for amide derivatives.
 
Lastly, in order to assess the reliability of the proposed classical model one needs to confront macroscopic data based on MD simulations to experimental condensed phase properties, when available. In particular, for the pure system the density and heat of vaporization will be compared.

\subsection{Molecular dynamics simulation details and liquid-phase properties}+
 
 All MD simulations were carried out with the Polaris-MD code~\cite{polaris}, and considering periodic boundary conditions; first, in the isothermal-isobaric ensemble (NPT) to equilibrate the systems, and then the production part in the constant-volume ensemble (NVT). The temperature was maintained by a Nose-Hoover thermostat and the system pressure isotropically by an Andersen's barostat. Verlet leapfrog scheme was employed with a \SI{1}{fs} integration time step of the dynamic equations of motion. The equilibration period of our systems varies between \SI{5}{ns} for pure phases (alkanes, monoamides) and \SI{8}{ns} for alkanes/monoamides mixtures, with a temperature scaling interval each 10~steps, the equilibration period was fixed based on analysis of the time evolution of density, energy and structural analysis (see Figure S6 in the ESI). The production run is about \SI{20}{ns}, which was enough considering the size of the systems that we simulated. The simulation box size impact on ligands properties have also been analyzed and was shown to be minor (see ESI Table S4). More details about the systems are available in Table~\ref{results}. The C-H structural parameters were constrained to their gas-phase equilibrium values thanks to the RATTLE algorithm (the convergence criteria was set to \SI{1e-6}{\angstrom})~\cite{miyamoto1992settle}. The fast multipole method was used for computing Coulomb electrostatic forces and polarization interactions~\cite{coles2015fast}.

As mentioned previously, the density and heat of vaporization were computed for each pure systems of interest.
The average bulk density is computed by performing NPT simulations using the formula:
\begin{equation}\label{key}
\expval{\rho}=\frac{m}{\expval{V}}=\frac{N_{mol}M}{\expval{V}N_A}, 
\end{equation}
where $\expval{V}$ is the average volume of the simulation box, $ N_{mol}$ is the number of molecules in the simulation box, $M$ is the molar mass of the molecule and $N_A$ is the Avogadro constant.

The enthalpy of vaporization is computed by performing NVT simulations, using the formula: 
\begin{equation}
\label{P1} \Delta H_{vap}(T) = U_{g}^{pot}-U_{l}^{pot}+k_{B}T , \\ 
\end{equation}
where, $U_{l}^{pot}= \frac{U_{T}^{pot}}{N_{mol}}$ is the potential energy of a molecule in liquid phase and $U_{T}^{pot}$ is the total potential energy of the system, and $U_{g}^{pot}$ the potential energy of one molecule in the gas phase (in vacuum).

The binary monoamide/dodecane mixture are solutions that may not behave ideally. The deviations from ideal behavior, consequence of new interactions between the monoamides and dodecane within the mixture, can be characterized by excess quantities. The excess volume $\mathrm{\Delta V_{ex}}$ has been calculated by subtracting the pure component molar volume from the mixture molar volume, using the following equation:
\begin{equation}
  \label{Vex}
    \Delta V_{ex}= \frac{x_1M_1 + x_2M_2}{\rho} - \frac{x_1M_1}{\rho_1} - \frac{x_2M_2}{\rho_2},
\end{equation}
where $x_1$ and $x_2$ are mole fractions of monoamide and dodecane, respectively, $M_1$ and $M_2$ the respective molar masses, and $\rho_1$ and $\rho_2$ the densities of the respective pure components; $\rho$ is the density of the mixture.

The excess enthalpy $\Delta H_{ex}$ of the binary mixture corresponds to the thermodynamic activity of each component in the mixture. It is computed as: 
\begin{equation}
  \label{Hex}
    \Delta H_{ex}= H_{mix} - x_1H_1 - x_2H_2,
\end{equation}
where $H_{mix}$ is the enthalpy of the monoamide/dodecane mixture, $x_1$ and $x_2$ being defined above, and $H_1$ and $H_2$ are the enthalpies of pure monoamide and pure dodecane solvents.

\section{Physical properties of pure phases: FF validation}
All the figures labeled “S” below are provided as the supplementary material. 

\subsection{Density and heat of vaporization}
In Table~\ref{results}, we report the density and the enthalpy of vaporization for the alkane series (ethane, propane, butane, heptane, dodecane), the two monoamides DEHiBA and DEHBA (displayed in Figure~\ref{fig:molecules}) and the primary amide  N,N-Dimethylacetamide (DMA).

\begin{table*}[ht!]
\caption{Densities $\rho$ (in \si{\kilogram\per\meter\cubed}) and heats of vaporization $\Delta H_{vap}$ (in \si{\kcal}) of alkanes and monoamides (DEHiBA and DEHBA) using the parameters derived in this work. $^a$ Experimental values are taken from Haynes \etal~\cite{haynes2014crc}. $^b$ OPLS-AA values taken from Refs.~\citenum{thomas2006}, and \citenum{opls}.  $^c$ OPLS-L values taken from Ref.~\citenum{OPLS2012}. Error estimates were obtained by block averaging, considering the oscillations of the energy and temperature during the last \SI{100}{\ps} of the production trajectory. The temperatures were selected based on the available experimental data.}
\centering
\begin{tabular}{l*4{S[table-format=3.0]}{S[table-format=2.2(2)]}*2{S[table-format=2.2]}}
\toprule
&&  \multicolumn{3}{c}{$\rho$(\si{\kilogram\per\meter\cubed})} & \multicolumn{3}{c}{$\Delta H_{vap}$ (\si{\kcal})}  \\ \cmidrule(lr){3-5}\cmidrule(lr){6-8}
{Name} & {T(\si{\kelvin})} & {This work } & {{\expe}$^a$} & {OPLS }  & {This work } &{{\expe}$^a$}& {OPLS }  \\ \midrule
Ethane     & 185    & 566 & 544  & 538 $^b$  & 3.65  \pm 0.06  & 3.60  &  3.44 $^b$   \\ 
Propane    & 225    & 604 & 587  & 580$^b$   & 3.57  \pm 0.07  & 4.40  &  4.50  $^b$  \\ 
Butane     & 273    & 596 & 601  & 589$^b$   & 4.66  \pm 0.10  & 5.36   &  5.00 $^b$\\ 
Heptane    & 298    & 700 & 677  & 679$^b$   & 7.69  \pm 0.10  & 8.60   &  9.58 $^b$\\ 
           & 371    & 637 & 616  &     & 7.84  \pm 0.05  & 7.60   &    \\ 
Decane     & 298    & 761 & 730  & {727$^b$, 726$^c$}  & 12.90 \pm 0.10  & 12.30  &  13.35$^b$ \\ 
Dodecane   & 298    & 784 & 745  & {750$^b$, 744$^c$}   & 15.43 \pm 0.19  & 14.70  &  {22.3$^c$} \\
           & 490    & 604 & 590  &    & 11.51 \pm 0.20  & 10.54  &    \\ 
Tridecane  & 300    & 781 & 756  & 840 $^b$  & 14.90 \pm 0.10  & 15.70  &  {12.9$^b$}\\ 
Isopentane & 301    & 648 & 617  &    & 6.32  \pm 0.04  & 5.88   &    \\  
DMA        & 298    & 937 & 936  & 911$^b$   & 10.81 \pm 0.11  & 11.75  & 11.99   \\
DEHiBA     & 298    & 898 & 865  &    & 23.34 \pm 0.15    & {}        & {}     \\
           & 308    & 892 & 858  &    & 23.13 \pm 0.15    & {}        & {}     \\
           & 318    & 883 & 851  &    & 21.46 \pm 0.15    & {}        & {}     \\
DEHBA      & 298    & 916 & 861  &    & 24.93 \pm 0.21   & {}        & {}     \\
           & 308    & 905 & 854  &    & 24.44 \pm 0.21   & {}        & {}     \\
           & 318    & 895 & 847  &    & 22.56 \pm 0.21   & {}        & {}     \\
\bottomrule 
\end{tabular}
\label{results}
 \end{table*}
We can clearly notice that the bulk density values agree well with the experimental ones, as the largest deviation is 5\% for dodecane and 6\% for DEHiBA. This confirms the correct description of the phase equilibria, connected the accurate positioning of the minima of potential wells by our FFs. For the heat of vaporization, we can solely consider the alkanes for the comparison to experimental data. The largest deviation is less than \SI{1}{\kcal} for dodecane. For the simulated monoamides no experimental heat of vaporization is available in the literature, but since the same parametrization approach was used for both systems, we can safely trust our predicted values. To our knowledge, we are the first group to report the value of the heat of vaporization of DEHiBA and DEHBA via computational methods. The heats of vaporization of DEHiBA and DEHBA come out quite close for both molecules ($\simeq$~\SI{23.5}{\kcal}  for DEHiBA and $\simeq$~\SI{25}{\kcal} for DEHBA), this was expected since both molecules are quite similar as they only differ by the branching of the alkyl chain attached to the carbonyl group (butyl vs. iso-butyl). Overall, the computed properties yield a great agreement with experiment, and the errors are within the error bars of alkanes traditional FF developed based on density and heat of vaporization, about 5\% for the densities and up to \SI{1}{\kcal} for the heats of vaporization~\cite{HV-MD,OPLS2012,opls,amber}.

Some FFs reported in the literature~\cite{gromos,MM4,charmm,amber,opls,OPLS2012}, may give very accurate densities and heats of vaporization for some carbon chain lengths. This is not surprising as the latter properties were used in their parameterization process. However, as already mentioned, these FFs are not always transferable for all molecular groups. For instance, the dodecane heat of vaporization obtained with OPLS-AA FF is about $\simeq$~\SI{22.3}{\kcal} with a deviation of $\simeq$~\SI{7.6}{\kcal} from the experiment, and a density of \SI{839}{\kilogram\per\meter\cubed} overestimating the experimental value of \SI{745}{\kilogram\per\meter\cubed}~\cite{OPLS2012}. Moreover, some studies in literature have reported the gel-phase formation at room temperature for long-chain hydrocarbons of more than eight carbons (including dodecane) using the OPLS-AA force field~\cite{vo2015computational,opls2013effect}. These deviations pair with the fact that the OPLS-AA FF was adjusted to reproduce liquid densities and enthalpies of vaporization of short alkanes (ethane, propane, and butane). As a result, Siu~{\etal} have re-optimized these parameters for longer alkanes (labeled L-OPLS~\cite{OPLS2012}), following the same parameterization approach of the original OPLS (based on experimental densities and vaporization heats). Still, this may lead to the same disadvantages for much longer molecules than the ones used in the parameterization. It should also be mentioned that, even though cis-9-octadecene molecule was used in the parameterization process of L-OPLS, a deviation of $\approx$ \SI{2}{\kcal} from reference data was accepted which is more than the maximum deviation we report in this work (\SI{1}{\kcal}). All of this affirm the strength and reliability of our parametrization approach and the developed FFs in simulating/predicting thermodynamic properties of alkanes and monoamides.

\subsection{Structural properties: Chain conformation (trans-gauche populations)}

\begin{figure}[ht!]
    \centering
    \includegraphics[width=0.8\linewidth]{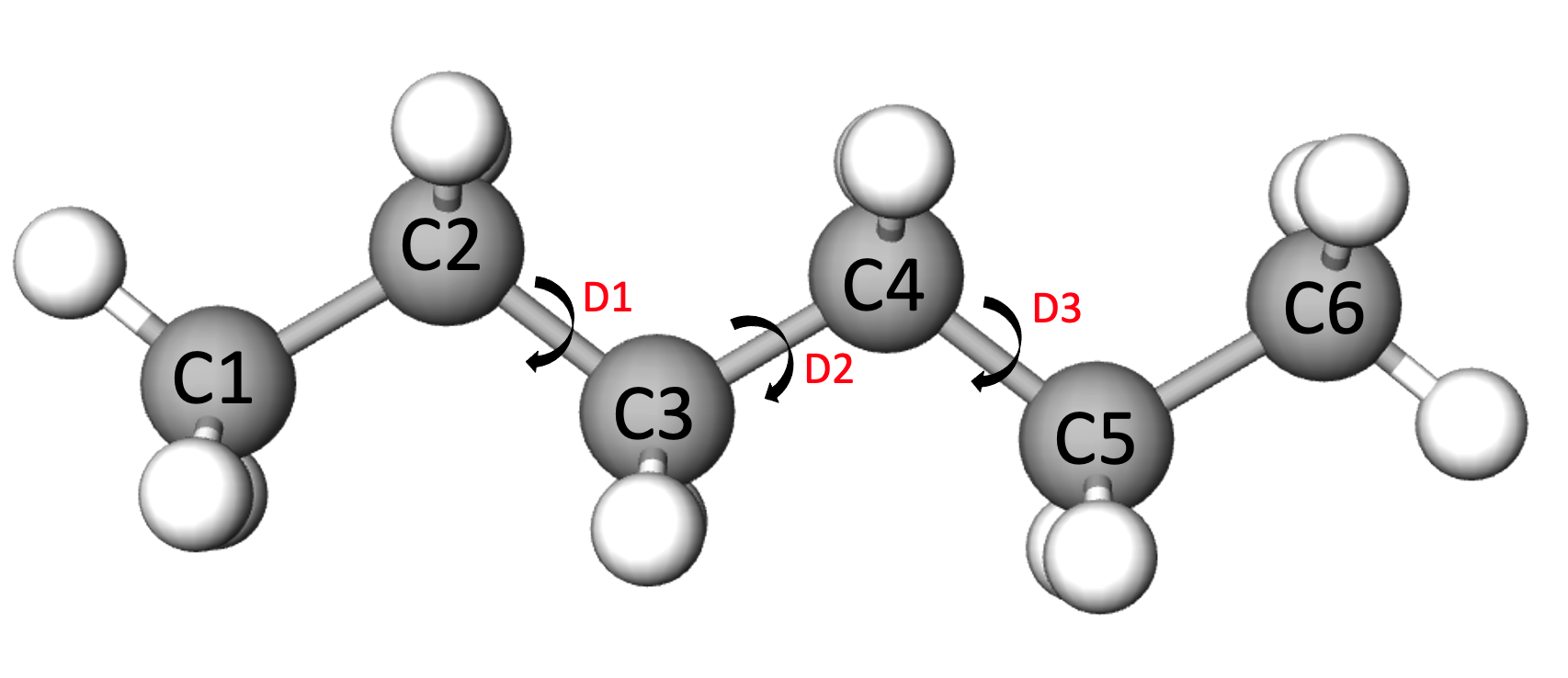}
    \caption{Representation of dihedral angle along the carbon chain for the hexane molecule.}
    \label{dih-pop}
\end{figure}

The chain conformation of alkane is analyzed in the light of the population of the gauche and trans states in the alkane series (heptane, decane, dodecane and tridecane), which is characterized by the \textbf{D} ratio defined as $\mathrm{\frac{population ~ of~ trans}{population~(gauche+trans)}}$ for each specific torsion angle labelled as in Figure~\ref{dih-pop}. The gauche and trans conformations were classified based on the torsional barriers at \ang{120} and \ang{240} between the two states (Figure~\ref{fig:dih}). This kind of analysis serves to determine the preferred conformations in the liquid phases, and among the fundamental structural units in biology and chemistry in general~\cite{dettenmaier1978conformation,bartell1963structure,menger1988does}.

Table~\ref{G-T-pop} compiles the trans-gauche population ratio as a function of the dihedral angle along the carbon chain obtained from our FF and some other FFs (OPLS-AA and L-OPLS), the only available experimental values in literature are those of tridecane. We can clearly observe the good agreement of our work with the experimental data for tridecane, a difference of 6~\% for the terminal torsion populations (D1 and D9) and 2~\% for the torsion in the middle of the chain (Dx with x$\in{[3..7]}$) population was noted. The largest deviation was observed for heptane, about 7~\% for the terminal torsion population and 1~\% for the middle torsion ones.

It is worth mentioning here that the results obtained with the OPLS-AA FF for tridecane deviate by about 34~\% from the experimental ones for the chain terminal populations and 25~\% for the trans populations of the interior of the chain. This again confirms the inability of OPLS-AA FF to correctly describe the trans-gauche populations for alkanes. The more recent L-OPLS FF results are in quite good agreement with the experimental data reported by Casal~{\etal}~\cite{casal}, with an average deviation of about 3~\%, and quite close with the prediction of our FF (average deviation 4~\%) for the simulated alkanes.

\begin{table*}[ht!]
\caption{Trans populations in \% in heptane, decane, dodecane and tridecane solvents as a function of dihedral angle noted $\mathrm{Dn}$ along the carbon chain. $^a$ values from Monte Carlo (MC) simulations~\cite{thomas2006} with the original OPLS-AA FF~\cite{opls}; $^b$ values from MD simulations with the L-OPLS FF~\cite{OPLS2012}; $^c$ Experimental values for the first and the sixth carbon torsion of tridecane from Casal~{\etal}~\cite{casal}}
\centering
\label{G-T-pop}
\begin{tabular}{llllllllllll}
\toprule
Molecule & Dihedral  angle  & D1 & D2 & D3 & D4 & D5 & D6 & D7 & D8 & D9 & D10 \\
\midrule
heptane (\SI{298}{\kelvin})  & This work & 65 & 68 & 68 & 65 &  &  &  &  &  &  \\ 
                           & OPLS-AA $^a$  & 73 & 82 & 78 & 76 &  &  &  &  &  &   \\ 
decane (\SI{300}{\kelvin})   & This work    & 65 & 69 & 66 & 66 & 66 & 69 & 65 &  &  &     \\
                           & OPLS-AA $^a$   & 78 & 81 & 84 & 82 & 81 & 81 & 79 &  &  &     \\ 
dodecane (\SI{298}{\kelvin}) & This work & 64 & 68 & 65 & 65 & 65 & 65 & 65 & 68 & 64 & \\
                            & OPLS-AA $^a$ & 80 & 83 & 79 & 80 & 78 & 80 & 81 & 83 & 81 & \\ 
dodecane (\SI{490}{\kelvin})
& This work    & 56 & 61 & 57 & 57 & 57 & 57 & 57 & 61 & 56 &    \\ 
tridecane (\SI{298}{\kelvin}) & L-OPLS$^b$ & 55 & 66 & 64 & 64 & 64 & 64 & 64 & 64 & 66 & 55 \\
&OPLS-AA $^a$&	92 & 97& 98& 98 & 98 & 98 & 98 & 98& 97& 92 \\
& This work & 64 & 67 & 64 & 65 & 65 & 65 & 65 & 68 & 65 & 64 \\ 
& {\expe}$^c$         & 58 & -  & -  & -  & -  & 67 & -  & -  &  - & 58 \\
\bottomrule
\end{tabular}
\end{table*}

\subsection{Radial distribution functions}

\begin{figure}[ht!]
 \captionsetup[subfigure]{justification=centering}
   \centering
      \begin{subfigure}[b]{0.80\linewidth}
      \centering
       \includegraphics[width=\linewidth]{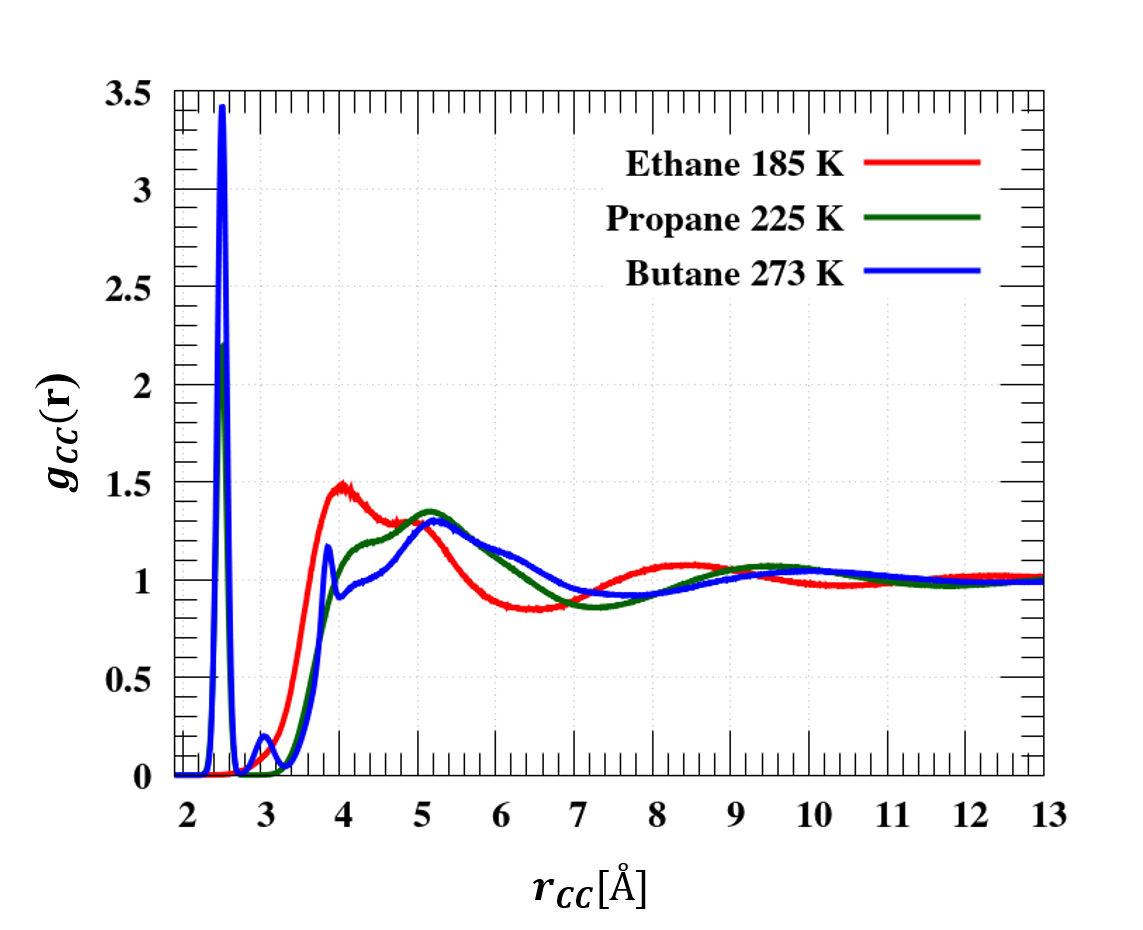}
      \caption{}
      \end{subfigure}
      \begin{subfigure}[b]{0.80\linewidth}
      \centering
      \includegraphics[width=\linewidth]{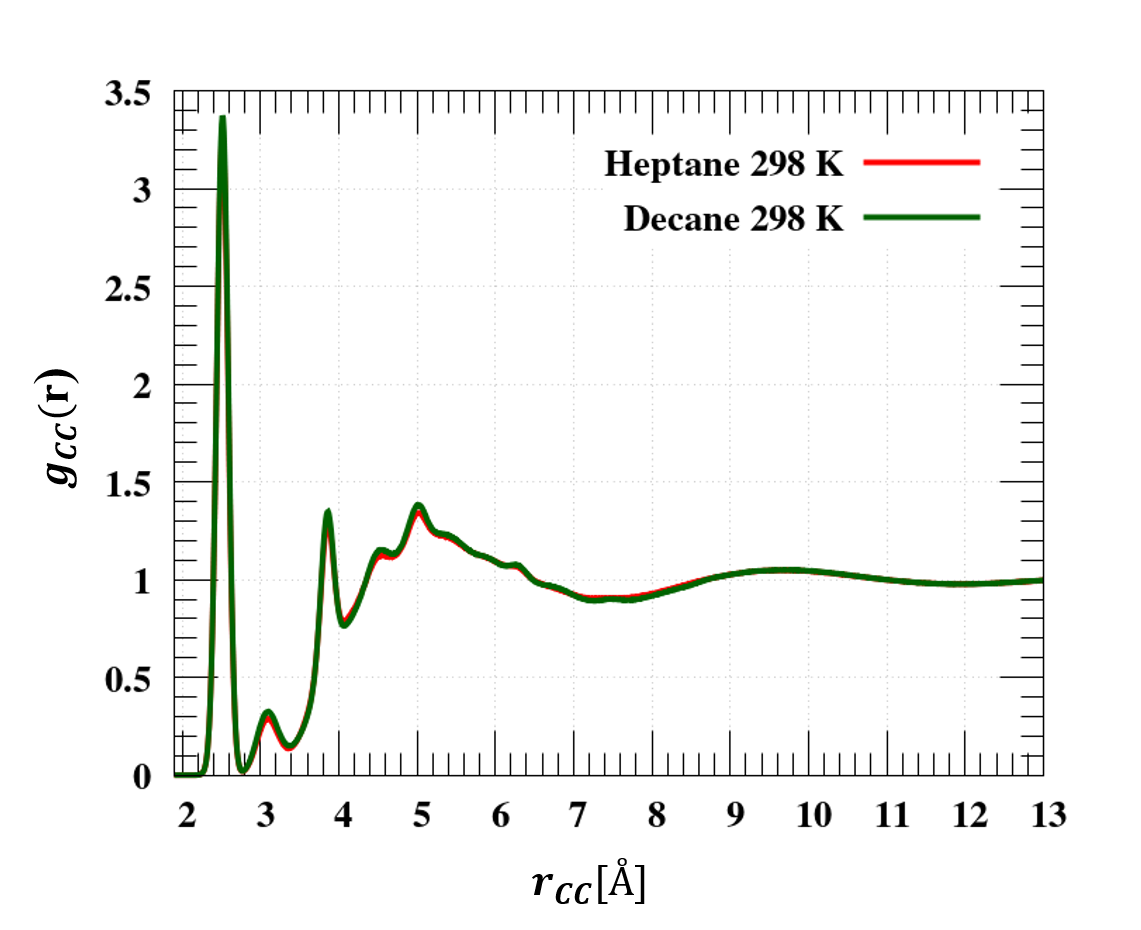}
     \caption{}
      \end{subfigure}
    \begin{subfigure}[b]{0.80\linewidth}
      \centering
      \includegraphics[width=\linewidth]{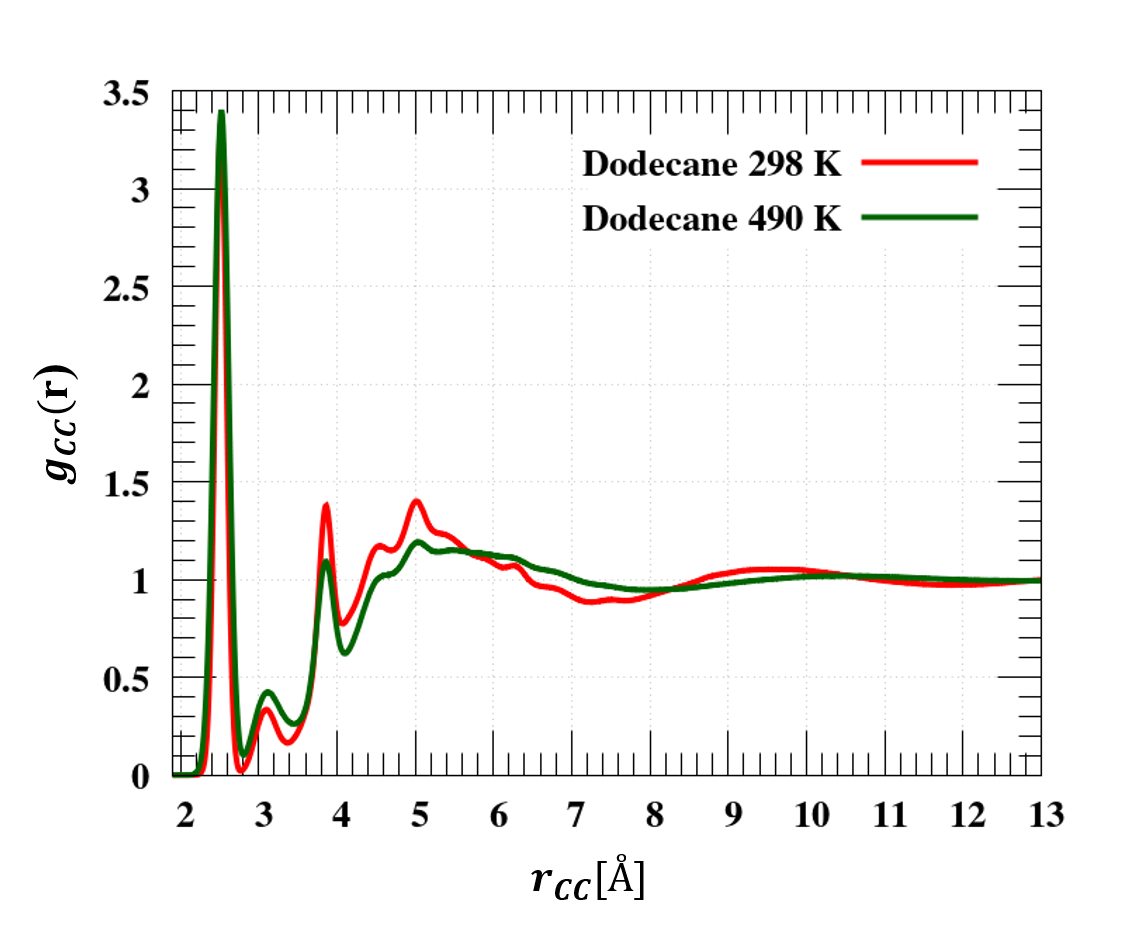}
     \caption{}
      \end{subfigure}
 \caption{Radial distribution functions of carbon atoms for selected alkanes: ethane, propane, butane (a), and heptane, decane (b) and dodecane (c).}
\label{fig:RDF}
 \end{figure}

The structural organization of the pure alkanes solvents was investigated by calculating the radial distribution functions (RDFs, $g(r)$), using VMD package~\cite{vmd}, as they reveal the distribution of neighboring molecules and the long-range solvent organization. The $g(r)$ for carbon atoms in alkanes are plotted in Figure~\ref{fig:RDF}. 

We have drawn the \ce{C-C} RDFs starting at \SI{2}{\angstrom}, hence, the first C-C intramolecular interaction at \SI{1.53}{\AA} is not visible. The peak at \SI{2.55}{\AA} corresponding to the intramolecular \ce{C1-C3} second-neighbor interaction is present for all alkanes longer than ethane. The third peak at $\simeq$\SI{3.1}{\AA} corresponds to the gauche \ce{C1-C4} molecular segments. At the distance of $\simeq$~\SI{3.88}{\AA} appears a peak that can be attributed to the trans \ce{C1-C4} molecular segments. The peak positions reported in Table~\ref{tab:XRP-RDF} are quite consistent with the experimental X-ray diffraction results of Habenschuss~{\etal}~\cite{habenschuss1990x}. We have also calculated the structure factor for heptane and the results are in good agreement with experiment (see ESI Figure S7).

We can also notice the influence of temperature upon the RDFs, for dodecane at two different temperature (\SI{298}{\kelvin} and \SI{490}{\kelvin}). It is readily apparent in Figure \ref{fig:RDF}~(c), that the intensity of the 3rd peak increases while the 4th peak simultaneously decreases, implying that the gauche population increases with temperature and the reverse for the trans population. This was confirmed by the calculation of the populations as presented in Table~\ref{G-T-pop}. Indeed as the temperature rises, it becomes more likely to populate conformations with higher energies (gauche). This observation further support the strength and reliability of our FF in describing the structural organization of the pure alkanes.

\begin{table}[ht!]
\caption{Comparison of the positions in \si{\angstrom} of the local maxima of the \ce{C-C} RDFs computed in this work to X‐ray diffraction data~\cite{habenschuss1990x}.}
\centering
\begin{tabular}{l*8{S[table-format=1.2]}}
\toprule
 & \multicolumn{2}{c}{\ce{C1-C2}} & \multicolumn{2}{c}{\ce{C1-C3}} & \multicolumn{2}{c}{\ce{gC1-C4}} & \multicolumn{2}{c}{\ce{tC1-C4}} \\
 \cmidrule(lr){2-3}\cmidrule(lr){4-5}\cmidrule(lr){6-7}\cmidrule(lr){8-9}
& {MD}  & {\expe}  & {MD}  & {\expe}  & {MD}  & {\expe}  & {MD}  & {\expe}  \\
\midrule
Butane \SI{273}{\kelvin}   &  1.54   & 1.55     & 2.55   & 2.56    &  3.10     & 3.07      &  3.88          & 3.93      \\
Heptane \SI{298}{\kelvin}  &  1.53   & 1.55     & 2.54           & 2.56     &  3.10          & 3.14      &  3.87          & 3.92      \\
Decane  \SI{298}{\kelvin}          &  1.53         & 1.54     & 2.55           & 2.57     &  3.10          & 3.15      &  3.88          & 3.95     \\
Dodecane \SI{298}{\kelvin}          &  1.53         &  {}       & 2.55           &  {}      &  3.10          &  {}        &  3.88          &  {}     \\
\bottomrule
\end{tabular}

\label{tab:XRP-RDF}
\end{table}

\section{Molecular properties of dodecane/monoamides mixtures}
After having validated our FFs on pure alkane solvent properties, herein we present the results obtained for DEHBA and DEHiBA mixtures with dodecane. This study was performed to gain insight and to visualize molecular-level behavior of these extraction molecules (alkane-monoamides mixtures), for instance, to understand the impact of the monoamide structure, in particular the branching of the alkyl chain on the molecular organization of alkane-monoamides mixtures. At the end, this may give insights about the sharp difference measured for Pu(IV) extraction between DEHiBA and DEHBA~\cite{prabhu1997,recycling}.

In order to approach as much as possible experimental conditions, the simulations boxes where constructed based on available experimental data for DEHiBA (densities) at room temperature and atmospheric pressure (\SI{298}{\kelvin}, \SI{1}{\atm}). Since the experimental data were only available for DEHiBA~\cite{COQUIL2021}, the same simulation conditions were considered for the DEHBA/dodecane mixtures to be able to compare the solvent mixtures, as a function of the nature of the monoamide. Molecular compositions for the simulations are given in ESI Table~S5. The extractant mole fraction $x_{mono}$ was varied from \SIrange{10}{62}{\percent}.

\subsection{Density}

\begin{figure}[ht!]
    \centering
    \includegraphics[width=\linewidth]{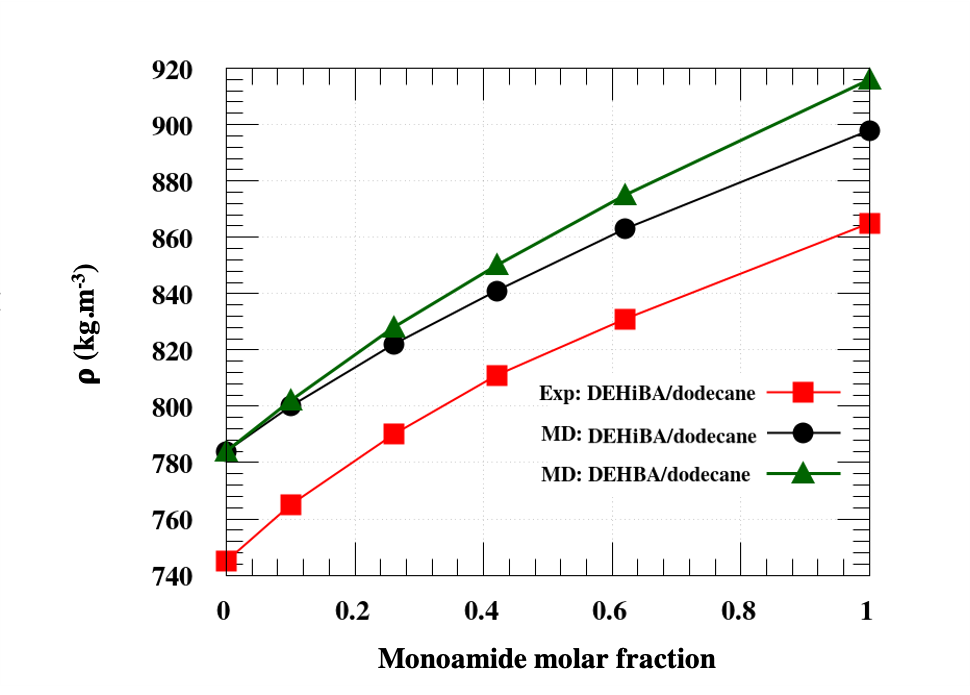}
    \caption{Mass density of the DEHiBA/dodecane and DEHBA/dodecane mixtures as a function of the mole fraction. Dark blue circles and green diamonds correspond to the simulated densities for the DEHiBA/dodecane and DEHBA/dodecane, respectively. Red squares correspond to experimental measurements by Coquil~{\etal}~\cite{COQUIL2021}.}
    \label{fig:mix-density}
\end{figure}

The computed densities of different DEHiBA/dodecane and DEHBA/dodecane mixtures are displayed in Figure~\ref{fig:mix-density}. The calculated values for the former compare favorably to experimental data, with an average percent error of \SI{5}{\percent} across the range of solvents (\SI{33}{\kilogram\per\meter\cubed}). The densities of the DEHBA/dodecane mixtures appear to be close to the DEHiBA/dodecane ones; this was expected since the structures of the two ligands are quite alike and also from the fact that density of both monoamides in pure phase are nearly the same, within \SI{4}{\kilogram\per\meter\cubed}.

\subsection{Radial distribution functions and aggregation numbers}

\begin{figure*}[ht!]
\captionsetup[subfigure]{justification=centering}
    \begin{subfigure}[t]{0.40\linewidth}
        \centering
        \includegraphics[width=0.95\linewidth]{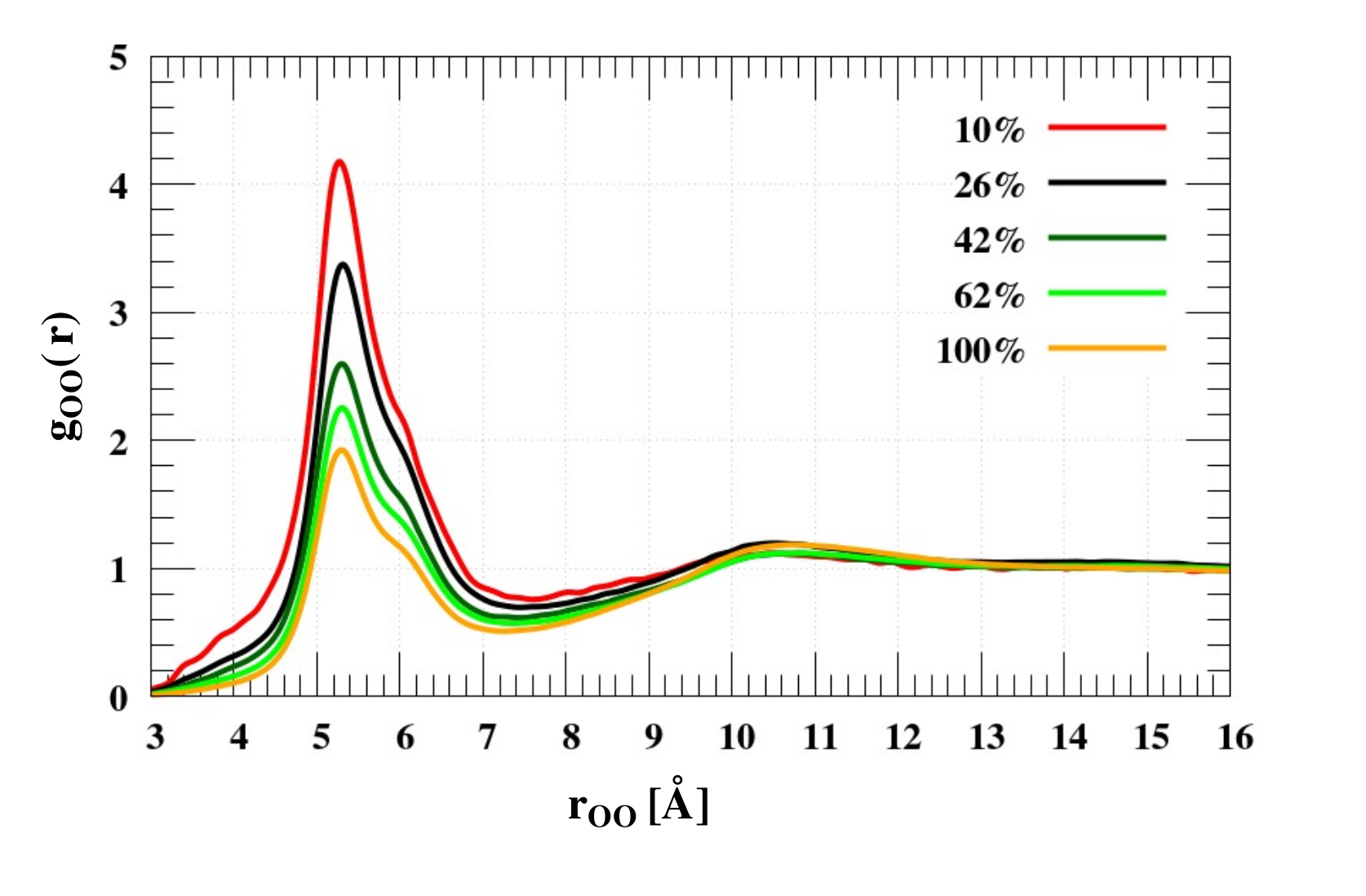}
        \caption{}
    \end{subfigure}
        \begin{subfigure}[t]{0.40\linewidth}
        \centering
        \includegraphics[width=0.95\linewidth]{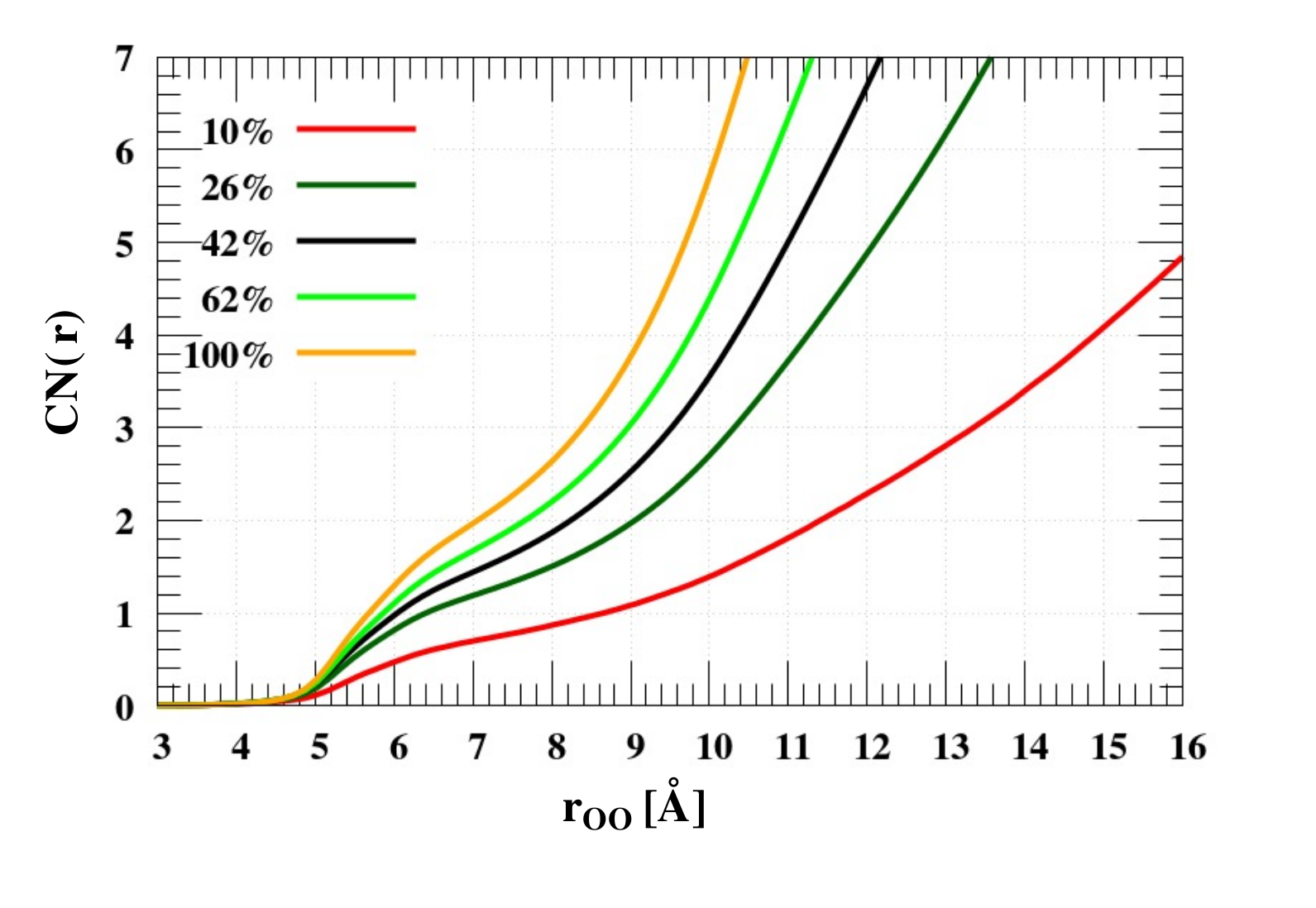}
        \caption{}
    \end{subfigure}
        \begin{subfigure}[t]{0.40\linewidth}
        \centering
        \includegraphics[width=0.95\linewidth]{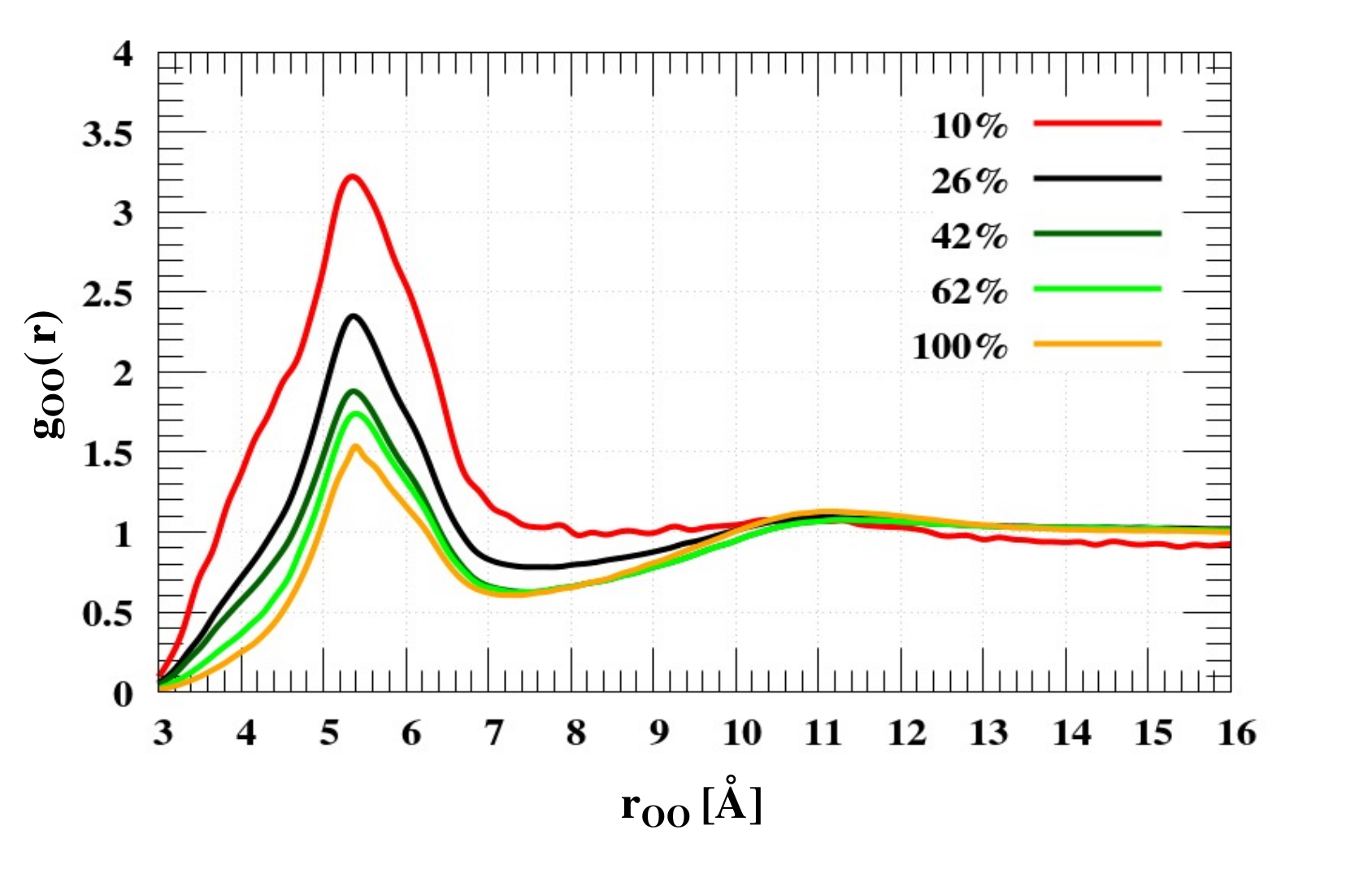}
        \caption{}
    \end{subfigure}
        \begin{subfigure}[t]{0.40\linewidth}
        \centering
        \includegraphics[width=0.95\linewidth]{RDF-oo-dehiba-int.pdf}
        \caption{}
    \end{subfigure}
\caption{Radial distribution functions and the coordination number (CN) of oxygen atoms for DEHiBA/dodecane (a--b) and DEHBA/dodecane mixtures (c--d) at different mole fractions.}
\label{rdf-mix-1}
\end{figure*}

\begin{figure*}[ht!]
\centering
\includegraphics[width=0.85\linewidth]{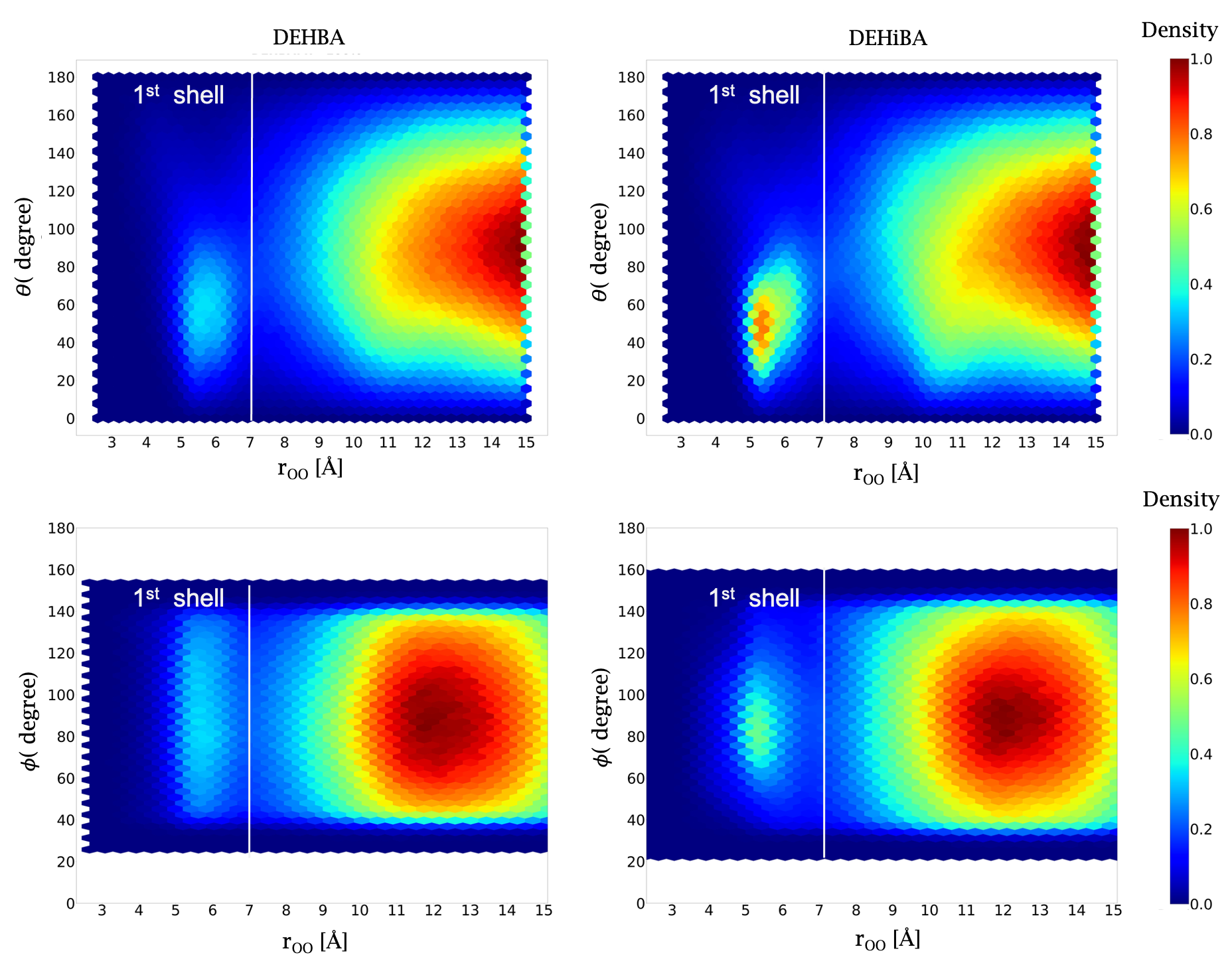}
\caption{Analysis of the carbonyl group C=O orientation, $\theta (r_{OO})$. and relative orientations of the NCO planes, $\phi(r_{OO})$, as defined in Figure \ref{fig:CON-illu}), as a function of the distance between the oxygen of the carbonyl group for the pure DEHiBA and DEHBA solvents. The color bar indicates the normalized count density.}
\label{fig:CO-analysis}
\end{figure*}

To gain insights into the structures of DEHiBA/DEHBA in the liquid phase as well as the organization of DEHiBA/DEHBA within the dodecane solvent, the RDFs between nitrogen and/or oxygen belonging to different monoamide were computed using a bin width of \SI{0.10}{\angstrom} and a cutoff distance of \SI{25}{\angstrom} (about half of the simulation box length), averaged over the \SI{20}{\ns} of the MD trajectories. The \ce{O-O} RDFs of the monoamide/dodecane mixtures at different concentrations are reported in Figure~\ref{rdf-mix-1}. For the pure monoamide phases, the RDF profiles are equivalent, suggesting that the molecular organizations for these two alike monoamides are quite similar (displayed in Figure~\ref{fig:molecules}). However, the peak positions are shifted by about \SI{+0.1}{\AA} for DEHBA as compared to DEHiBA. The relative steric hindrance increases by the branching of the alkyl group bonded to carbonyl site is probably responsible for the shift (Figures~S10-S11-S12-S13 of the ESI). 

Using Figure~\ref{rdf-mix-1}, one can discuss the evolution of the \ce{O-O} RDFs of DEHiBA/DEHBA for various monoamide/dodecane ratios. In both sets of mixtures, the peak heights of the RDFs gradually decrease as the concentration of monoamide increases. However, the peak positions are insensitive to the concentration indicating that the dilution of the extractant in dodecane has no significant impact on the molecular organization. These observations also apply to the \ce{O - N} and \ce{N - N} RDFs showed in Figures~S10 and S11 of the ESI.

The (\ce{O-O}) and (\ce{N-N}) RDFs suggest a "first coordination shell" ranging out to \SI{7.2}{\angstrom} (Figure \ref{rdf-mix-1} and \ref{fig:CO-analysis}), which corresponds to self-assembly of monoamides. The average coordination numbers were estimated by integrating (\ce{O-O}) and (\ce{N-N}) RDFs (see Figure \ref{rdf-mix-1}). For $x_{mono}$=\SI{10}{\percent}, it amounts to about 0.6 and about 1.2 for $x_{mono}$=\SI{26}{\percent}, for both DEHiBA and DEHBA, which suggests the possibility of monoamides dimer formations in these phases. The coordination number steadily increases as the monoamide concentration grows, in a similar way for both monoamides; it reaches 1.5 for $x_{mono}$=\SI{42}{\percent}, about 1.9 for $x_{mono}$=\SI{62}{\percent}, and tops at 2.4 for the pure phases.

The snapshots showed in Figure~S8 of the ESI, exemplify the formation of aggregates in dodecane. We suspect that aggregation occurs because the monoamide molecules are polar, and prefer interacting with each other rather than with alkanes. In general, alkanes are known to be "insoluble" in polar solvent (such as water). They are soluble only in non-polar and slightly polar solvents. In our case, DEHiBA and DEHBA are polar molecules with a dipole moment\footnote{Obtained with quantum calculations at the MP2 level and aug-cc-pVTZ basis set.} of $\simeq$\SI{3.64}{\debye} (water $\simeq$ \SI{1.85}{\debye}), yet, the fact that they are long molecules with lipophilic tails makes them soluble in alkanes.

\subsection{Relative orientations in monoamide dimers}
\begin{figure}[ht!]
      \centering
      \includegraphics[width=0.8\linewidth]{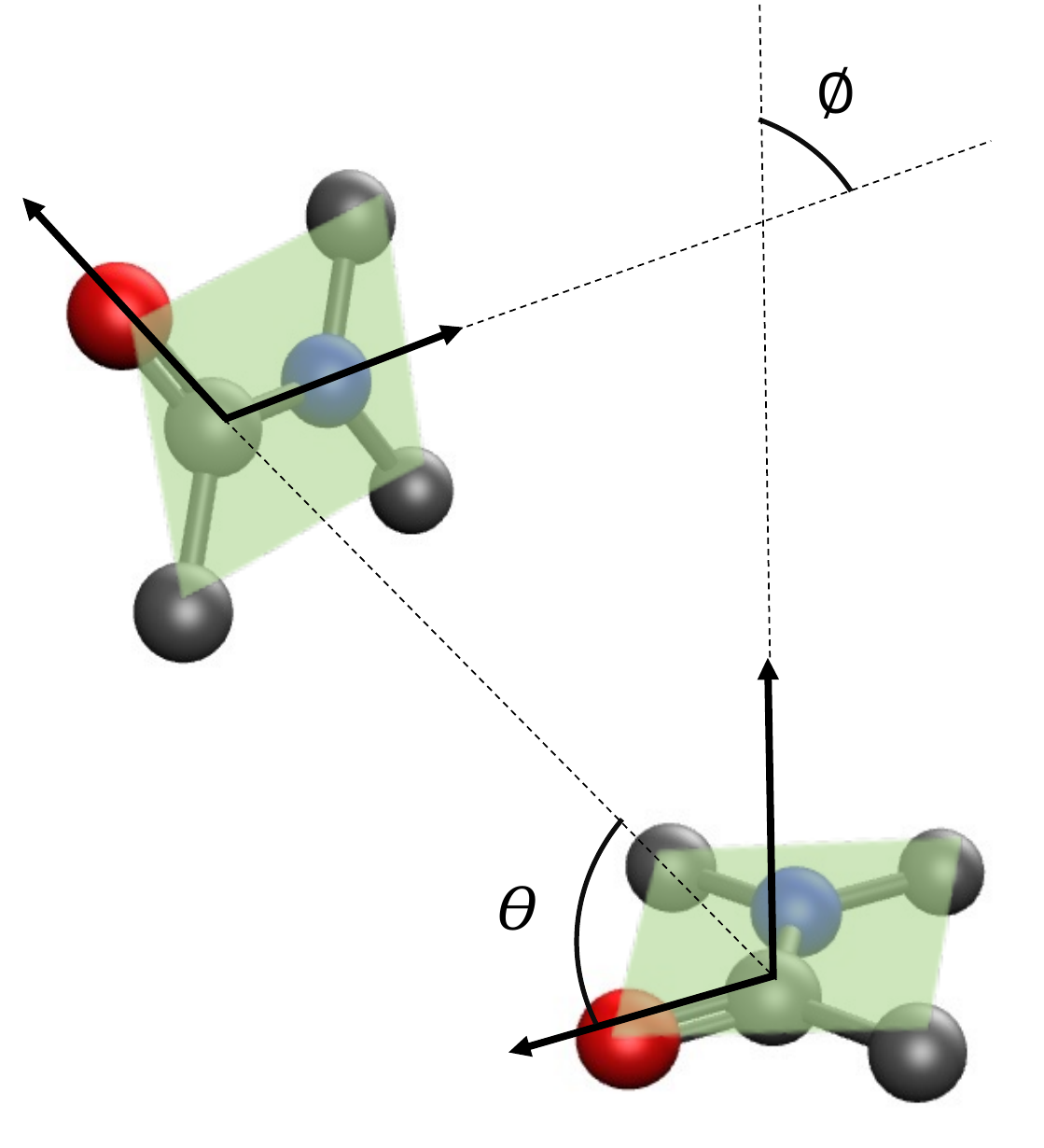}    
 \caption{Illustration of the orientation convention used to analyse the molecular organization for monoamide/dodecane mixtures. The carbonyl group C=O orientation ($\theta$) and the orientation of the flat surface of the amide function "NCO" ($\phi$).}
\label{fig:CON-illu}
\end{figure}

RDFs solely provide a 1D picture of the molecular organization for the DEHBA/DEHiBA pure and dodecane mixed solutions. To further characterize it and have a clearer vision in 3D, we have analysed the relative orientations of the carbonyl groups belonging to two clustered monoamide molecules, defined by the angle $\theta$  as well as the angle $\phi$ between the two NCO planes (See Figure \ref{fig:CON-illu}), as a function of the mole fraction. Figure~\ref{fig:CO-analysis} shows the angles $\phi$ and $\theta$ as a function of the distance between the oxygen of the carbonyl group for both extractants for pure systems (DEHBA and DEHiBA); the results for x=\SIlist{10;26;42;62;100}{\percent} mole fractions are available in ESI (see Figure S14) but do follow exactly the same trend that will be discussed below, as well as a time evolution of the relative orientations as a function of time (see Figure S15) that indicate that the property is well converged after \SI{8}{ns}. 

For DEHBA, the angular analysis of  $\phi$  shows that, at $d_{OO}$ =\SI{5.5}{\AA}, a vast range of $\phi$ angles (\SIrange{43}{135}{\degree}) is preferred in mixtures and pure phases. Contrary to DEHBA, DEHiBA seems to orient perpendicularly (\SIrange{70}{100}{\degree}) to another DEHiBA molecule while making dimers. As for long $d_{OO}$ distances ($d_{OO}$=\SIrange{9}{16}{\angstrom}) corresponding to the "second coordination sphere", one can notice that both monoamides molecules prefer perpendicular conformations. 

\begin{figure}[ht!]
\captionsetup[subfigure]{justification=centering}
  \centering
    \begin{subfigure}[t]{0.75\linewidth}
        \centering
        \includegraphics[width=0.9\linewidth]{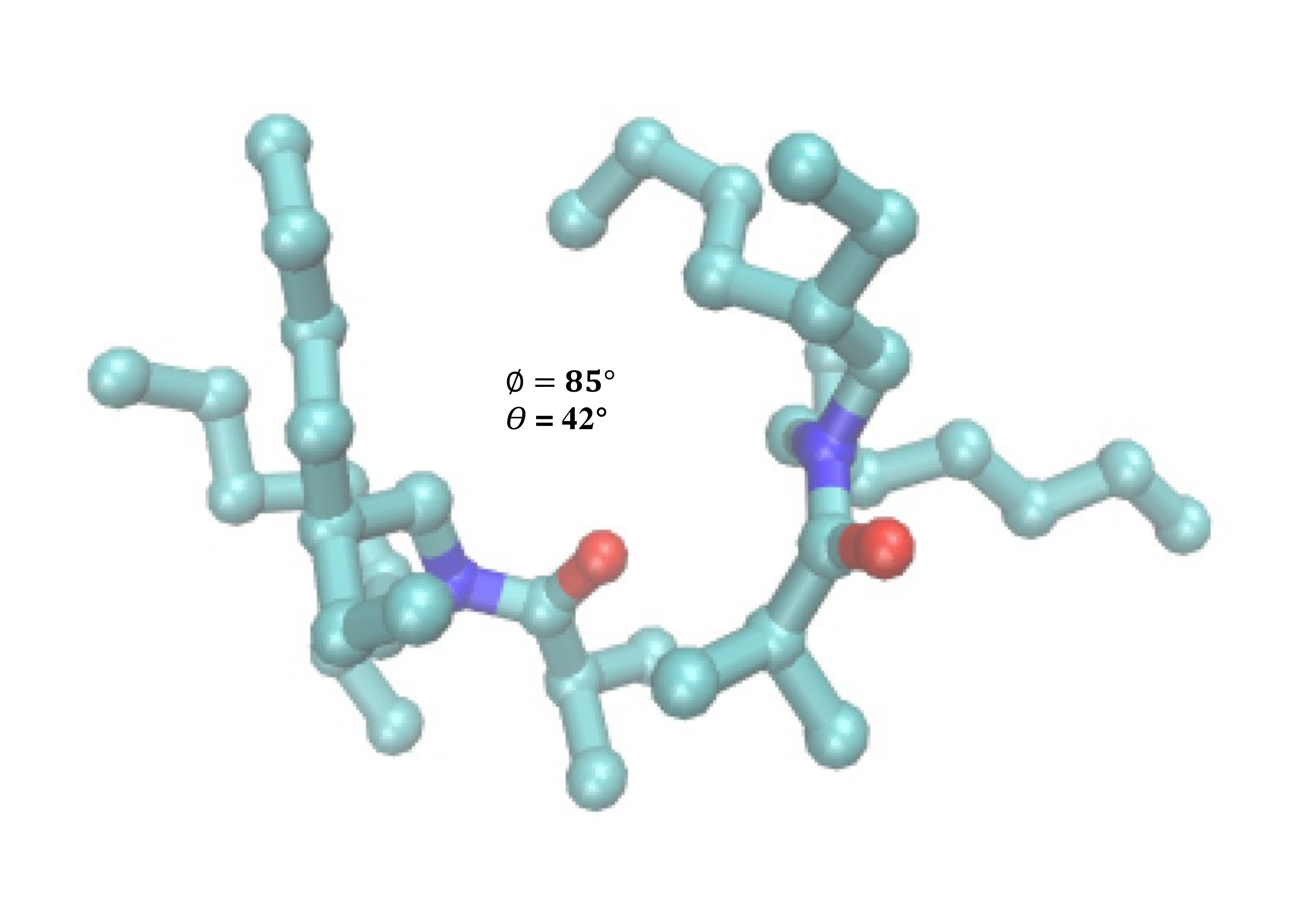}
        \caption{DEHiBA dimers}
    \end{subfigure}
   \begin{subfigure}[t]{0.75\linewidth}
        \centering
        \includegraphics[width=0.9\linewidth]{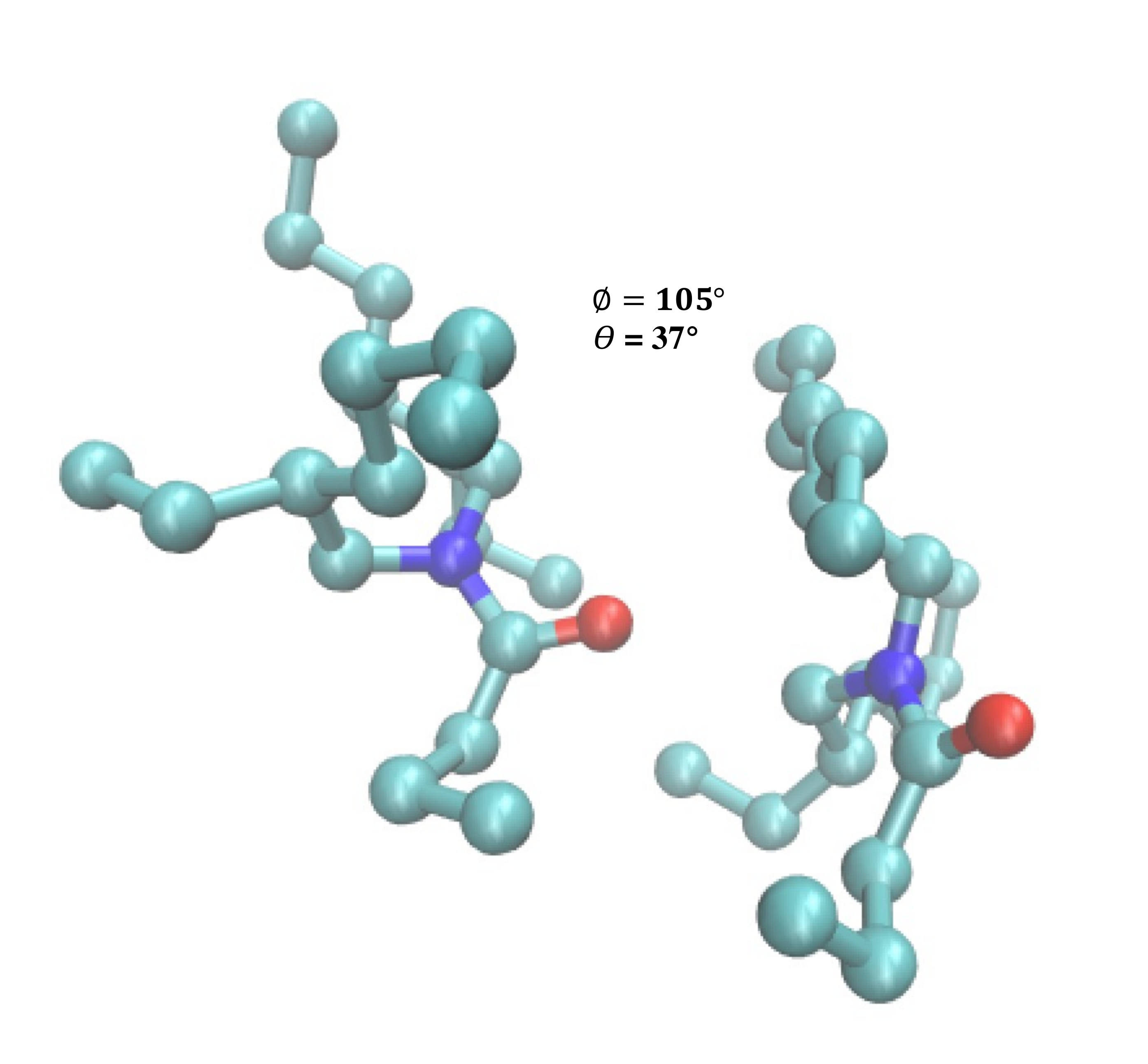}
        \caption{DEHBA dimers}
  \end{subfigure}
\caption{The preferred DEHiBA (a) and DEHBA (b) dimer structures according to the \ce{C=O} and \ce{N-C=O} relative orientation analysis.}
\label{fig:dimers}
\end{figure}

As for the analysis of the carbonyl group orientation, for both systems (DEHBA and DEHiBA) diluted in dodecane or in pure phase, the top panels of Figure~\ref{fig:CO-analysis} suggest that the monoamides preferably interact with each other in a conformation where the angle $\theta$ between carbonyl groups is in the range of \SIrange{20}{80}{\degree}, see Figure~\ref{fig:dimers}.

\subsection{Excess volume and excess enthalpy}

\begin{figure}[ht!]
\captionsetup[subfigure]{justification=centering}
  \begin{subfigure}[t]{0.9\linewidth}
      \centering
 \includegraphics[width=0.9\linewidth]{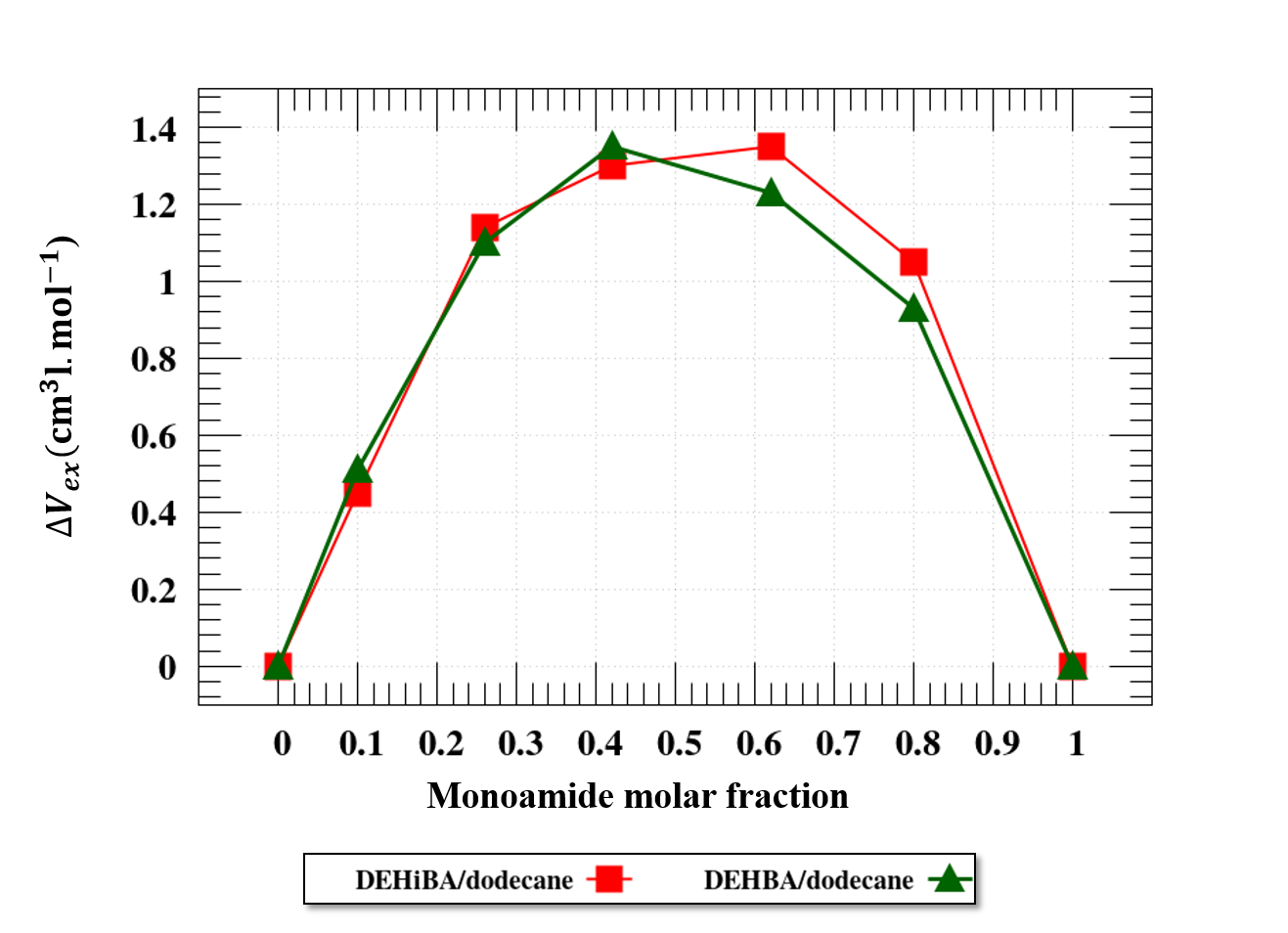}
  \end{subfigure}   \\   
  \begin{subfigure}[b]{0.9\linewidth}
      \centering
      \includegraphics[width=0.9\linewidth]{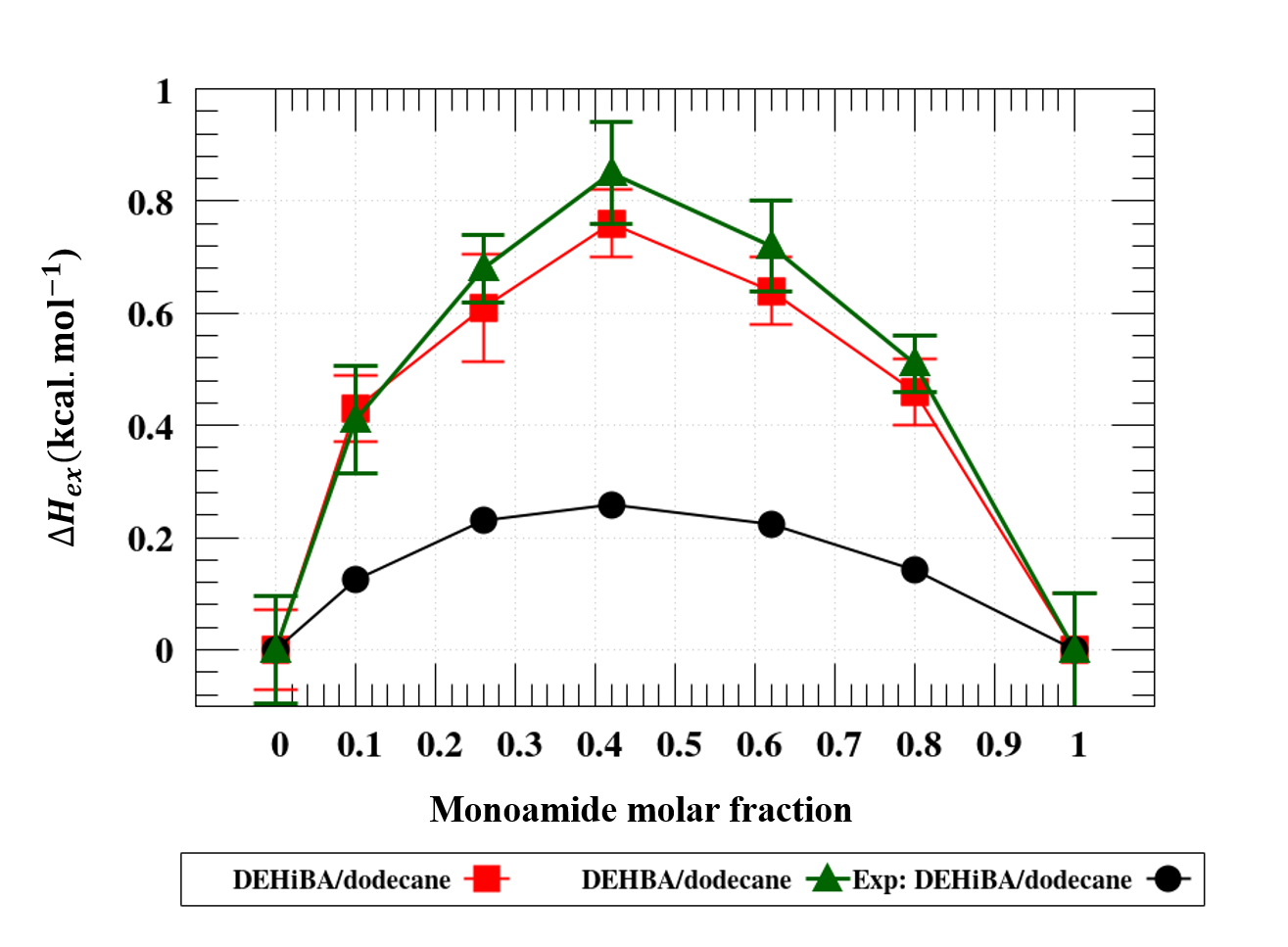}
  \end{subfigure}
\caption{Excess Volume ($\Delta V_{ex}$) and Excess Enthalpy ($\Delta H_{ex}$) of mixing of the DEHiBA/dodecane and DEHBA/dodecane binary mixtures.}
\label{excess} 
 \end{figure}

The excess volume of mixing and excess enthalpy of mixture are a result of complex combination of molecular properties between two molecular species such size, shape, as well as the inherent molecular interaction. These properties have been extensively calculated for TBP/n-alkanes mixtures in  the literature~\cite{cui2014molecular,servis2018role}. It was used as a measure of the force-field accuracy for such solutions and also to give insight into extractant-solvent interactions. Such data have been recently measured experimentally by Coquil~{\etal}~\cite{COQUIL2021}, but it remains of great interest to predict them with MD simulations. Our predictions are drawn in Figure~\ref{excess}.

The excess volume of mixing is found to be positive for the entire range of mole fraction, which indicates looser packing in the mixtures compared to the pure phases. The excess enthalpy of mixing appears to be endothermic for both the DEHBA/dodecane and DEHiBA/dodecane mixtures and reaches a maximum for the \SI{42}{\percent} mole fraction, hence dodecane shows an unfavorable mixing for all mole fractions and for both extractant ligands. Comparing to experimental data, our values are bit higher than expected, but the tendency and the endothermic behavior is well captured with a maximum around the \SI{42}{\percent} mole fraction.

The uncertainties for both properties were also estimated (see details in the ESI) and the average errors are about \SI{0.1}{\kcal} for the excess enthalpies and about \SI{0.01}{\cm\cubed\per\mol} for the excess volumes. Only the uncertainties for $\Delta H_{ex}$ are displayed in Figure~\ref{excess}. 

Recently, Coquil~{\etal}~\cite{COQUIL2021} have measured the enthalpy of mixing for DEHiBA with dodecane. A comparison to these experimental data with our values seem to overestimate this properties, however this does not affect the accuracy of our FF since, in the work of Servis{\etal}~\cite{servis2018role} and also the one of Arya Das {\etal}~\cite{das2018} have showed that it is important to get the correct sign for the excess enthalpy of mixing which proves the accuracy of the force field. 

The comparison of the properties of monoamide/dodecane mixtures to that reported by Servis~{\etal}~\cite{servis2018role} for  phosphoric extractants, such as tributyl phosphate (TBP), triamyl phosphate (TAP), dibutyl butyl phosphonate (DBBP) and diamyl amyl phosphonate (DAAP), shows that the enthalpy of mixing is overall larger, \SI{0.8}{\kcal}, for monoamides than for the organophosphorus extractant structures, e.g.~\SI{0.35}{\kcal} for TBP.

\section{Conclusion}
In this work, \textit{ab initio} based force-field model for short- and long-alkane chains and amide derivatives was successfully constructed by solely considering quantum chemical calculations (at the MP2 level of theory) and taking explicitly into account polarization effects. The different intra-molecular and inter-molecular parameters were derived and validated by performing MD simulations to calculate liquid-phase thermodynamic and structural properties. The reported simulation results for the alkanes are in great agreement with experimental data. These high-quality results for both molecular families suggest that this purely \textit{ab initio} parametrization methodology is promising and can be applied to other molecules of interest.

The developed FFs were also used to simulate and calculate properties of monoamide/dodecane mixtures, namely DEHiBA/dodecane and DEHBA/dodecane. The structural analysis revealed that for both mixtures, amide ligands tend to self-assemble (aggregate) in the organic solution. However, The RDFs calculations showed that the molecular organization for both systems is quite similar. The relative orientation analysis for the C=O carbonyl group and the amide function for both systems showed a small difference on the molecular organization for both monoamides. Excess volume and excess enthalpy have been also calculated and seemed to be quite similar for both systems.
All the results and predictions are the first step towards a more realistic description of more complex phenomena in which these mixtures could interact with strongly ionic complexes, such as plutonium nitrate ions, and be in contact with other types of solvent, namely water. In particular, mimicking liquid-liquid extraction remains a challenging task in term of FF development but also computational resources, since way longer simulations as well as larger systems will be required.

\section*{Conflicts}
There are no conflicts to declare.

\section*{Acknowledgements}
We first would like to thank M. Masella and P. Guilbaud for fruitful discussions. We acknowledge support by the French government through the Program “Investissement d’avenir” (Grants LABEX CaPPA/ANR- 11-LABX-0005-01 and I-SITE ULNE/ANR16-IDEX-0004 ULNE), as well as by the Ministry of Higher Education and Research, Hauts de France council and European Regional Development Fund (ERDF) through the Contrat de Projets {\'E}tat Region (CPER CLIMIBIO). We further acknowledge the "Groupement de recherche" GDR 2035 SolvATE. In addition, this work was granted access to the HPC resources of CINES/IDRIS/TGCC under the allocation
2019--2020 (A0070801859) made by GENCI. We also acknowledge the CEA for the Ph.D. grant given to A.F.





\clearpage
\bibliography{refs}


\end{document}


\clearpage
\section{Force-field parameters}
\begin{table}
\centering
\begin{tabular}{c|c|S[table-format=2.2]S[table-format=8.0]|S[table-format=1.1]|S[table-format=4.0]|S[table-format=2.2]S[table-format=7.0]}
\toprule
i     & j    & {$A_{ij}$}   & {$B_{ij}$}       & {damping} & {$C_{ij}$} & \multicolumn{2}{c}{correc tor (1-4) }  \\
\midrule
CT3    & CT3  & 3.86     & 320643     & 0.3  & 1850 & 4.82   & 1000000 \\
CT3    & CT2  & 3.80     & 320000     & 0.3  & 1500 & 4.82   & 1000000 \\
CT3    & NCT3 & 4.45     & 320000     & 0.3  & 1500 & 5.30   & 1000000 \\
CT3    & C    & 3.33     & 71347      & 0.3  & 1550 & 4.70   & 1000000 \\
CT3    & O    & 4.25     & 300000     & 0.3  & 0    & 4.16   & 1000000 \\
CT3    & HA   & 3.84     & 5440       & 0.5  & 0    & 7.50   & 75000   \\
CT3    & N    & 4.65     & 600000     & 0.3  & 700  & 4.90   & 1000000 \\
CT2    & CT2  & 4.55     & 320000     & 0.3  & 1400 & 4.82   & 1000000 \\
CT2    & NCT3 & 4.50     & 1004868    & 0.3  & 500  & 4.82   & 1000000 \\
CT2    & C    & 3.65     & 72000      & 0.3  & 1500 & 4.82   & 1000000 \\
CT2    & O    & 4.40     & 1000000    & 0.3  & 500  & 4.16   & 1000000 \\
CT2    & HA   & 3.70     & 5700       & 0.5  & 0    & 7.50   & 75000   \\
CT2    & N    & 4.15     & 304000     & 0.3  & 700  & 4.82   & 1000000 \\
NCT3   & C    & 3.43     & 101347     & 0.3  & 700  & 4.82   & 1000000 \\
NCT3   & O    & 4.25     & 500000     & 0.3  & 700  & 4.16   & 1000000 \\
NCT3   & HA   & 3.23     & 1476       & 0.5  & 0    & 7.50   & 75000   \\
NCT3   & N    & 4.75     & 500000     & 0.3  & 700  & 5.30   & 1000000 \\
C      &  C    & 4.50    & 304000     & 0.3  & 500  & 4.82   & 1000000 \\
C      &  O     & 4.20   & 500000     & 0.4  & 1000 & 5.30   & 1000000 \\
C      & HA     & 3.23   & 1476       & 0.5  & 0    & 7.50   & 75000   \\
C      &  N     & 4.30   & 1000000    & 0.3  & 700  & 5.30   & 800000  \\
O      &  O     & 5.10   & 900000     & 0.3  & 0    & 4.80   & 1000000 \\
O      & HA    & 7.00    & 75000      & 0.5  & 0    & 8.00    & 50000   \\
O      & N     & 5.20    & 500000     & 0.5  & 1200 & 5.10    & 1000000 \\
HA     & HA    & 3.65    & 1000       & 0.5  & 0    & 8.00    & 50000   \\
HA     &  N    & 5.50    & 350000     & 0.5  & 0    & 8.00    & 50000   \\
N      &  N    & 4.70    & 9000000    & 0.3  & 700  & 5.60    & 500000 \\
\bottomrule
\end{tabular}
\caption{The Buckingham, 1-4 interactions parameters derived and used in this work.}
\label{tabs:param}
\end{table}

\begin{table}[htp]
\caption{Partial CM5 charges (q(CM5) a.u.) and atomic polarizabilities ($\alpha$ in \si{\angstrom\cubed}) used for alkanes and amides derivatives.}
\label{CM5charges}
\begin{tabular}{lS[table-format=2.2]S[table-format=1.1]}
\toprule
{Atom Type} & {q(CM5)} & {$\alpha (\si{\angstrom\cubed})$}\\ 
\midrule
CT3     &   -0.21 & 2.0\\
CT2     &   -0.14 & 2.0\\
CT1     &   -0.07 & 2.0\\
CT2-N   &   -0.02 & 2.0\\
HN      &    0.10 & 0.0\\
HA      &   +0.07 & 0.0\\
N       &   -0.34 & 1.3\\
O       &   -0.42 & 1.3 \\
C       &    0.32 & 1.0\\
\bottomrule
\end{tabular}
\end{table}

\begin{table}
\caption{Fitted torsion coefficients for alkanes and alkyls groups.}
\label{dih-para}
\begin{tabular}{l|*{3}{S[table-format=1.2]}S[table-format=1.0]}
\toprule
Dihedral torsion & {$K_1$} & {$K_2$} & {$K_3$} & {$\Phi_0$} \\
\midrule
CT3-CT2-CT2-CT3  & 2     & 0.61  & 2.42  & 0        \\
CT3-CT2-CT2-CT2  & 1.5   & 0.45  & 2.16  & 0        \\
CT2-CT2-CT2-CT2  & 1.3   & 0.36  & 1.97  & 0        \\
CT2-CT1-CT2-CT2  & 1.15  & 0.36  & 2     & 0        \\
HA-CT-CT-HA      & 0.25  & 0     & 0     & 0       \\
\bottomrule
\end{tabular}
\end{table}

\clearpage
\section{Polarization interaction term}
The corresponding polarization energy term is defined as
\begin{equation}	
U_{pol}  = \frac{1}{2}\sum_{i=1}^{N_{\mu}}\frac{\bm{\mu_i}^2}{\alpha_i} - \sum_{i=1}^{N_{\mu}}{\bm{\mu_i}.\bm{E_{i}^{q}}} - \frac{1}{2}\sum_{i=1}^{N_{\mu}}\sum_{j=1,j\neq i}^{N_{\mu}} \bm{\mu_i}\mathrm{T}_{ij}\bm{\mu_j},
\end{equation}
with the dipolar tensor defined as:
\begin{equation}
\mathrm{T}_{ij}=\frac{1}{4\pi\epsilon_0}\left( \frac{f_5(r_{ij})}{r_{ij}^5} \begin{bmatrix}
x^2 & xy & xz \\ 
xy & y^2 & yz \\ 
xz & yz & z^2
\end{bmatrix}  -\frac{f_3(r_{ij})}{r_{ij}^3}I_3\right) ,
\end{equation}
Here, $I_3$ is the identity matrix. The $f_5$ and $f_3$ functions are introduced to account for short-range damping effects to prevent the so-called "polarization catastrophe", proposed by Thole \cite{thole} as:
\begin{equation}
f_3(r_{ij})=1-exp(- {a_{ij}}r_{ij}^3),
\end{equation}
\begin{equation}
f_5(r_{ij})=1-(1+ {a_{ij}}r_{ij}^3)exp(- {a_{ij}}r_{ij}^3),
\end{equation}
where $a_{ij}$ is an adjustable damping factor, a parameter which depends on the nature of  atoms $i$ and $j$. In the present study, the damping factor was fixed at 0.3 at the beginning for all polarizable sites and fitted when needed for a better reproduction of the long-range interaction (See Table~\ref{tabs:param}), the 0.3 value was chosen arbitrary.
\clearpage

\section{Intramolecular potential and Interaction energy curves}
Herein, we reported the energy profiles of the dihedral angle scans  in the docedane and  just some of the dimer structures with the corresponding interaction energy curves as a function of distance between the two molecules. In total, we fitted over 40 other structures and overall the MM energy curves fits well all the QM reference data.

\begin{figure*}[ht]
\centering
\includegraphics[width=\linewidth]{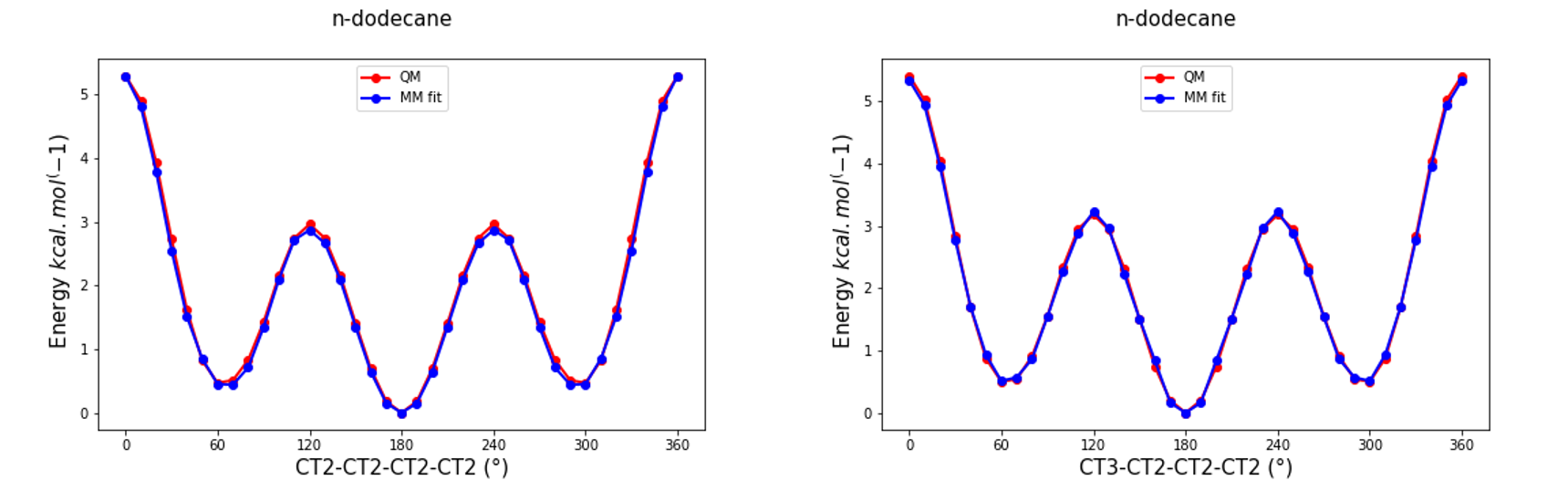}
\caption{Relative energy profiles of the dihedral angle scans in dodecane (\ce{{CT2}-{CT2}-{CT2}-{CT2}} middle torsion on the left and \ce{{CT3}-{CT2}-{CT2}-{CT2}} terminal torsion on the right). The red lines represent the MP2 QC curves, while the blue ones the fitted FFs.}
\label{fig:dih-dod}
 \end{figure*}

 \begin{figure}
 \centering
      \includegraphics[width=.9\linewidth]{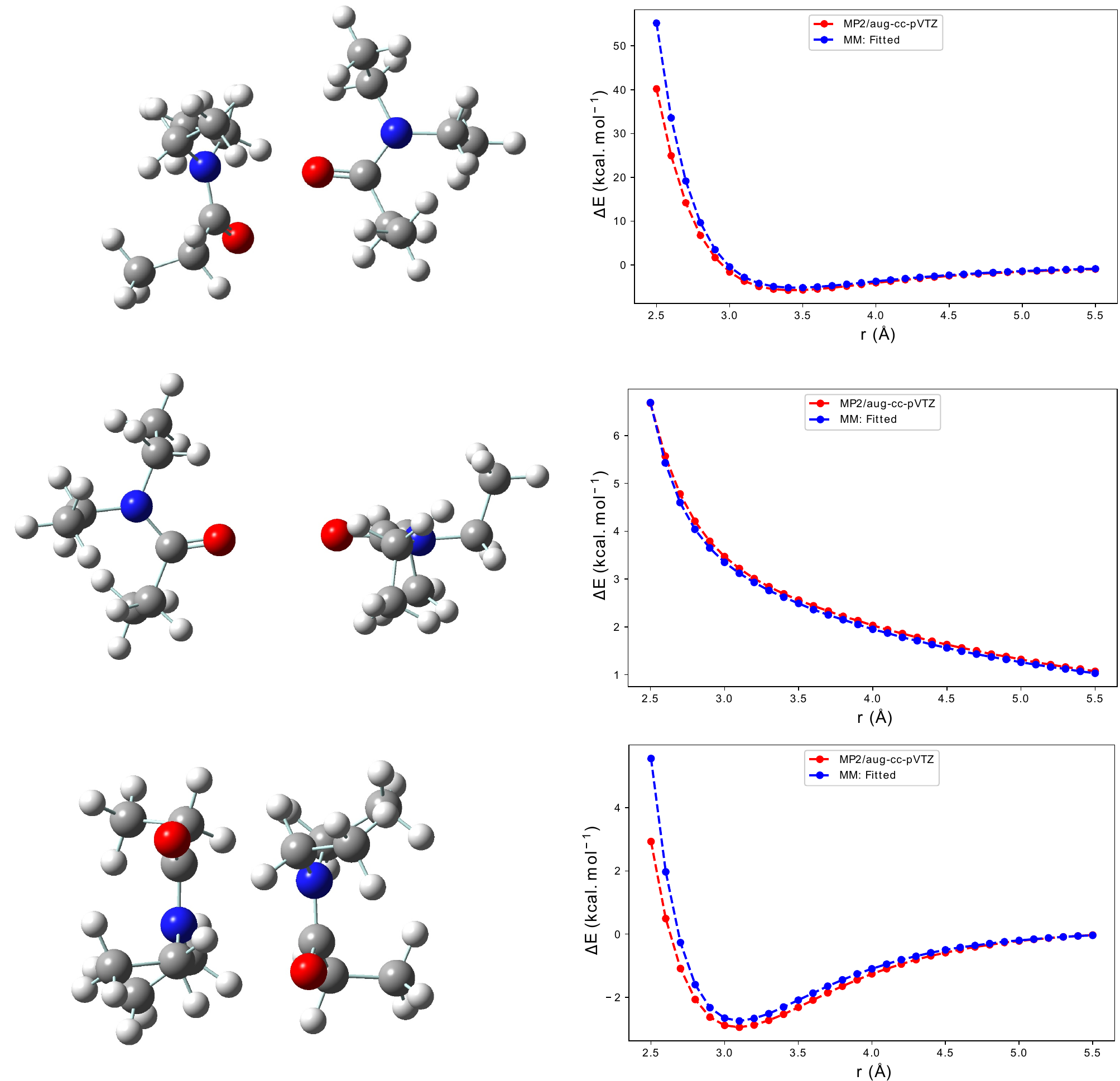}
\caption{Selected DEPA dimer structures on the left with the corresponding interaction energy curves as a function of distance between the two molecules. MP2 values with aug-cc-pVTZ basis sets are in red and the fitted FF in blue.}
\label{fig:SI-IE1}
 \end{figure}

 \begin{figure}
 \centering
      \includegraphics[width=.9\linewidth]{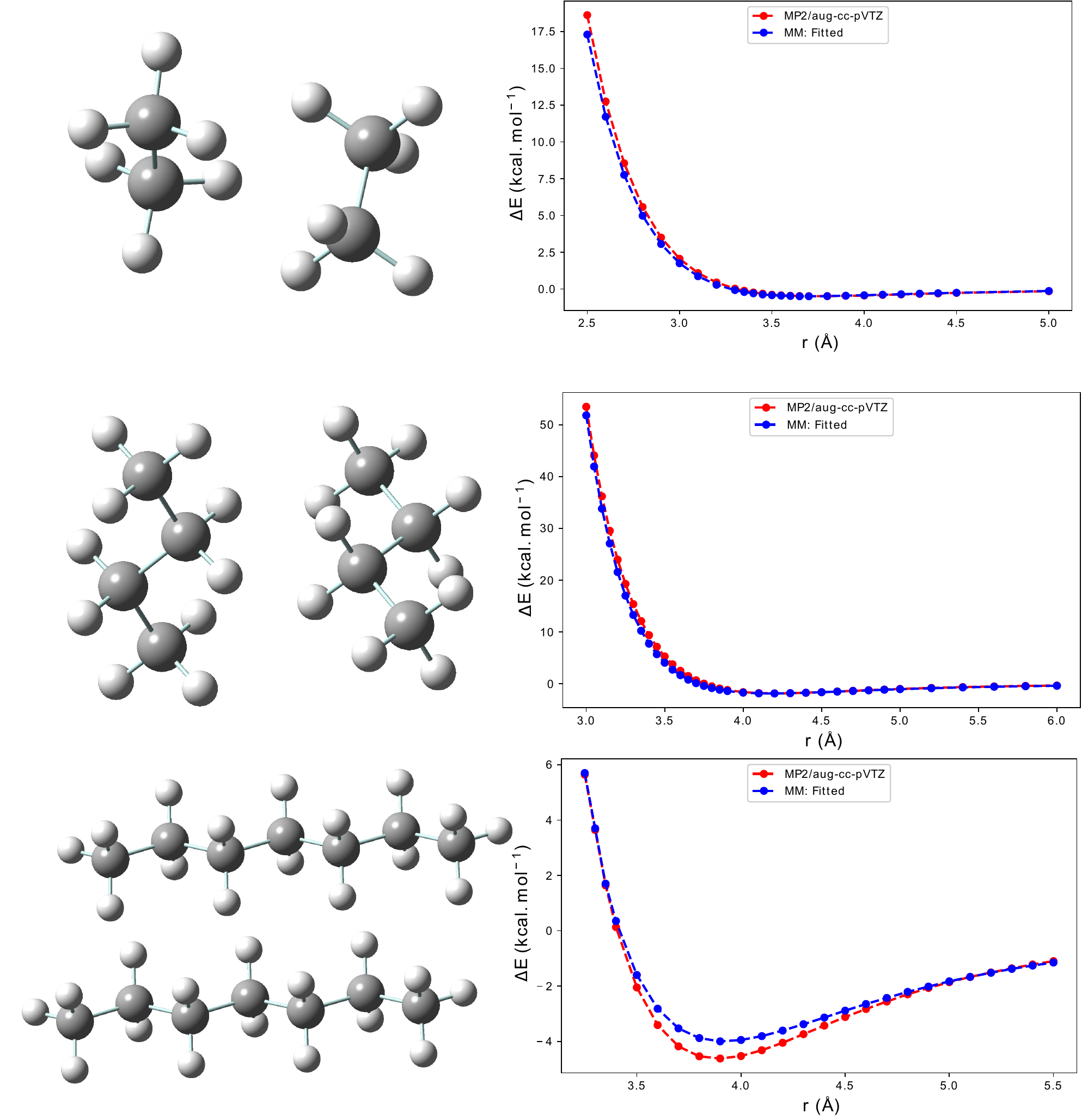}
\caption{Selected dimer structures (for ethane, butane, and heptane) opposite to their interaction energy curves as a function of distance between the two molecules. MP2 values with aug-cc-pVTZ basis sets are in red and the fitted FF in blue.}
\label{fig:SI-IE2}
 \end{figure}
 \begin{figure}
 \centering
      \includegraphics[width=.9\linewidth]{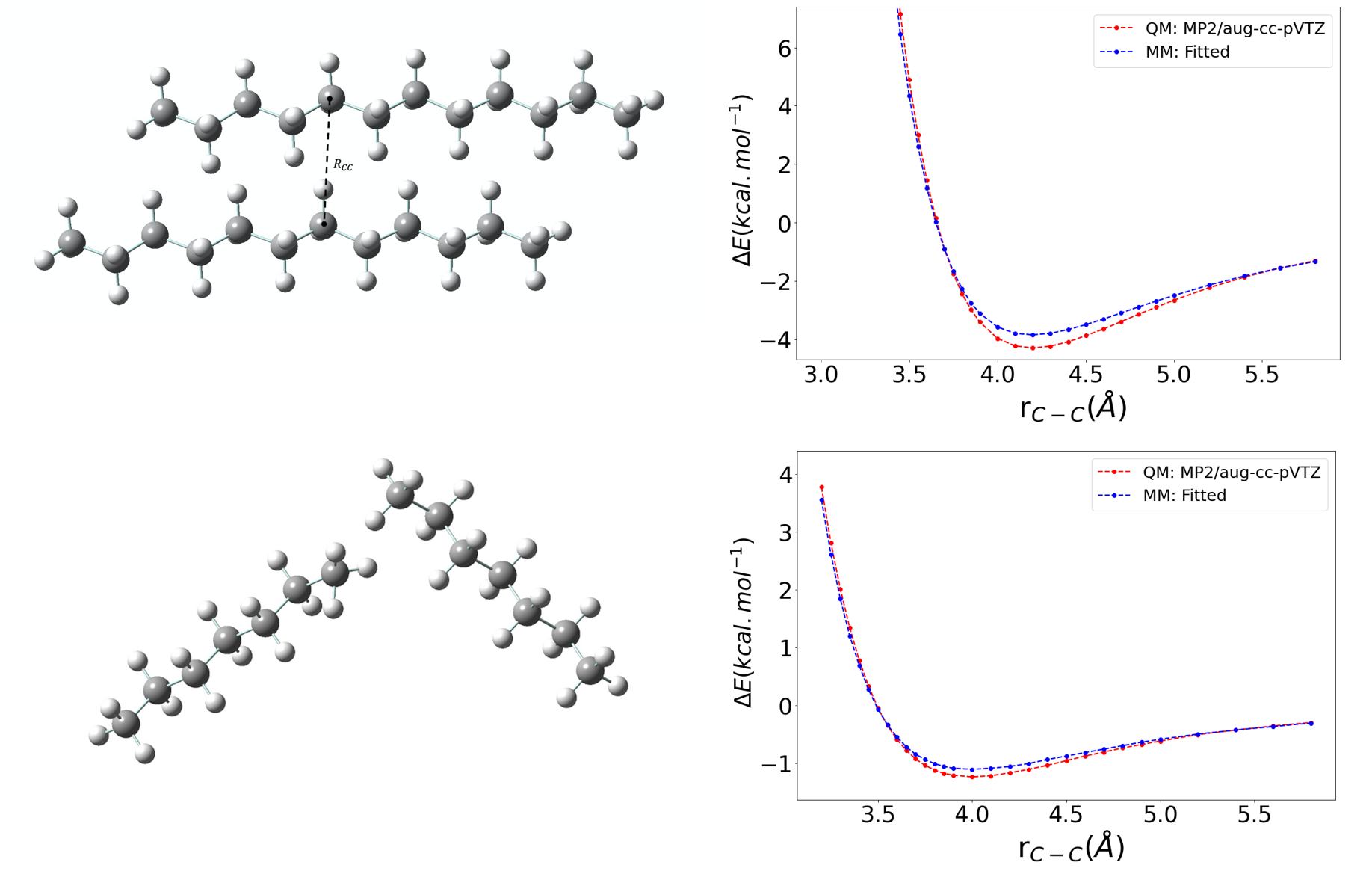}
\caption{Selected dimer structures for dodecane opposite to their interaction energy curves as a function of distance between the two molecules. MP2 values with aug-cc-pVTZ basis sets are in red and the fitted FF in blue.}
\label{fig:SI-IE3}
 \end{figure}
 
  \begin{figure}
 \centering
      \includegraphics[width=.9\linewidth]{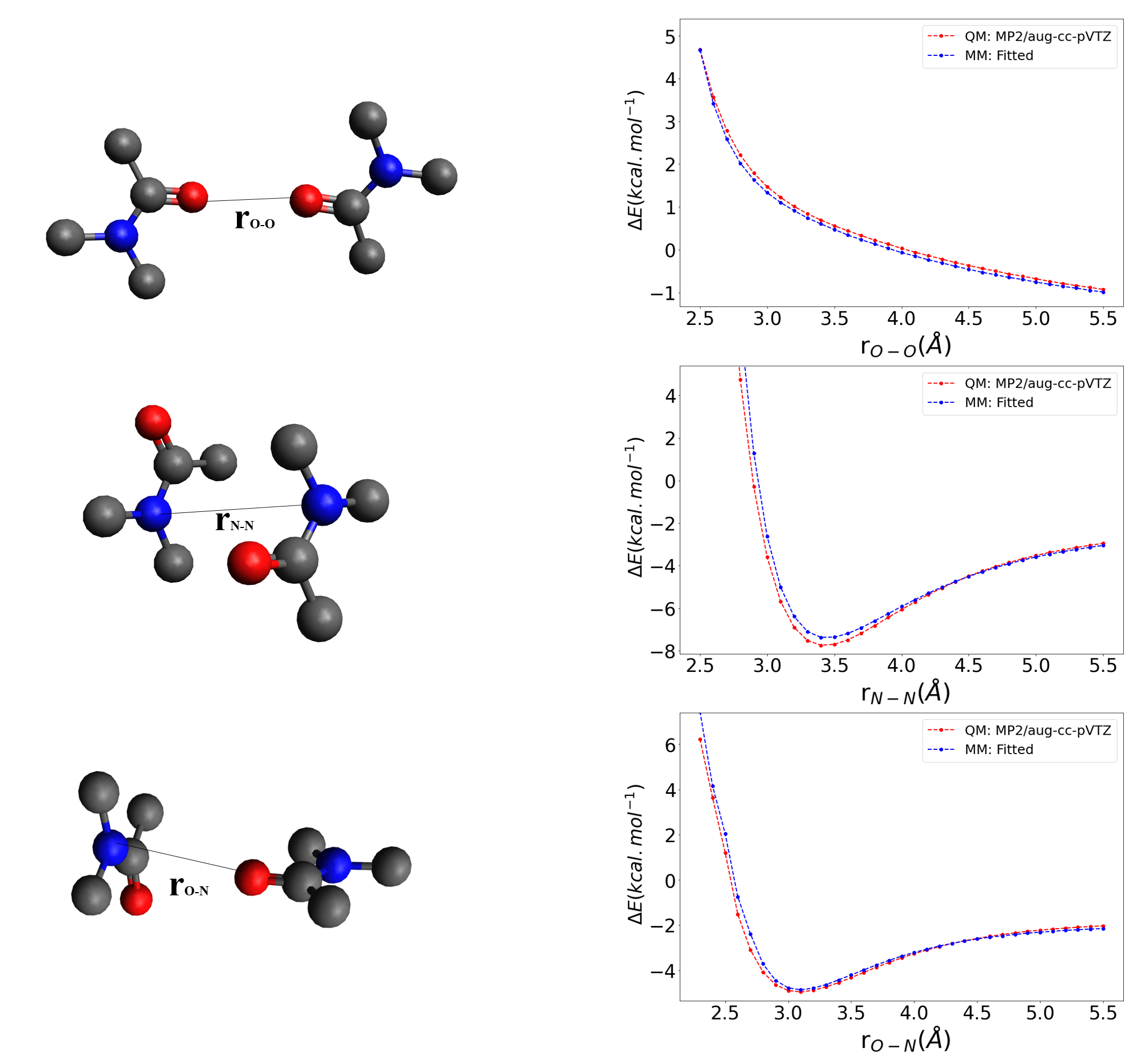}
\caption{Selected dimer structures for DEHBA opposite to their interaction energy curves as a function of distance between the two molecules. MP2 values with aug-cc-pVTZ basis sets are in red and the fitted FF in blue.}
\label{fig:SI-IE4}
 \end{figure}

\begin{figure}
\centering
\includegraphics[width=0.95\linewidth]{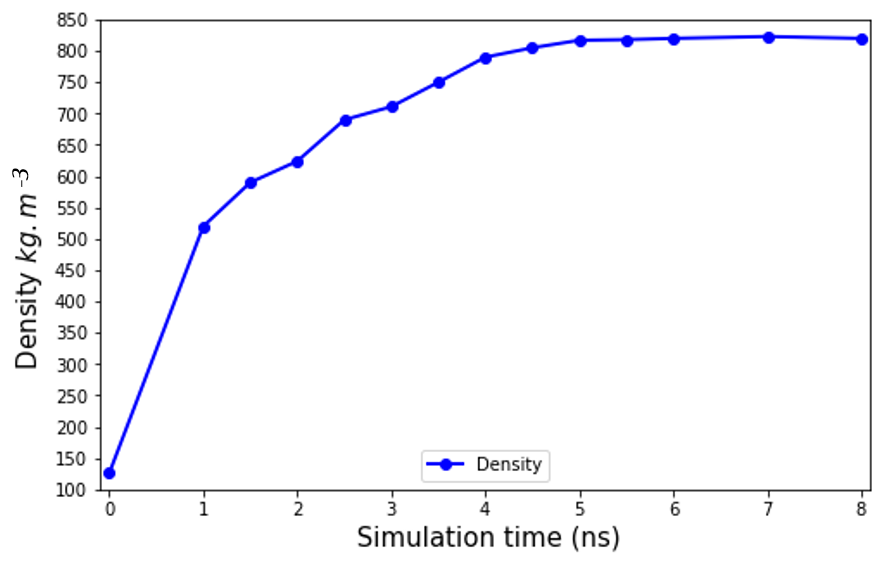}
\caption{Time evolution of the density  for 26\% DEHBA in dodecane during the equilibration phase up to \SI{8}{\ns}.}
\label{Sfig:density-analysis-time}
\end{figure}

\newpage 
\section{Error estimation}
The approach used to estimate the errors in this work is based entirely on the statistical nature of the results. Assuming that a MD simulation is performed for a total period of $t_{tot}$ (production time). We select the last 20\% of the production (noted $t_{st}$, we can divide $t_{st}$ into $N$ segments with the end point of each segment being $t_i = i \Delta t$ (with $i = 1, 2, . . . , N $) where $\Delta t = t_{st}/N $. Any time-averaged property can be calculated for each of the time intervals $\Delta t_i = t_i - t_{i-1} = \Delta t$, and as a result, each MD simulation will produce $N$ values of the property $P$. If we denote each estimate of $P$ to be $P_i$ (with $i = 1,2,...,N$), the best estimate of the property can be calculated as :
\begin{equation}
    \overline{P} = \frac{\sum_{i}{P_i}}{N}
\end{equation}
The uncertainty of the samples $P_i$ was quantified "in our work" by the sample standard deviation defined as :
\begin{equation}
   \sigma = \sqrt{\frac{\sum_{i}{(P_i - \overline{P})^2}}{N-1}}.
\end{equation}

The estimation of $P$ also associated with an uncertainty $\Delta P = \sigma $ is written as :
 \begin{equation}
     P = \overline{P} \pm \Delta P.
\end{equation}
In our case, the uncertainties for the density ($\rho$), heat of vaporization ($\Delta H_{vap}$), the excess volume ($\Delta V_{ex}$) and excess enthalpy  ($\Delta H_{ex}$) were evaluated by considering the uncertainty on volumes and energies. The two latter being weighted sums of several variables $x, y, z,$ the errors can be estimated by : 
\begin{equation}
\sigma_f = \sqrt{\left(\pdv{f}{x}.\sigma_x\right)^2 + \left(\pdv{f}{y}.\sigma_y\right)^2 + \left(\pdv{f}{x}.\sigma_z\right)^2 },
\end{equation}
where $\sigma_f$, $\sigma_x$, $\sigma_y$ and $\sigma_z$ are the absolute errors associated to the observable $f, x, y$ and $z$, respectively.

\clearpage

\begin{table}
\caption{Impact of the simulation box size on ligands properties, namely the density ($\rho$) and heat of vaporization ($\Delta H_{vap}$). Experimental values are taken from Haynes \etal~\cite{haynes2014crc}. Relative deviations $\delta$ (in \%) and absolute deviation $\Delta$ (in \si{\kcal}) with respect to experimental values are reported.}
\label{tab:boxsize}
\begin{tabular}{l*4{S[table-format=3.0]}*4{S[table-format=2.1]}}

\toprule
Ligand & \multicolumn{4}{c}{$\rho$ (\si{\kg\per\cubic\meter})} & \multicolumn{4}{c}{$\Delta H_{vap}$ (\si{\kcal})}\\
 \cmidrule(lr){1-1}\cmidrule(lr){2-5}\cmidrule(lr){6-9}
                        & \multicolumn{2}{c}{\# Molecules} & & & \multicolumn{2}{c}{\# Molecules} & &  \\
\cmidrule{2-3}\cmidrule{6-7}
& {343}  & {729}  &  {\expe} & {$\delta$(\%)}                    
& {343}  & {729}  &  {\expe}  & {$\Delta$(\si{\kcal})} \\
\midrule
DMA                     & 937           & 925           & 900                   & \numrange{4}{3}              & 10.2          & 10.4          & 10.9                   & 0.7                            \\
Butane                  & 596           & 592           & 601                   & \numrange{1}{2}              & 3.6           & 3.7           & 4.4                    & 0.8                            \\
Heptane                 & 700           & 690           & 677                   & \numrange{3}{2}              & 7.7           & 7.8           & 8.6                    & 0.9                            \\
DEHiBA                  & 898           & 892           & 865                   &  \numrange{4}{3}              & 23.3          & 23.3          &                       &                             \\
\bottomrule
\end{tabular}
\end{table}
The analysis of the data in Table~\ref{tab:boxsize} reveals that the maximum deviation between the two box sizes is 3\% for density and \SI{0.2}{\kcal} for the heat of vaporization, hence we conclude that the initial simulation boxes were adequately chosen, and that the box size effect is in our case is minor.

\begin{table}
\caption{Composition of DEHiBA/dodecane and DEHBA/dodecane mixtures, associated simulation boxes and densities $\rho$ (\si{\kilogram\per\meter\cubed}). L denotes the extractant ligand. Experimental values are taken from Ref.~\citenum{COQUIL2021}.}
\begin{tabular}{*4{S[table-format=1.2]}*7{S[table-format=3.0]}}
\toprule
\multicolumn{4}{c}{$C_{mono}$ (\si{\mol\per\liter})} &
\multicolumn{4}{c}{$\rho$ (\si{\kilogram\per\meter\cubed})} &
\multicolumn{3}{c}{\# Molecules}\\
\cmidrule(lr){1-4}\cmidrule(lr){5-8}\cmidrule(lr){9-11}
\multicolumn{2}{c}{DEHiBA} & \multicolumn{2}{c}{DEHBA} &
\multicolumn{2}{c}{DEHiBA} & \multicolumn{2}{c}{DEHBA} &\\
\cmidrule(lr){1-2}\cmidrule(lr){3-4}\cmidrule(lr){5-6}\cmidrule(lr){7-8}
{\expe} & {MD} & {\expe} & {MD} & {\expe} & {MD} & {\expe} & {MD} & {L} & {dodecane} &{\% L}\\
\midrule
0.40 & 0.42 & {-} & 0.43 & 765 & {-} & 800 & 802 & 34 & 316 & 10\\
1.00 & 1.04 & {-} & 1.07 & 790 & {-} & 822 & 828 & 92 & 258 & 26\\
1.50 & 1.44 & {-} & 1.47 & 811 & {-} & 841 & 850 & 149 & 201 & 42\\
2.00 & 2.07 & {-} & 2.03 & 832 & {-} & 863 & 875 & 217 & 133 & 62\\
\bottomrule 
\end{tabular}
\label{mix-density}
\end{table}

\begin{table}
\begin{tabular}{*5{S[table-format=1.2]}}
\toprule
   & \multicolumn{2}{c}{DEHiBA/dodecane}   & \multicolumn{2}{c}{DEHBA/dodecane}  \\ 
   \cmidrule(lr){2-3} \cmidrule(lr){4-5}
{$x_{mono}$}      & {(\ce{O-O})RDF}  &   {(\ce{N-N})RDF}  & {(\ce{O-O})RDF}  & {(\ce{N-N})RDF}  \\
\midrule
0.10        & 0.67    &  0.65    & 0.63    &  0.63  \\
0.26        & 1.23    & 1.22     & 1.17    & 1.20   \\
0.42        & 1.54    & 1.55    & 1.51    & 1.50   \\
0.62        & 1.97    &  1.93    & 1.92    & 1.91   \\
1.00        & 2.40    & 2.43     & 2.40    & 2.40  \\
\bottomrule
\end{tabular}
\caption{Coordination numbers derived from the radial distribution function of oxygen and nitrogen atoms, for the DEHiBA/dodecane and DEHBA/dodecane mixtures as a function of the $x_{mono}$ mole fraction of monoamide.}
\end{table}

\clearpage

\begin{figure}
\centering
\includegraphics[width=0.95\linewidth]{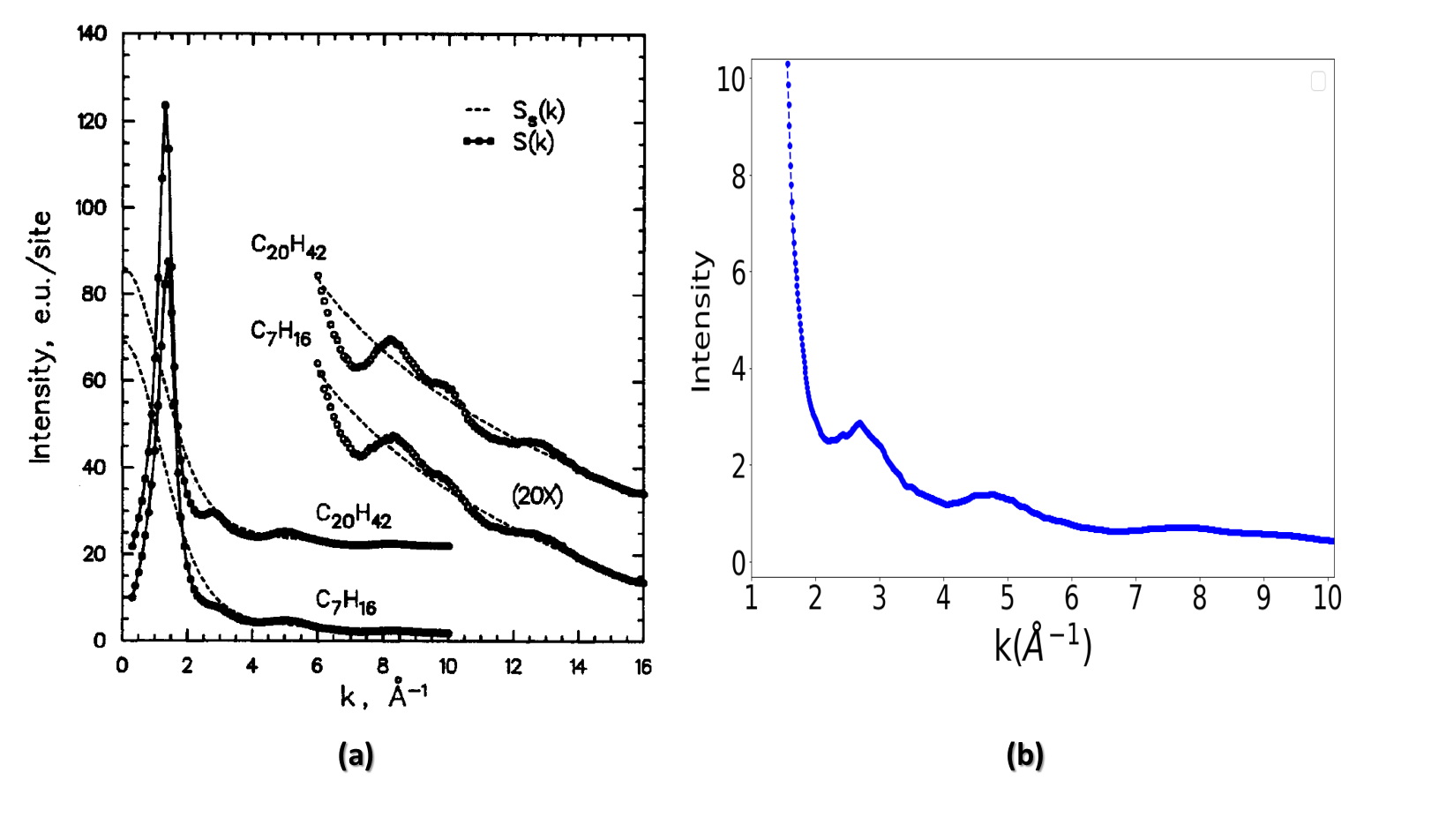}
\caption{Structure factors for heptane, (a) Experimental data from Habenschuss~{\etal}~\cite{habenschuss1990x}, (b) simulations using Debyer code\cite{Debyer}.}
\label{Sfig:SFheptane}
\end{figure}
\clearpage

\begin{figure}
    \begin{subfigure}[t]{0.47\linewidth}
        \centering
        \includegraphics[width=\linewidth]{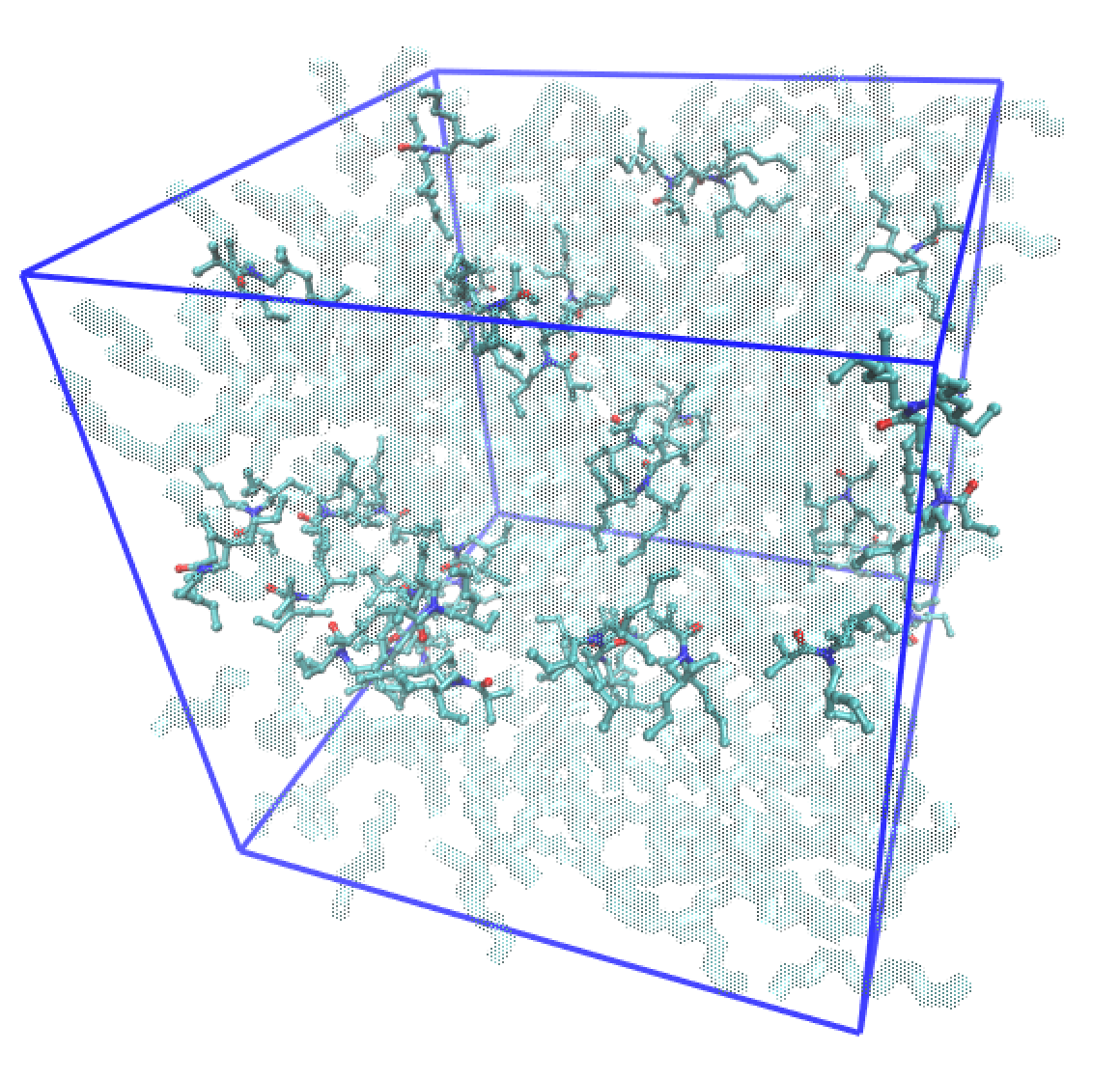}
        \caption{10\% DEHiBA}
        \label{fig:box-dehiba-10pp}
    \end{subfigure}
        \begin{subfigure}[t]{0.47\linewidth}
        \centering
        \includegraphics[width=\linewidth]{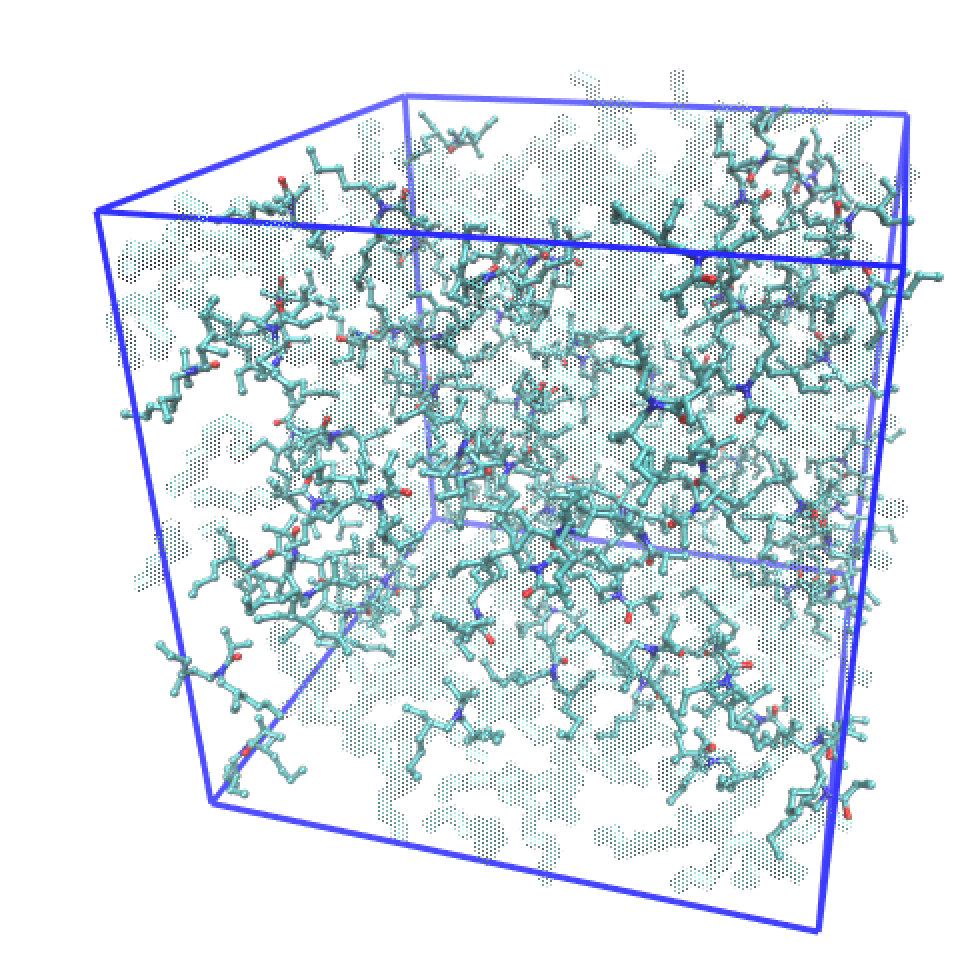}
      \caption{26\% DEHiBA}
        \label{fig:box-dehiba-26pp}
    \end{subfigure}
        \begin{subfigure}[t]{0.47\linewidth}
        \centering
        \includegraphics[width=\linewidth]{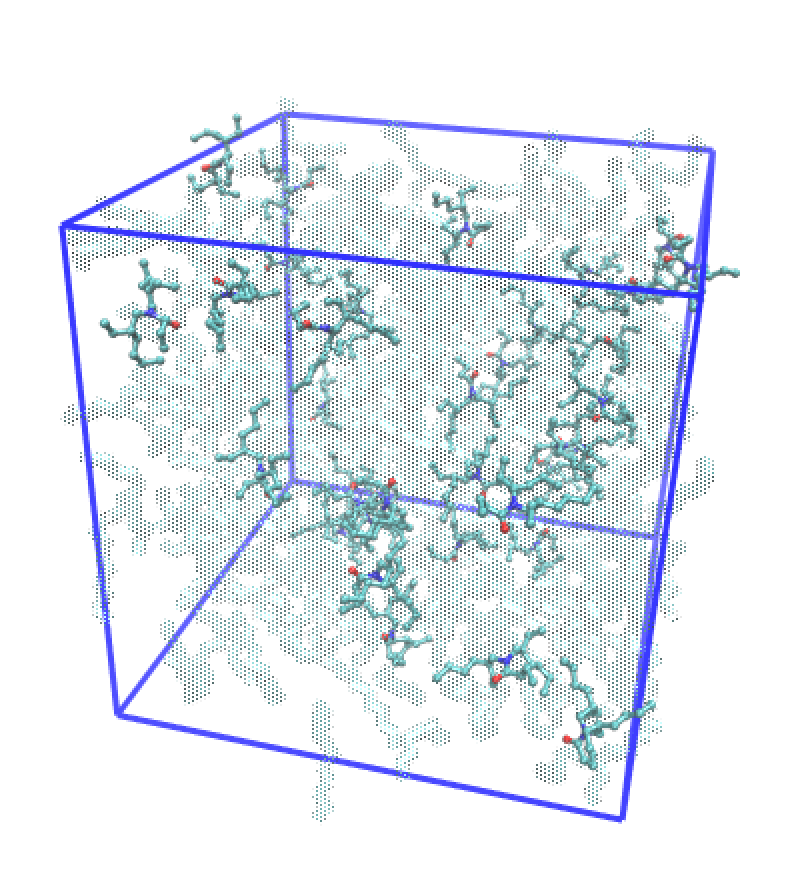}
      \caption{10\% DEHBA}
        \label{fig:box-dehba-10pp}
    \end{subfigure}
        \begin{subfigure}[t]{0.47\linewidth}
        \centering
        \includegraphics[width=\linewidth]{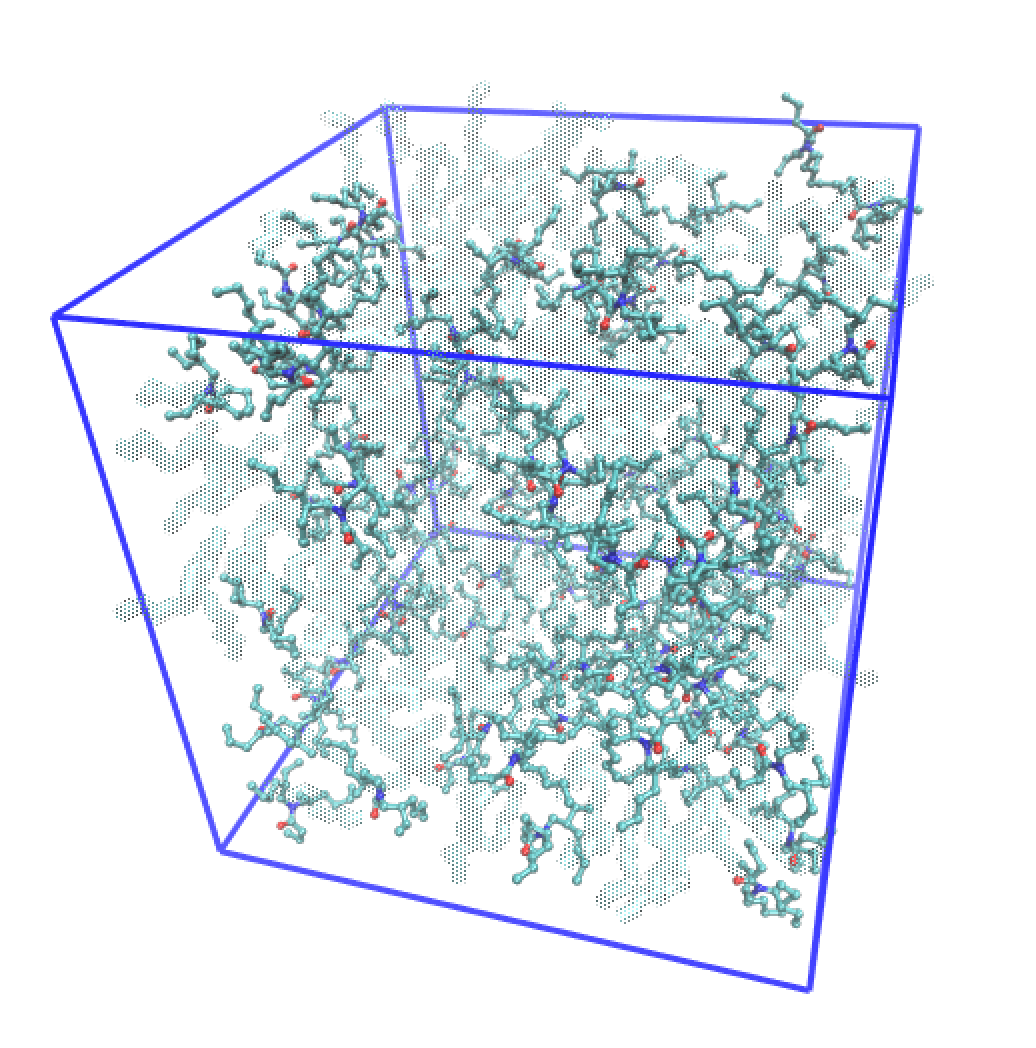}
     \caption{26\% DEHBA}
    \label{fig:box-dehba-26pp}
    \end{subfigure}
\caption{Perspective views illustrating self-assembly of the monoamides in DEHiBA/dodecane (a--b) and DEHBA/dodecane (c--d) mixtures, with \SIrange{10}{26}{\percent} of monoamide concentrations. Dodecane is greyed for clarity.}
\label{fig:box-mixtures}
\end{figure}

\begin{figure}[ht]
      \begin{subfigure}[b]{0.45\linewidth}
      \centering
       \includegraphics[width=\linewidth]{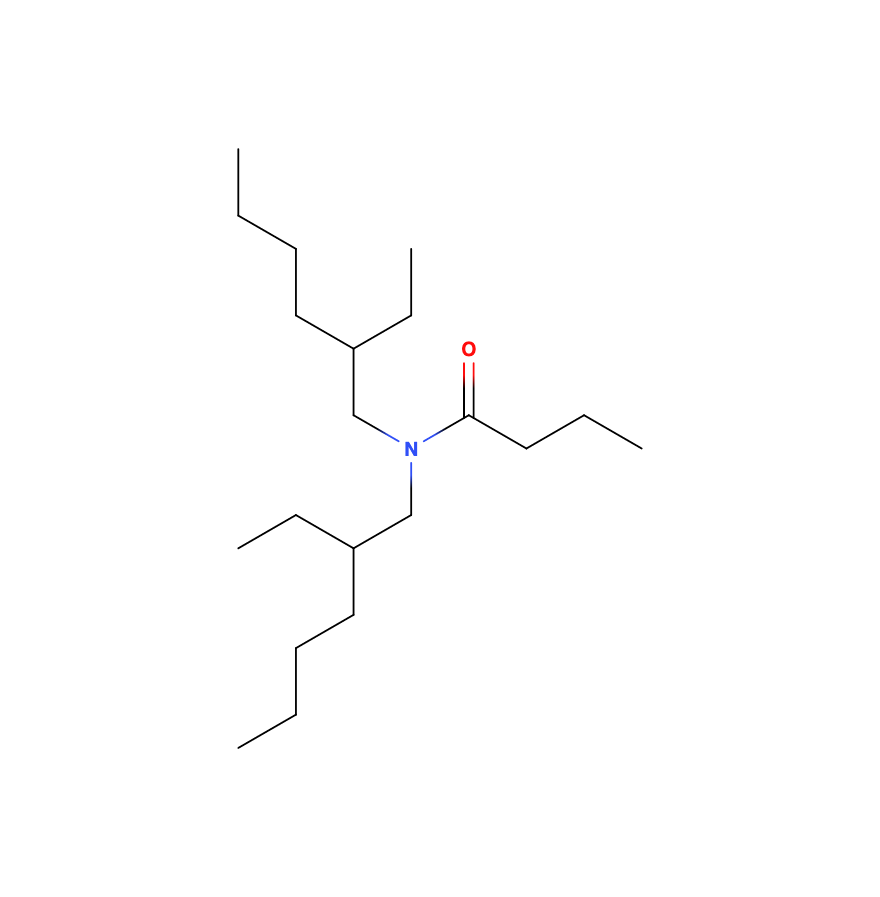}
       \caption*{(a)}
      \end{subfigure}
            \begin{subfigure}[b]{0.45\linewidth}
      \centering
       \includegraphics[width=\linewidth]{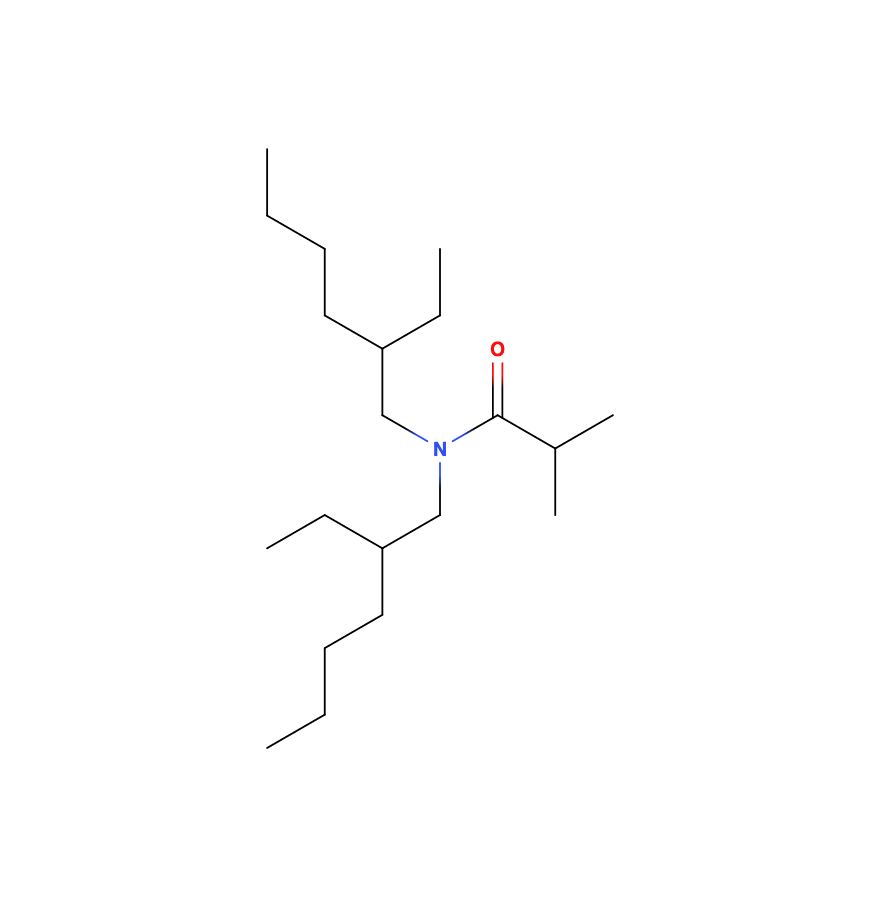}
       \caption*{(b)}
      \end{subfigure}
    \caption{Structure of (a) di-2-ethylhexyl-butyramide (DEHBA) and (b) di-2-ethylhexylisobutyramide (DEHiBA).}
    \label{dehibavsdehba}
\end{figure}

\begin{figure}
    \begin{subfigure}[t]{0.49\linewidth}
        \centering
        \includegraphics[width=\linewidth]{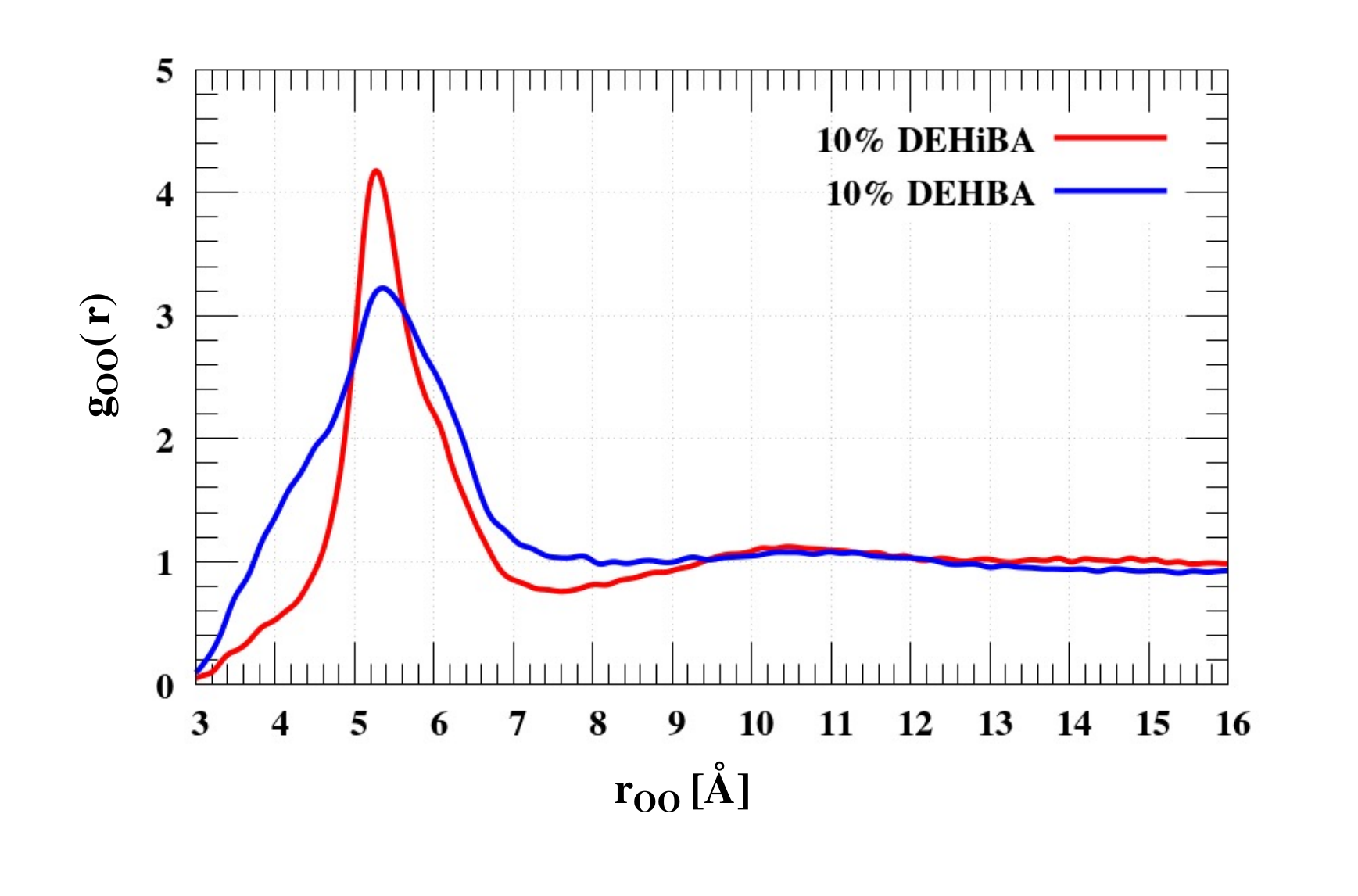}
       \caption{$x_{mono}$ = \SI{10}{\percent}}
       \label{fig:rdf-oo-10pp}
    \end{subfigure}
        \begin{subfigure}[t]{0.49\linewidth}
        \centering
        \includegraphics[width=\linewidth]{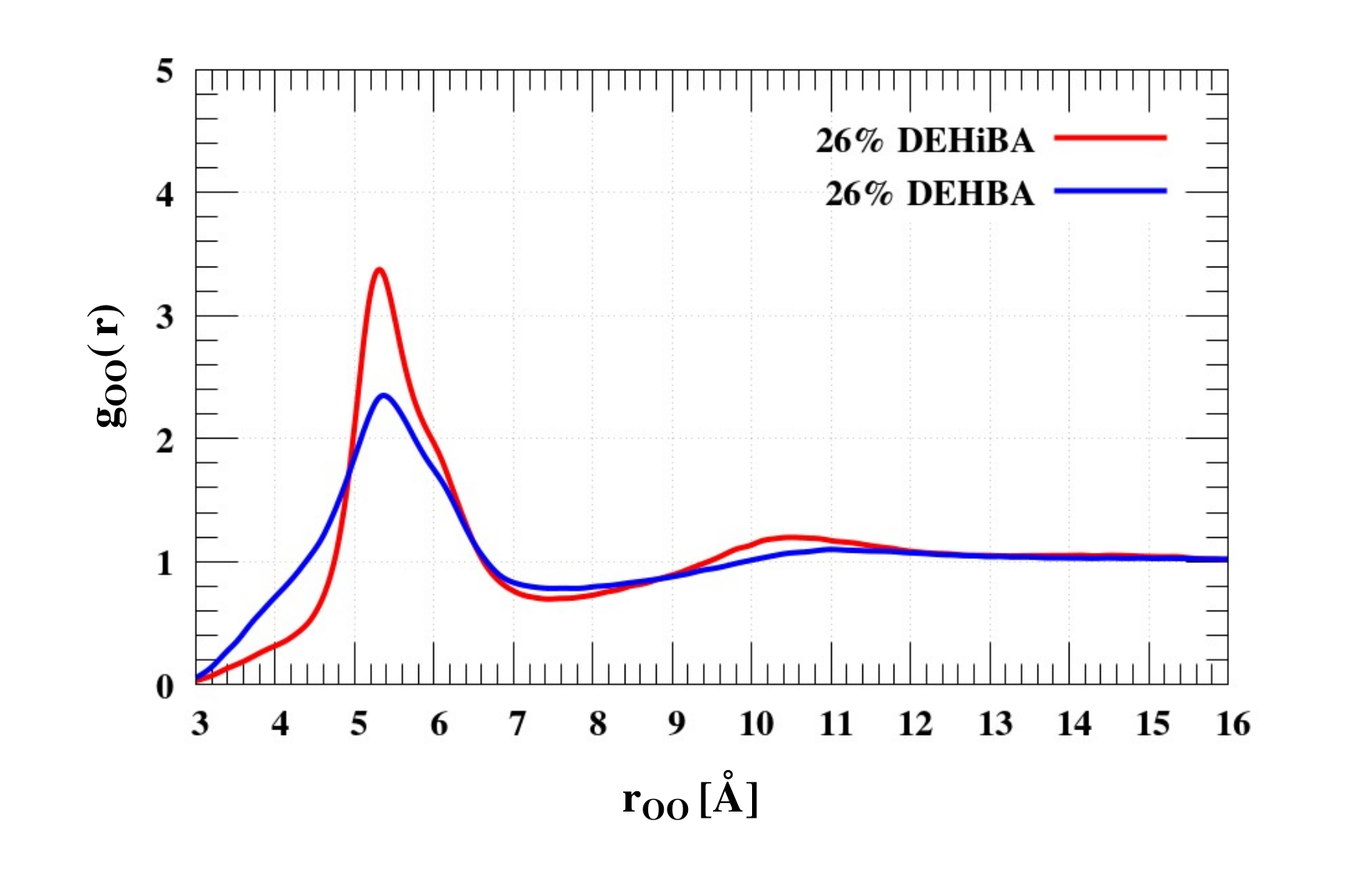}
       \caption{$x_{mono}$ = \SI{26}{\percent}}
       \label{fig:rdf-oo-26pp}
    \end{subfigure}
    \begin{subfigure}[t]{0.49\linewidth}
      \centering
      \includegraphics[width=\linewidth]{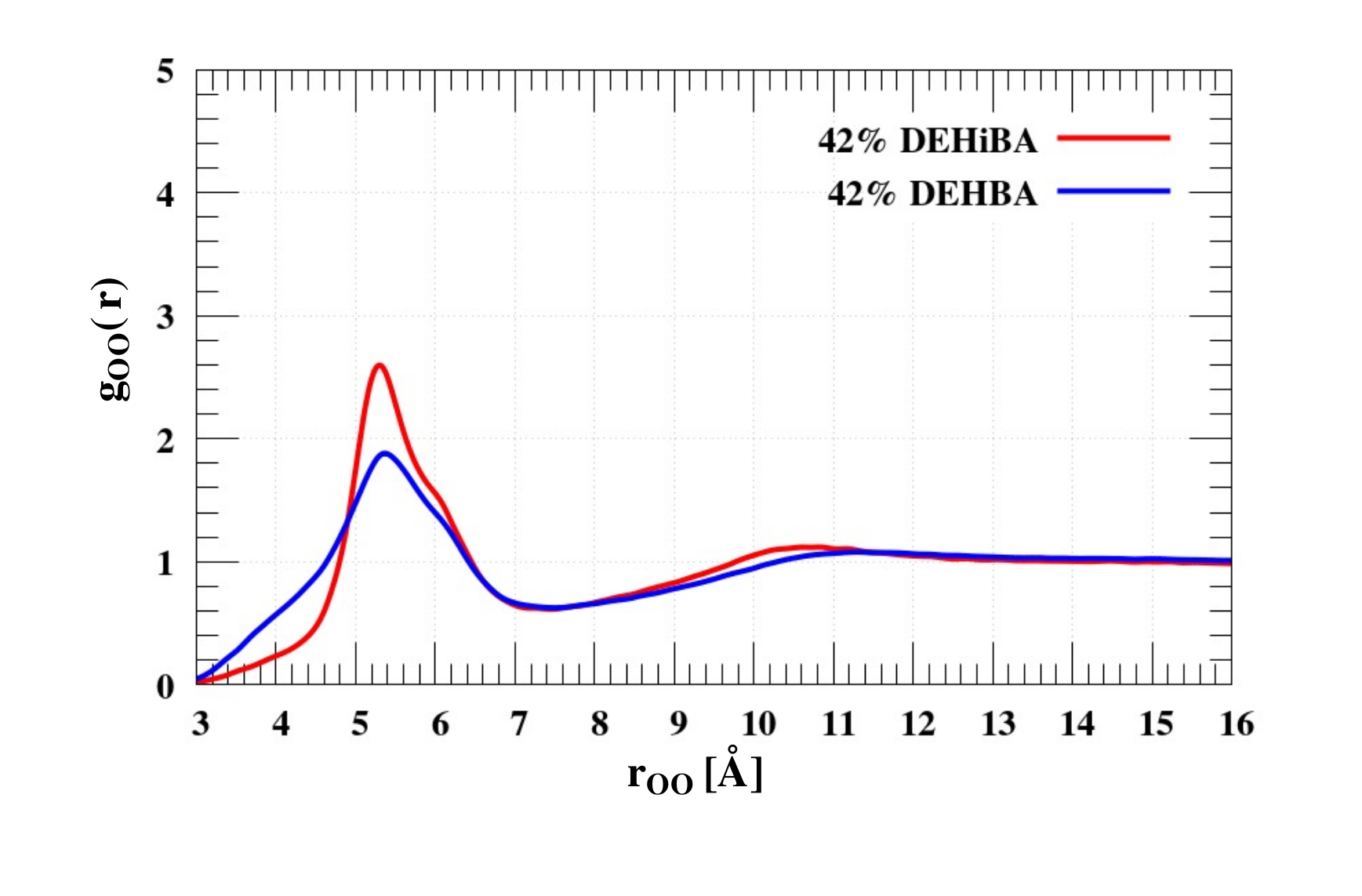}
       \caption{$x_{mono}$ = \SI{42}{\percent}}
       \label{fig:rdf-oo-42pp}
    \end{subfigure}
    \begin{subfigure}[t]{0.49\linewidth}
      \centering
      \includegraphics[width=\linewidth]{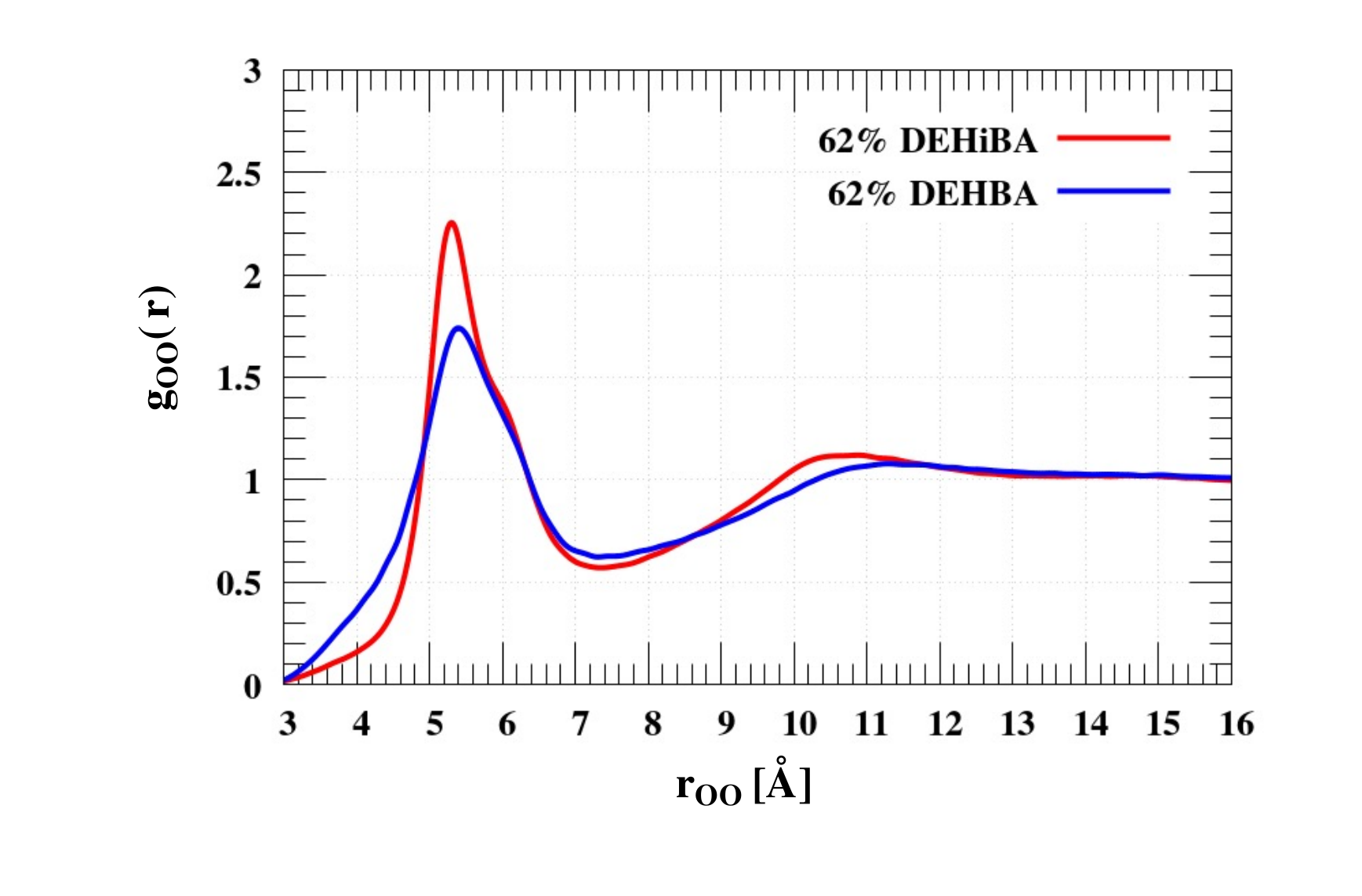}
       \caption{$x_{mono}$ = \SI{62}{\percent}}
       \label{fig:rdf-oo-62pp}
    \end{subfigure}
        \begin{subfigure}[t]{0.49\linewidth}
      \centering
      \includegraphics[width=\linewidth]{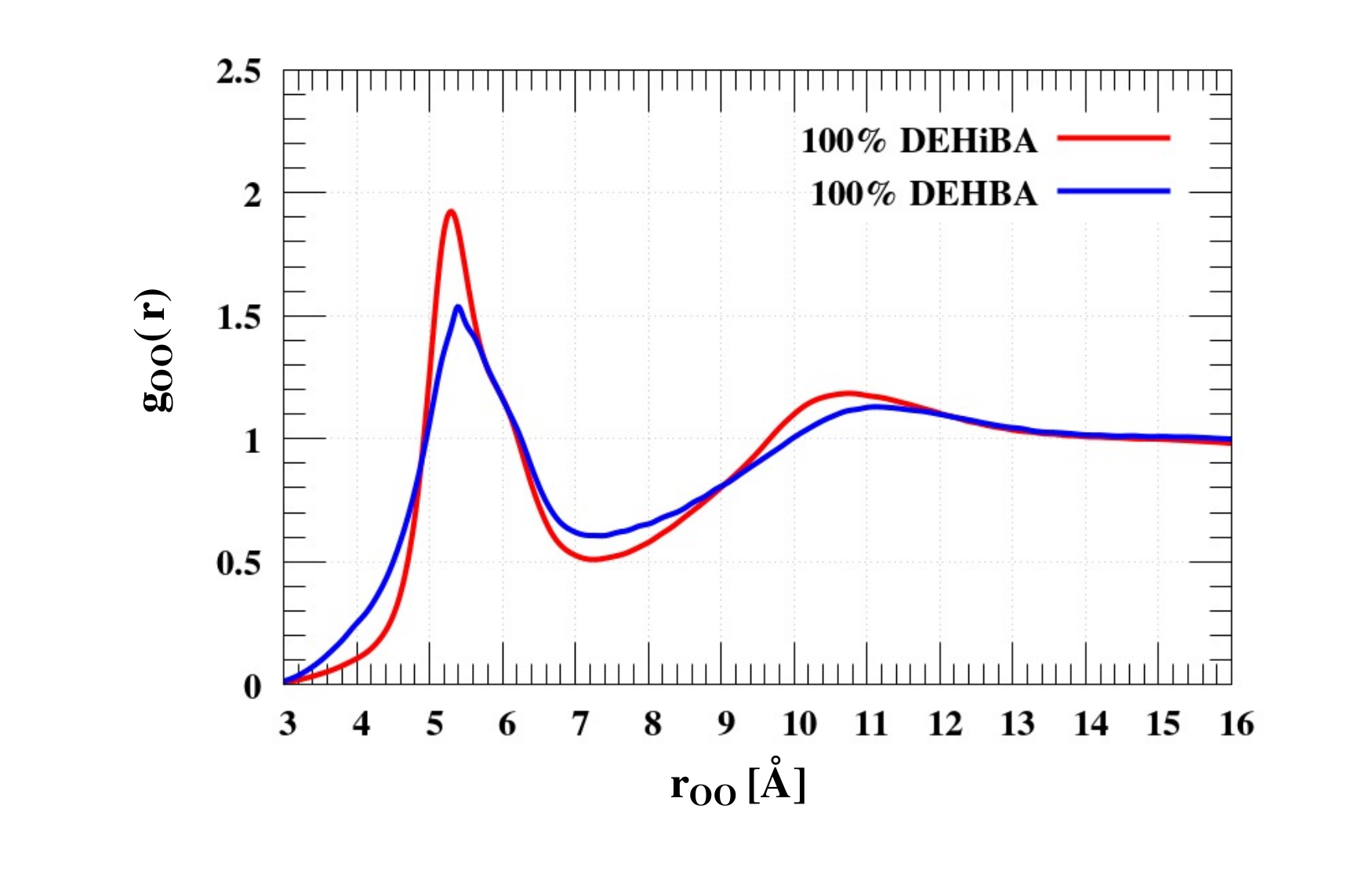}
       \caption{$x_{mono}$ = \SI{100}{\percent}}
       \label{fig:rdf-oo-100pp}
    \end{subfigure}
\caption{Radial distribution functions of oxygen atoms for DEHiBA/dodecane and DEHBA/dodecane mixtures at different monoamide $x_{mono}$ mole fractions.}
\label{rdf-oo-mix}
\end{figure}

\begin{figure}
    \begin{subfigure}[t]{0.49\linewidth}
        \centering
        \includegraphics[width=\linewidth]{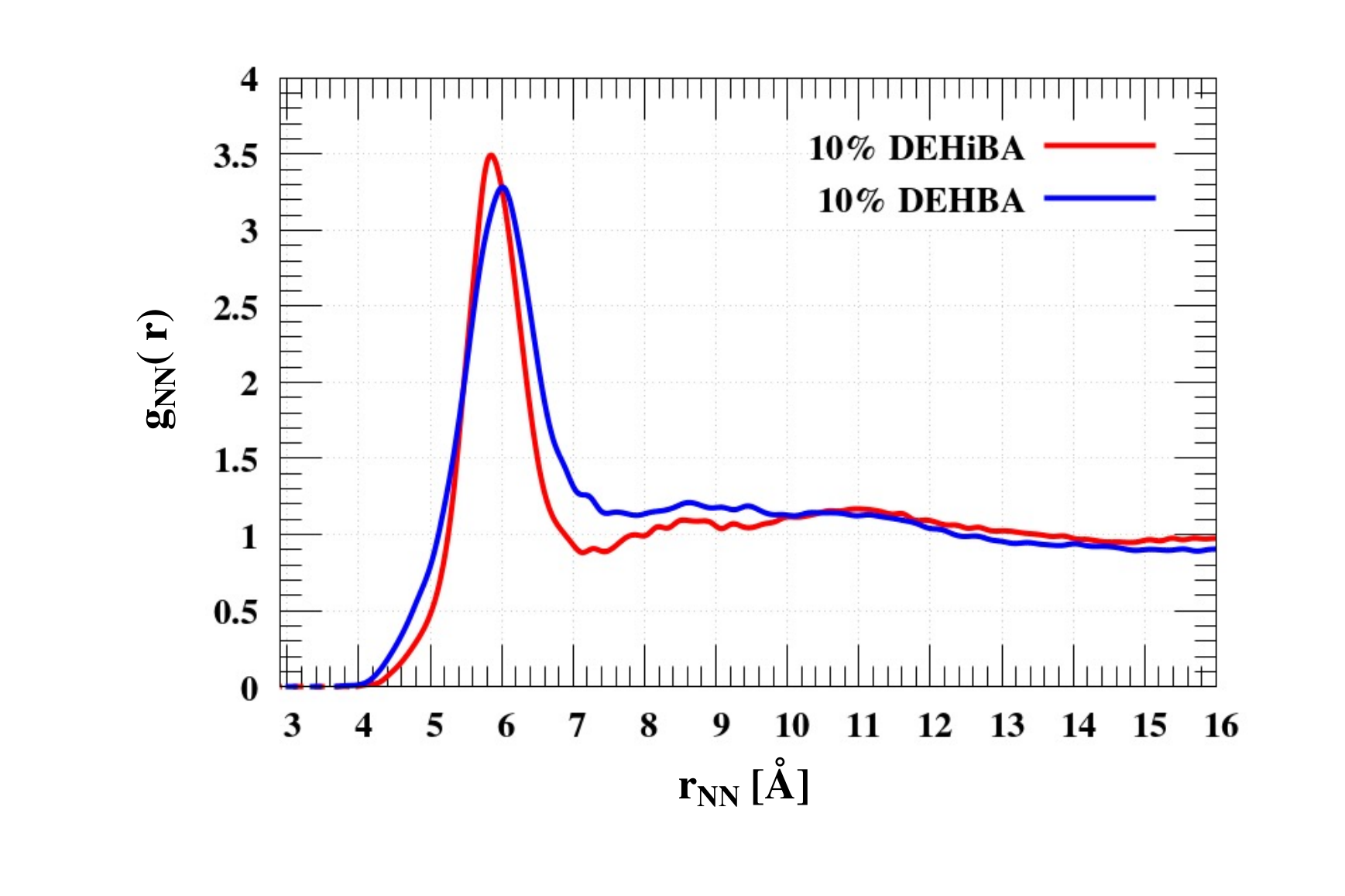}
       \caption{$x_{mono}$ = \SI{10}{\percent}}
       \label{fig:rdf-nn-10pp}
    \end{subfigure}
        \begin{subfigure}[t]{0.49\linewidth}
        \centering
        \includegraphics[width=\linewidth]{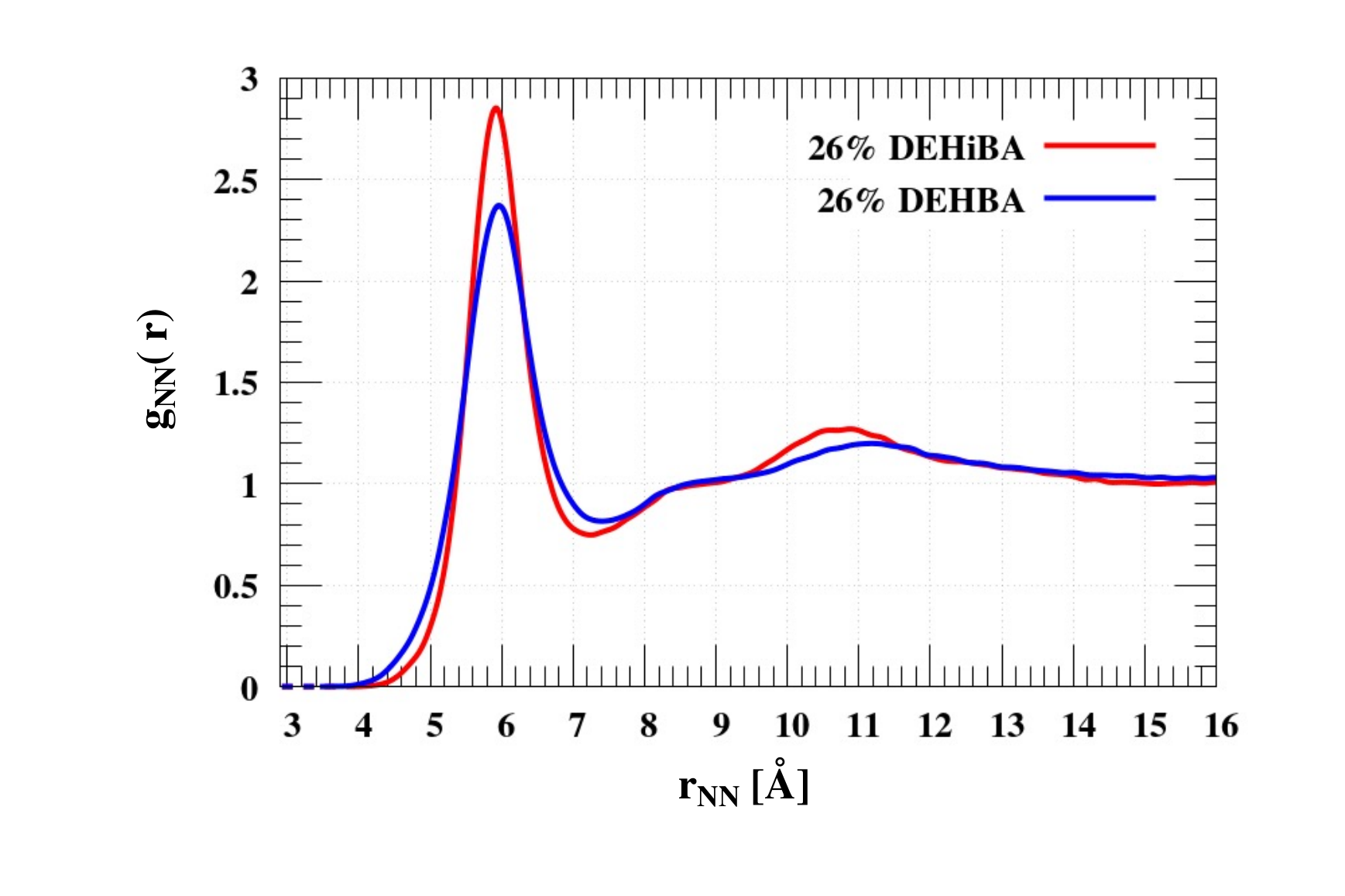}
       \caption{$x_{mono}$ = \SI{26}{\percent}}
       \label{fig:rdf-nn-26pp}
    \end{subfigure}
    \begin{subfigure}[t]{0.49\linewidth}
      \centering
      \includegraphics[width=\linewidth]{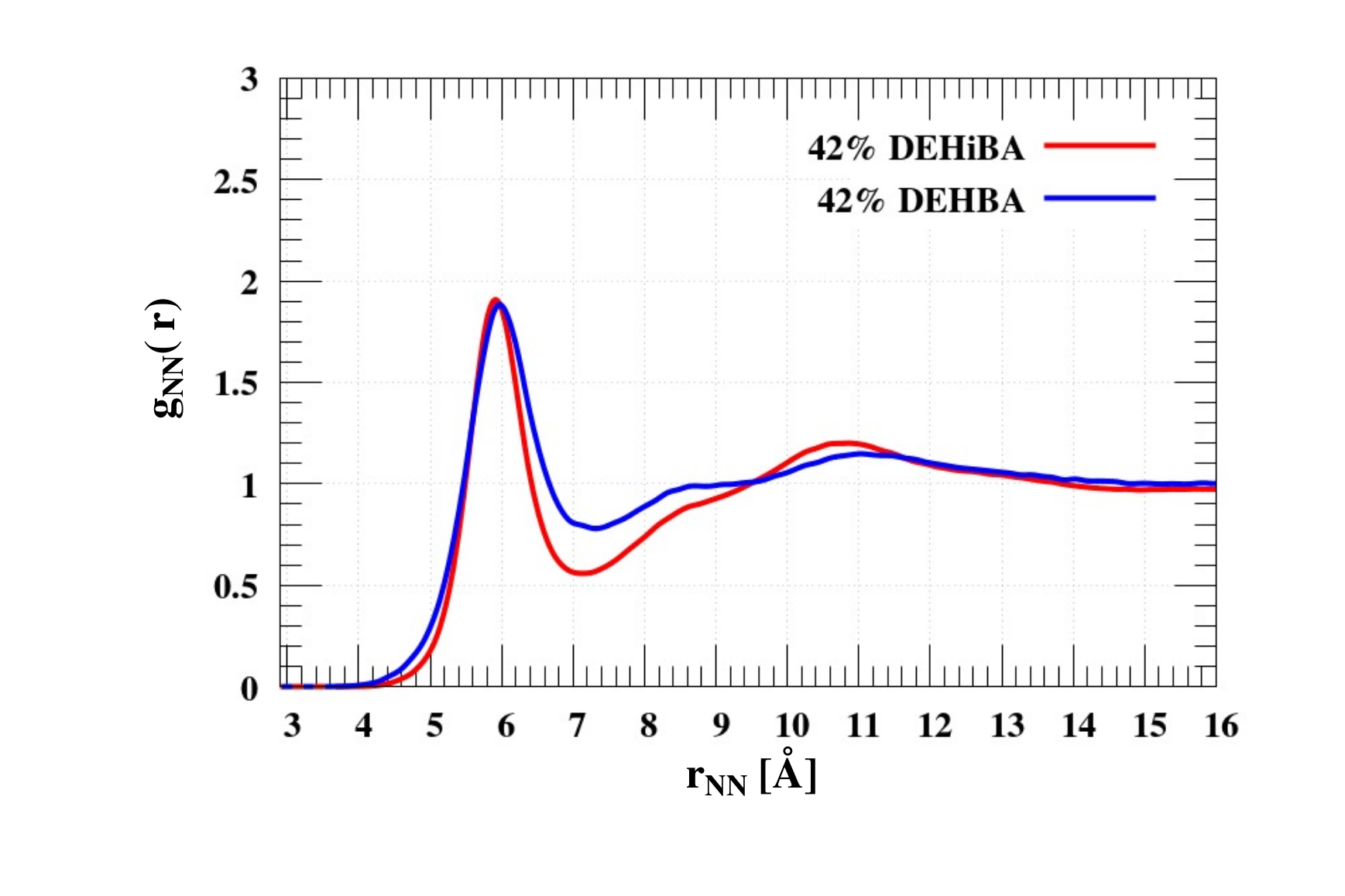}
       \caption{$x_{mono}$ = \SI{42}{\percent}}
       \label{fig:rdf-nn-42pp}
    \end{subfigure}
    \begin{subfigure}[t]{0.49\linewidth}
      \centering
      \includegraphics[width=\linewidth]{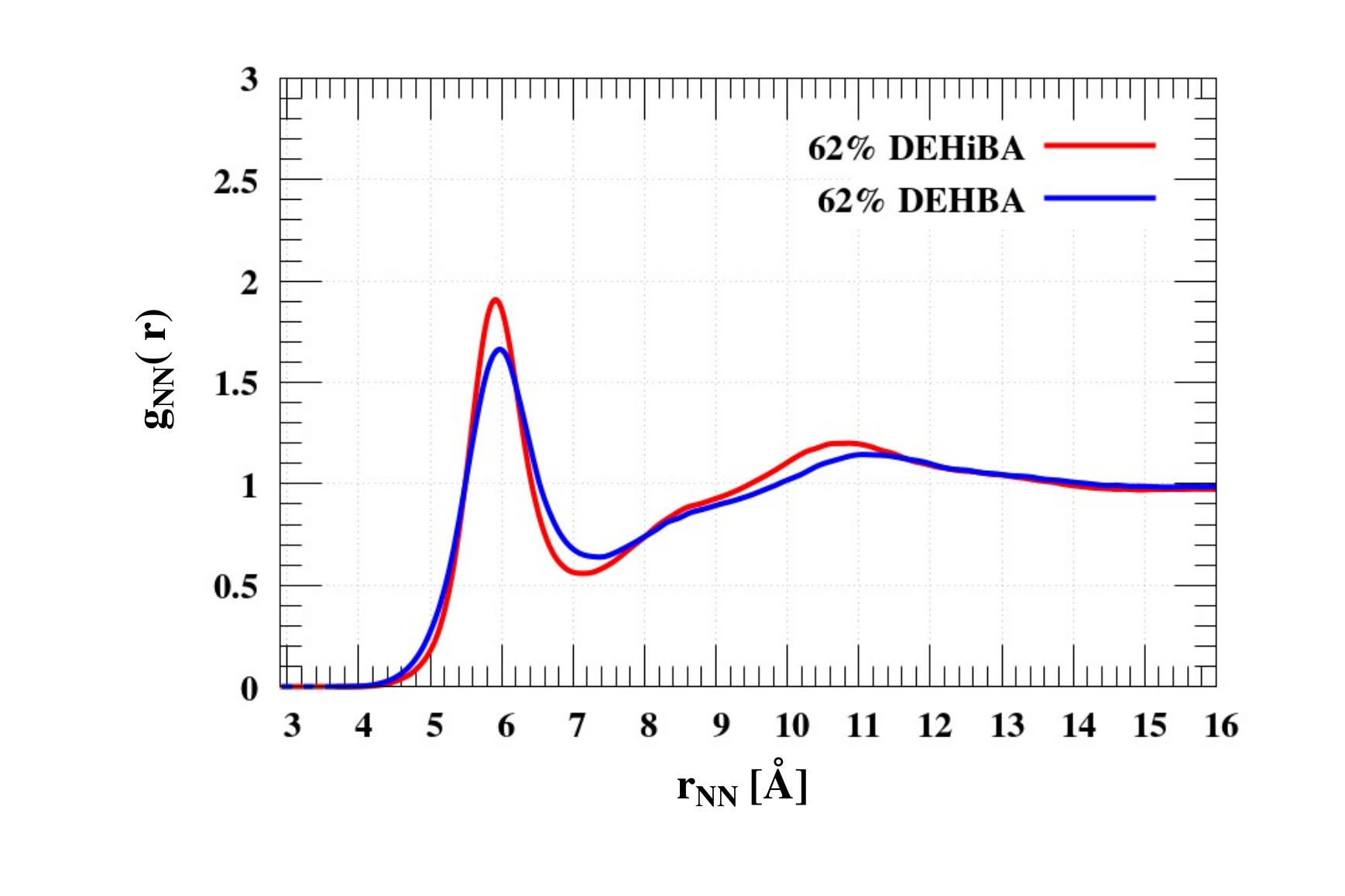}
       \caption{$x_{mono}$ = \SI{62}{\percent}}
       \label{fig:rdf-nn-62pp}
    \end{subfigure}
        \begin{subfigure}[t]{0.49\linewidth}
      \centering
      \includegraphics[width=\linewidth]{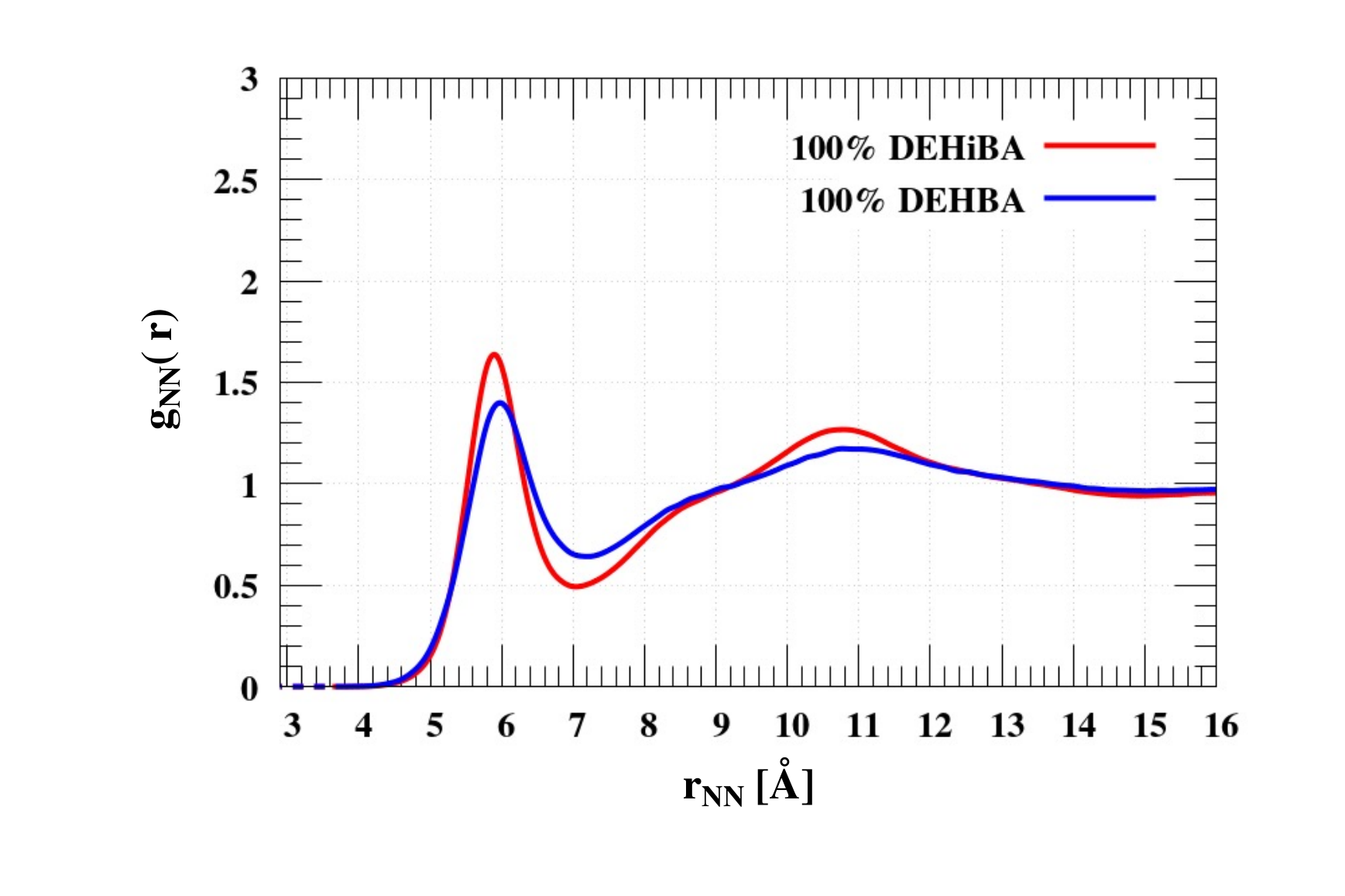}
       \caption{$x_{mono}$ = \SI{100}{\percent}}
       \label{fig:rdf-nn-100pp}
    \end{subfigure}
\caption{Radial distribution functions of nitrogen atoms for DEHiBA/dodecane and DEHBA/dodecane mixtures at different monoamide $x_{mono}$ mole fractions.}
\label{rdf-nn-mix}
\end{figure}
\begin{figure}
    \begin{subfigure}[t]{0.49\linewidth}
        \centering
        \includegraphics[width=\linewidth]{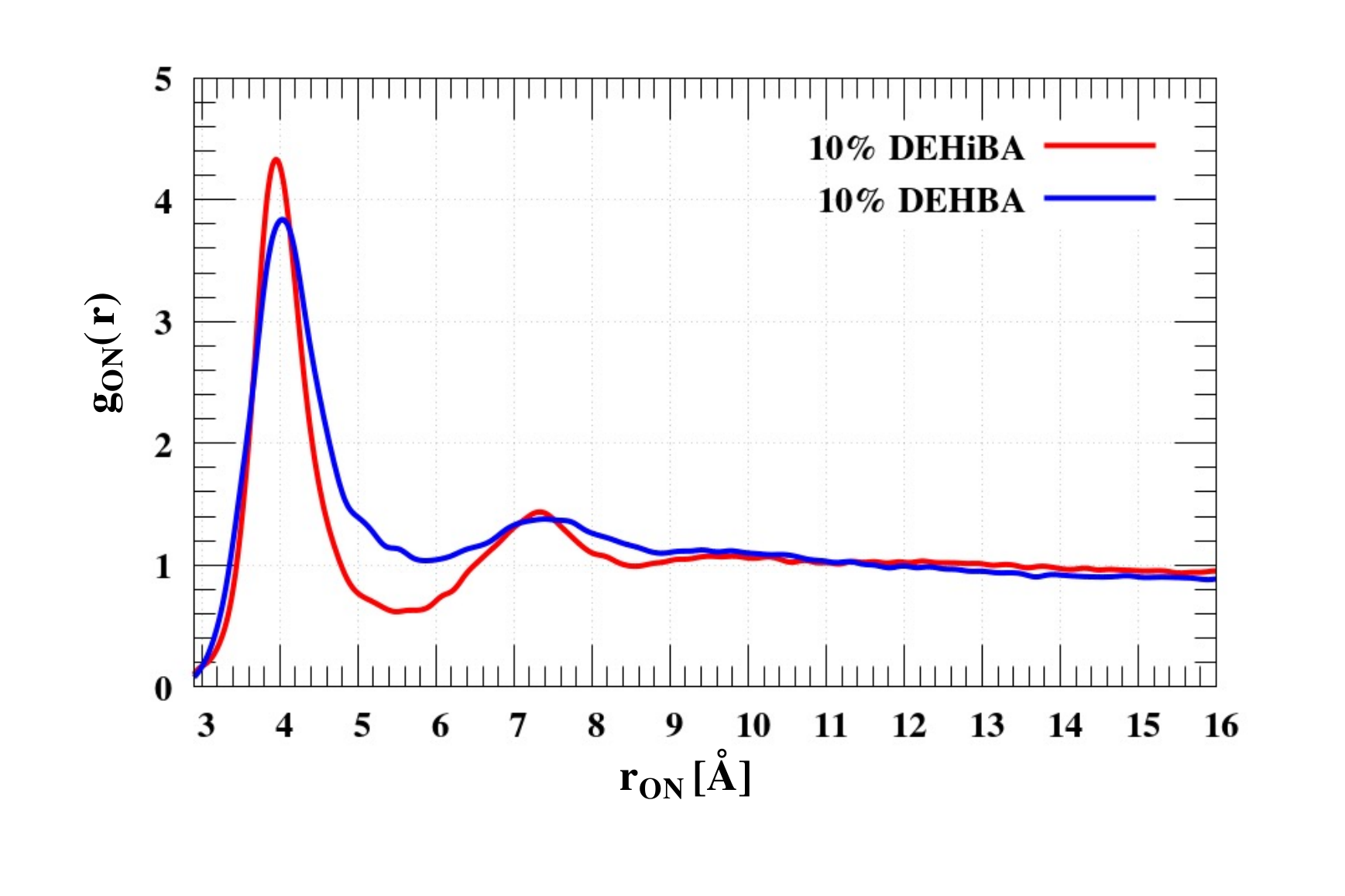}
       \caption{$x_{mono}$ = \SI{10}{\percent}}
       \label{fig:rdf-on-10pp}
    \end{subfigure}
        \begin{subfigure}[t]{0.49\linewidth}
        \centering
        \includegraphics[width=\linewidth]{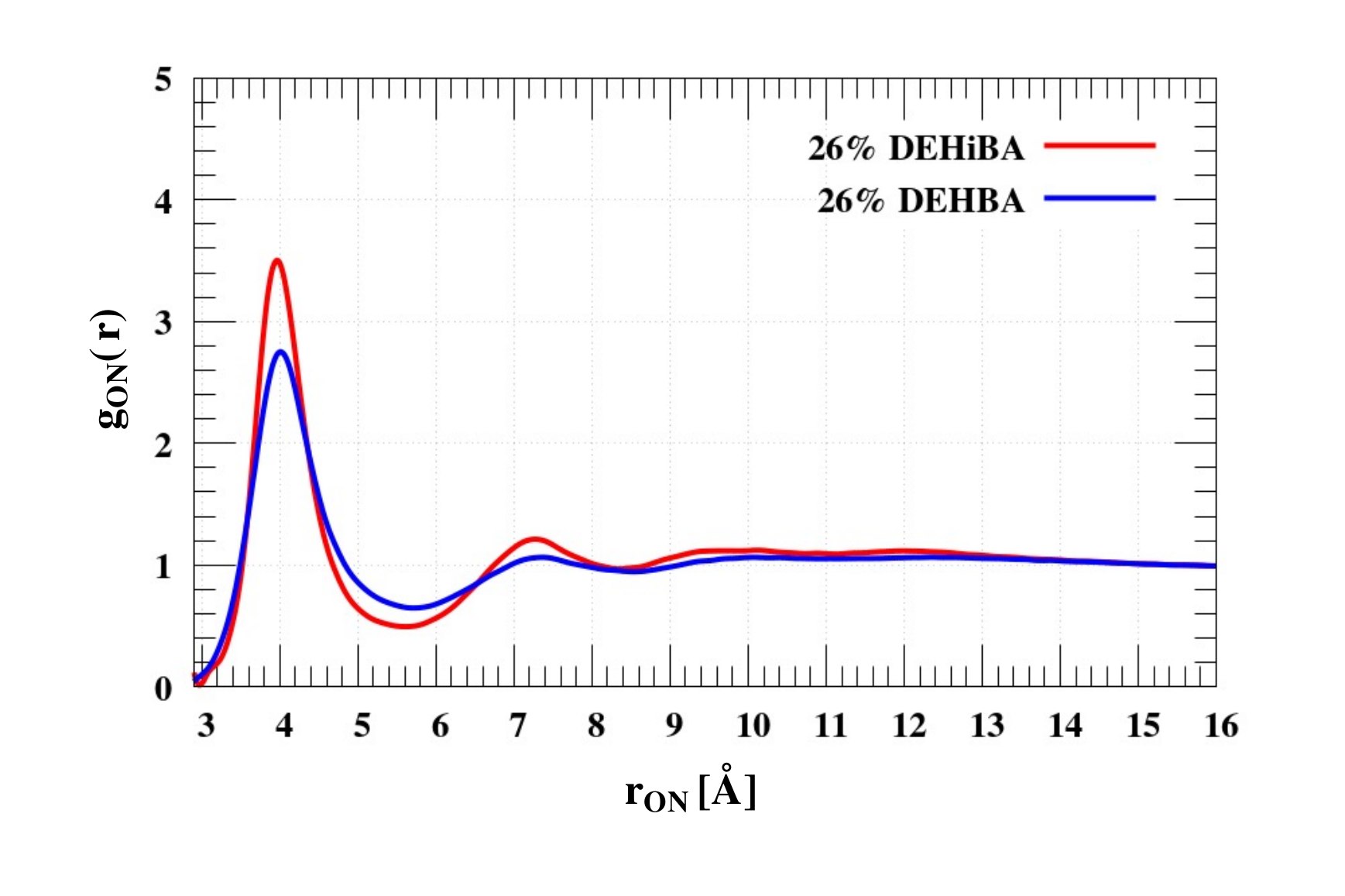}
       \caption{$x_{mono}$ = \SI{26}{\percent}}
       \label{fig:rdf-on-26pp}
    \end{subfigure}
    \begin{subfigure}[t]{0.49\linewidth}
      \centering
      \includegraphics[width=\linewidth]{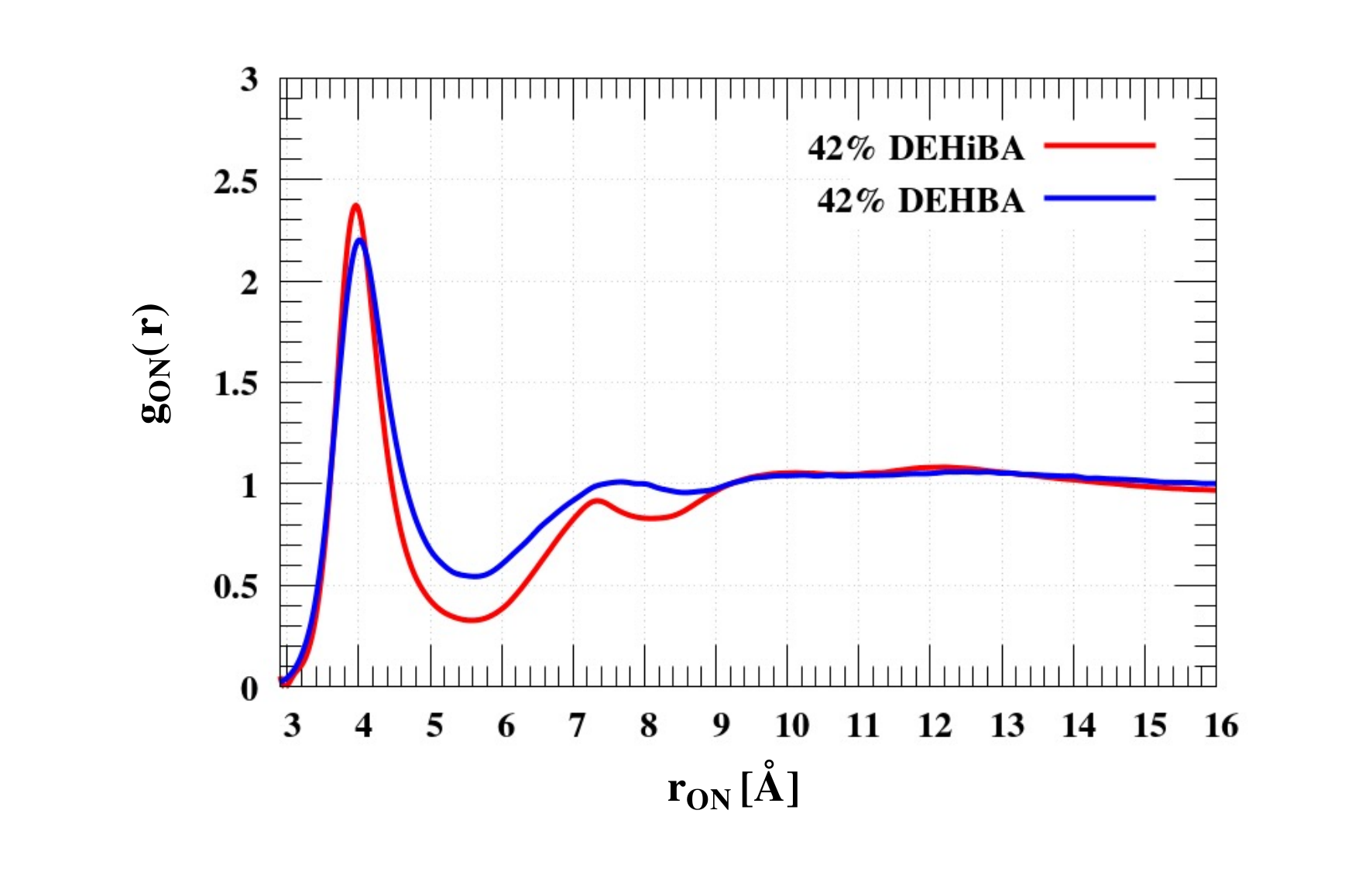}
       \caption{$x_{mono}$ = \SI{42}{\percent}}
       \label{fig:rdf-on-42pp}
    \end{subfigure}
    \begin{subfigure}[t]{0.49\linewidth}
      \centering
      \includegraphics[width=\linewidth]{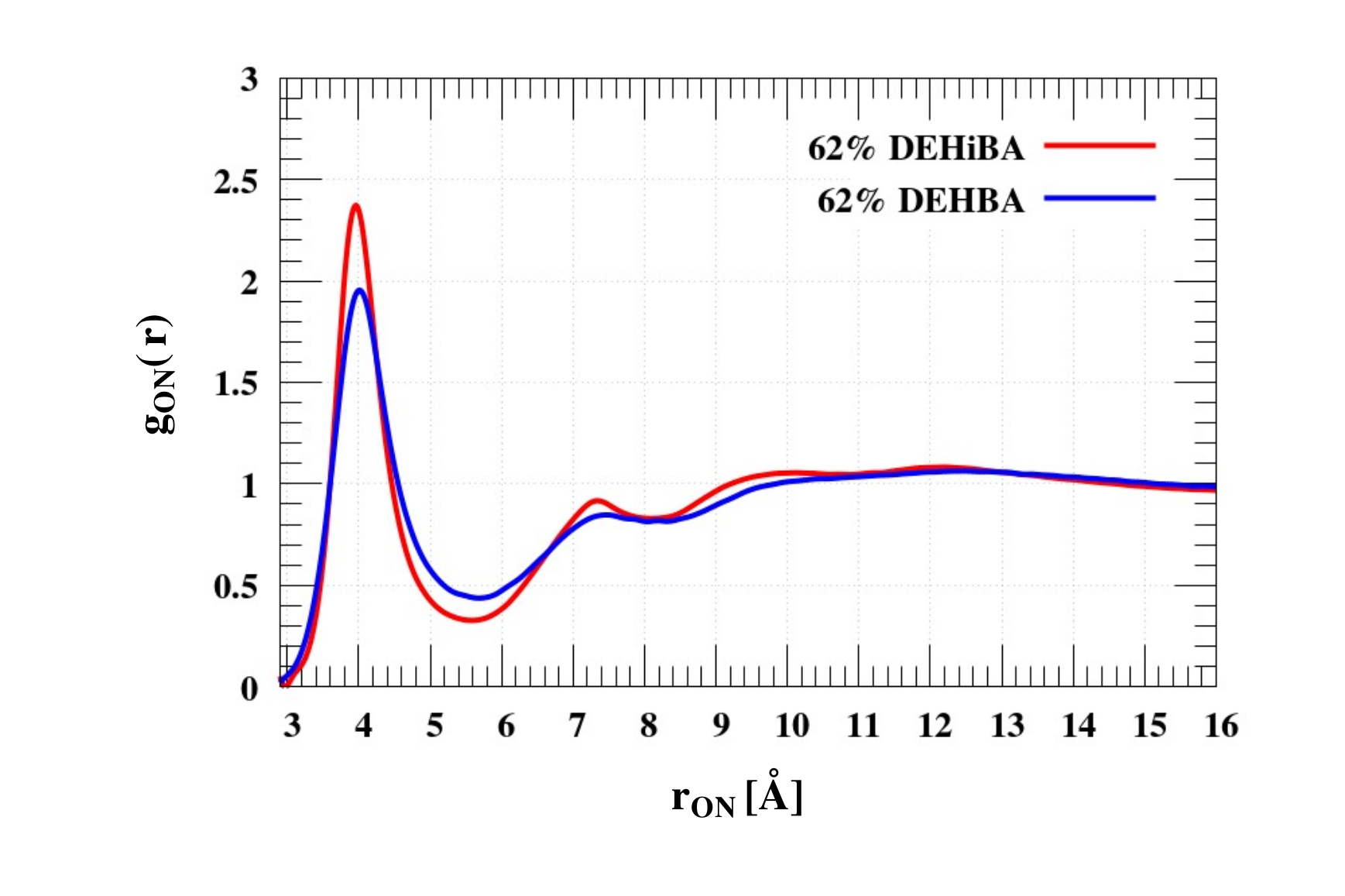}
       \caption{$x_{mono}$ = \SI{62}{\percent}}
       \label{fig:rdf-on-62pp}
    \end{subfigure}
        \begin{subfigure}[t]{0.49\linewidth}
      \centering
      \includegraphics[width=\linewidth]{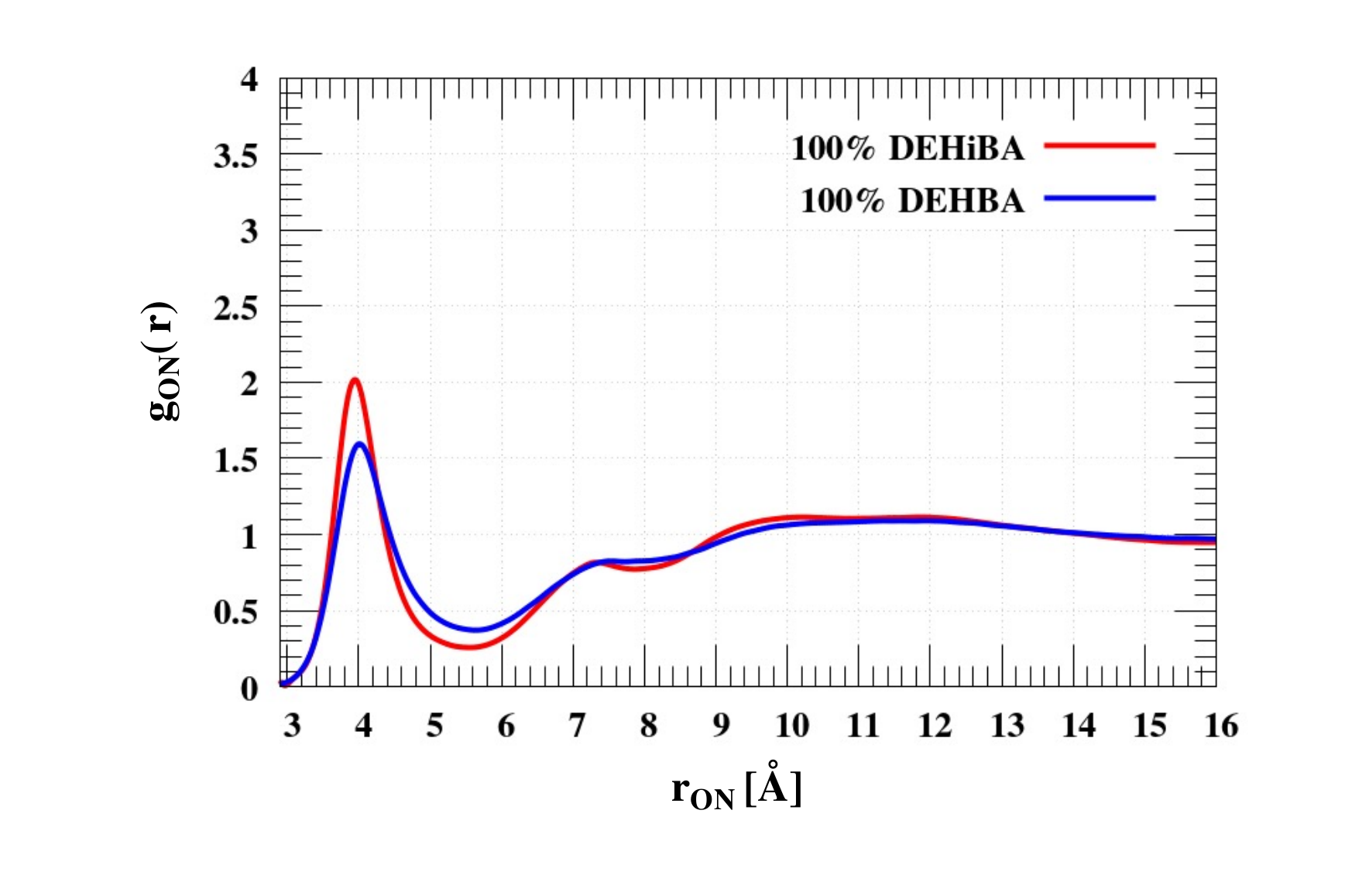}
       \caption{$x_{mono}$ = \SI{100}{\percent}}
       \label{fig:rdf-on-100pp}
    \end{subfigure}
\caption{Radial distribution functions between oxygen and nitrogen atoms for DEHiBA/dodecane and DEHBA/dodecane mixtures at different monoamide $x_{mono}$ mole fractions.}
\label{rdf-on-mix}
\end{figure}

\begin{figure}
\captionsetup[subfigure]{justification=centering}
    \begin{subfigure}[t]{0.47\linewidth}
        \centering
        \includegraphics[width=\linewidth]{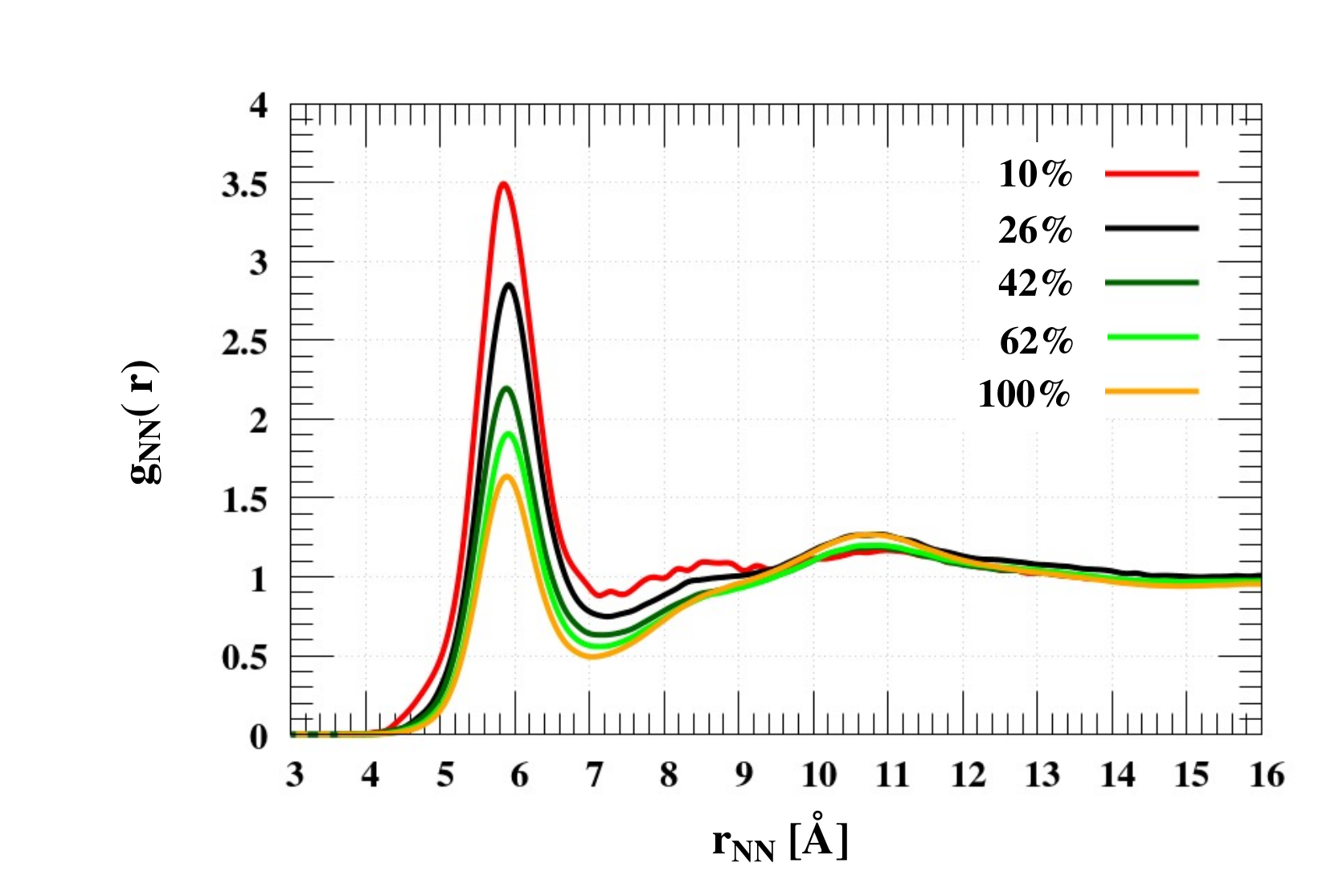}
        \caption{DEHiBA/dodecane mixtures}
    \end{subfigure}
        \begin{subfigure}[t]{0.48\linewidth}
        \centering
        \includegraphics[width=\linewidth]{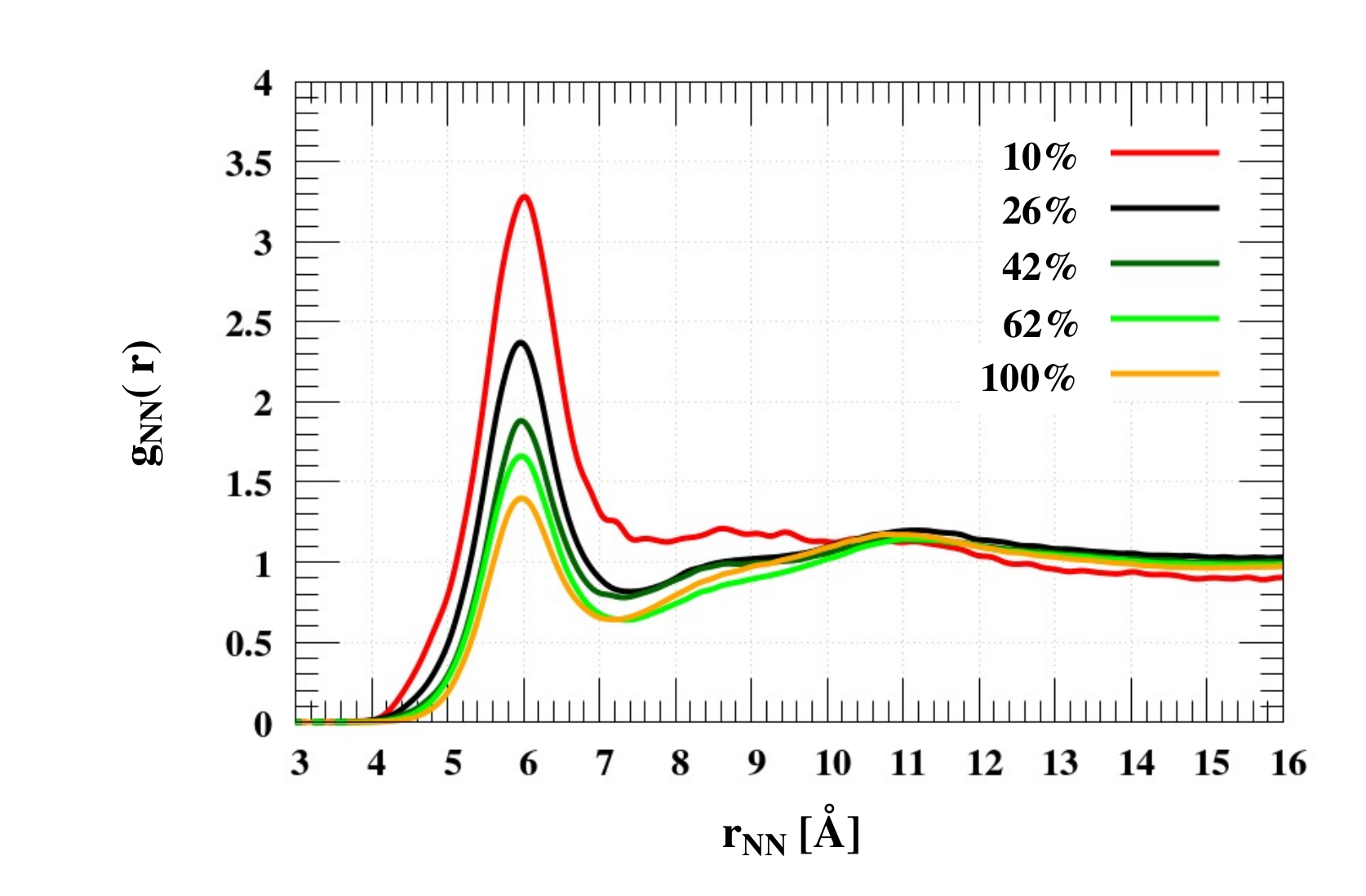}
        \caption{DEHBA/dodecane mixtures}
    \end{subfigure}
\caption{Radial distribution functions of nitrogen atoms for DEHiBA/dodecane and DEHBA/dodecane mixtures at at different monoamide mole fractions ($x_{mono}$=\SIlist{10;26;42;62;100}{\percent}).}
\label{rdf-mix-2}
\end{figure}

\begin{figure}
\captionsetup[subfigure]{justification=centering}
        \begin{subfigure}[t]{0.47\linewidth}
        \centering
        \includegraphics[width=\linewidth]{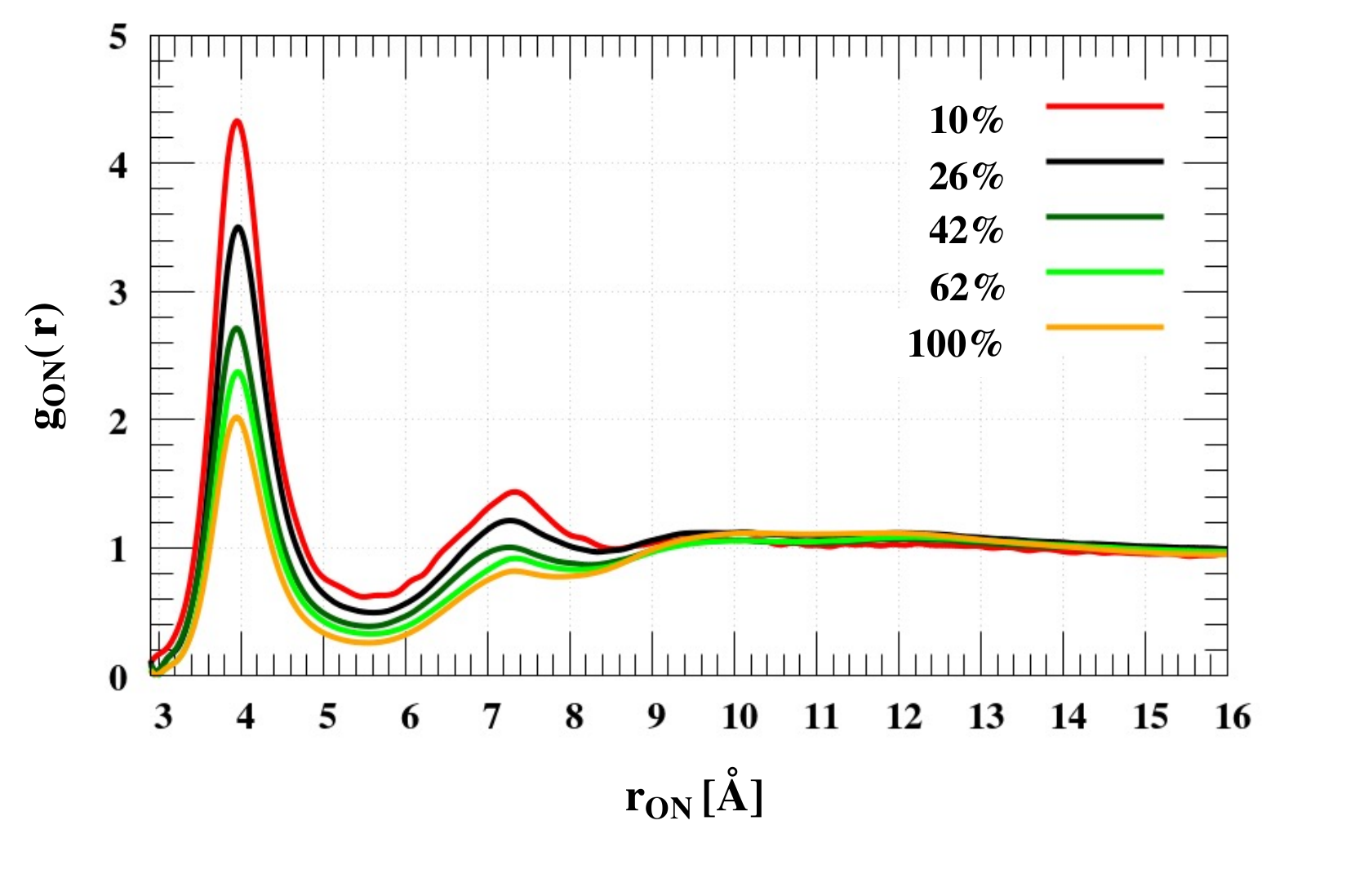}
        \caption{DEHiBA/dodecane mixtures}
    \end{subfigure}
        \begin{subfigure}[t]{0.47\linewidth}
        \centering
        \includegraphics[width=\linewidth]{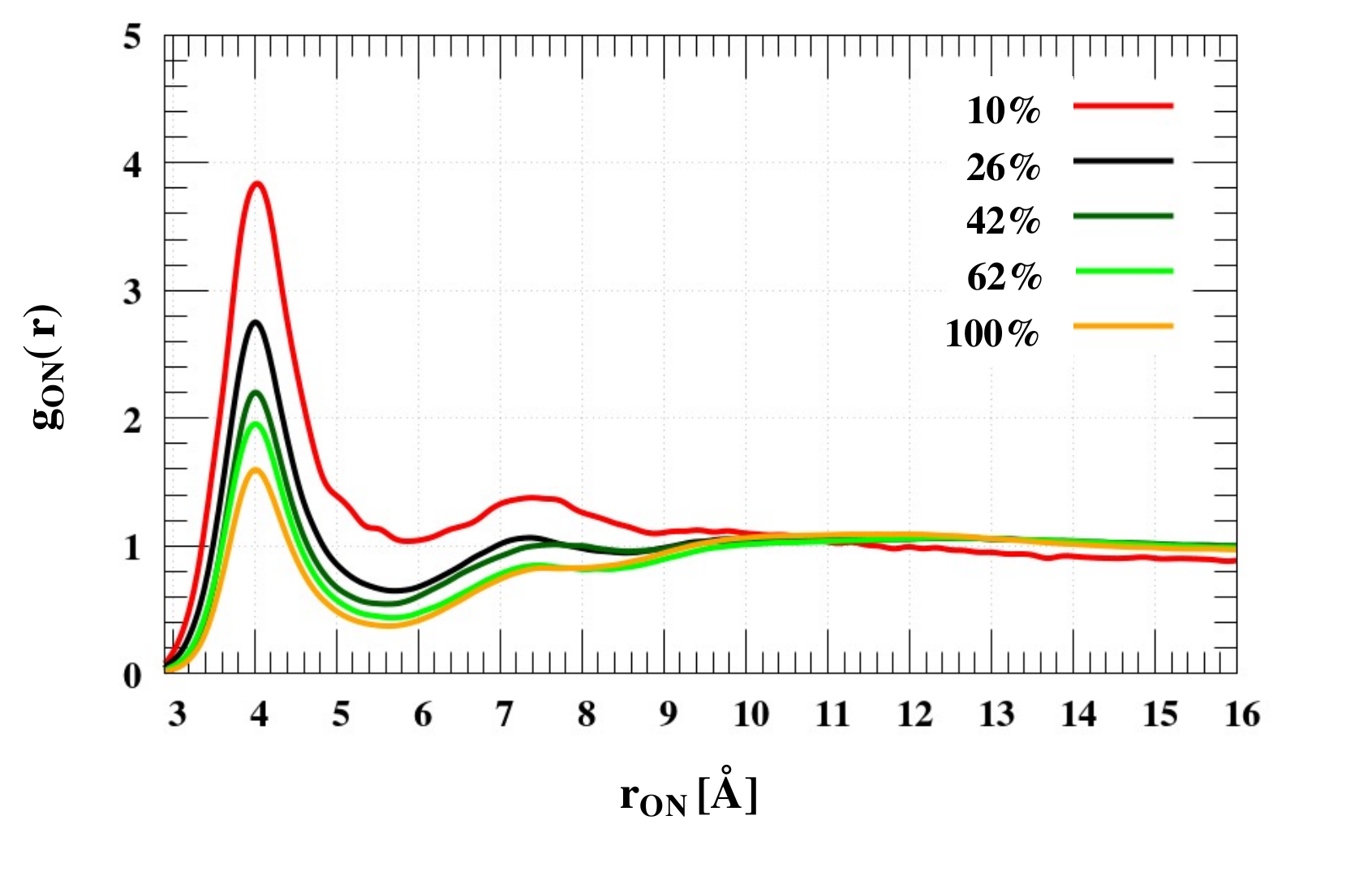}
        \caption{DEHBA/dodecane mixtures}
    \end{subfigure}
\caption{Radial distribution functions between oxygen and nitrogen atoms for DEHiBA/dodecane and DEHBA/dodecane mixtures at at different monoamide mole fractions ($x_{mono}$=\SIlist{10;26;42;62;100}{\percent}).}
\label{rdf-mix-3}
\end{figure}

\begin{figure}
\centering
\includegraphics[width=0.95\linewidth]{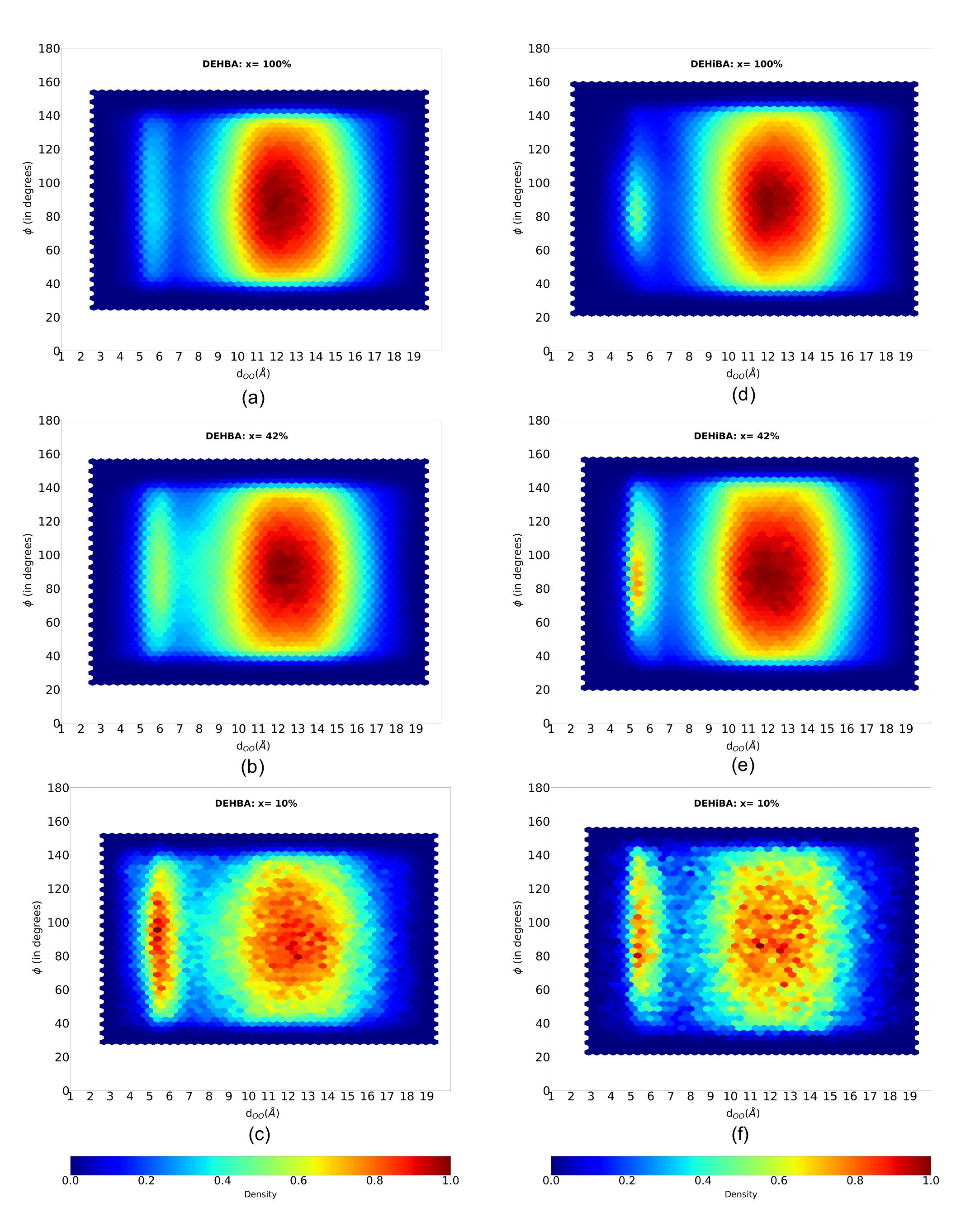}
\caption{Analysis of the relative orientations of the NCO planes, $\phi(r_{OO})$, as a function of the distance between the oxygen of the carbonyl group for DEHBA/dodecane (left column) and DEHiBA/dodecane mixtures (right column).  The color bar indicates the normalized count density. 100\% DEHBA,(b) 42\% DEHBA, (c) 10\% DEHBA, (d) 100\% DEHiBA,(e) 42\% DEHiBA and (f) 10\% DEHiBA. The color bar indicates the normalized count density.}
\label{Sfig:CO-analysis}
\end{figure}

\begin{figure}
\centering
\includegraphics[width=0.95\linewidth]{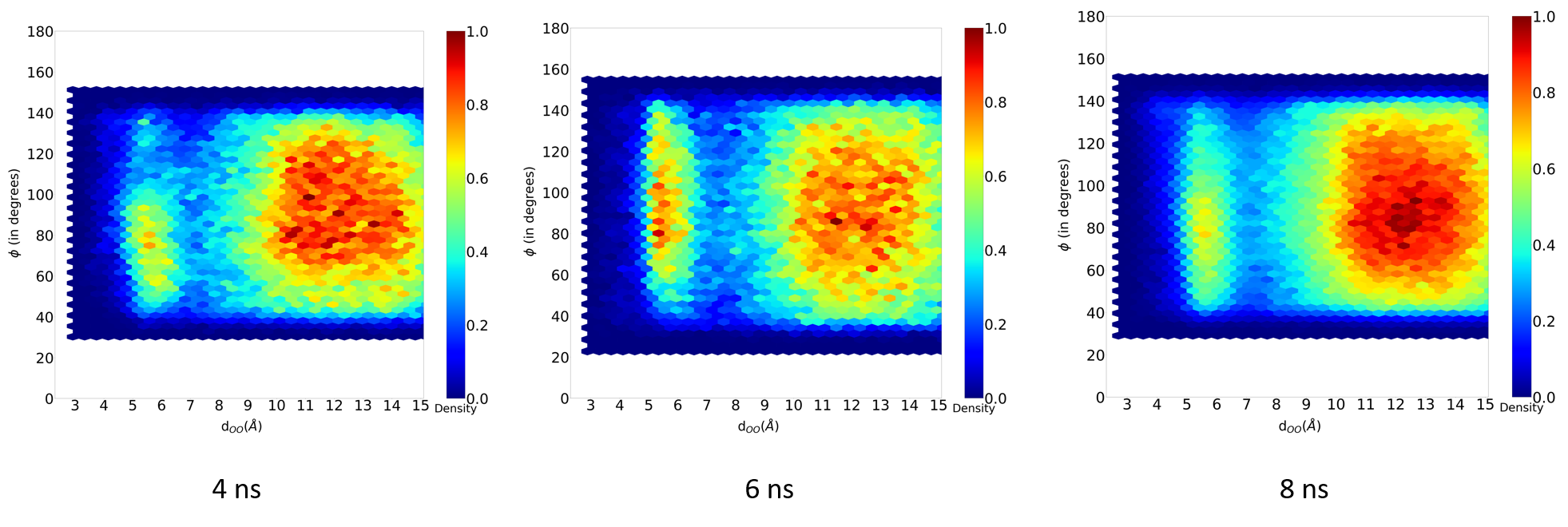}
\caption{Analysis of the relative orientations of the NCO planes, $\phi(r_{OO})$, as a function of the distance between the oxygen of the carbonyl group for 26\% DEHBA in dodecane averaged for \SI{4}{\ns}, \SI{6}{\ns} and  \SI{8}{\ns}  of the equilibration phase.  The color bar indicates the normalized count density. The color bar indicates the normalized count density.}
\label{Sfig:CO-analysis-time}
\end{figure}

\clearpage
\bibliography{refs.bib}